\documentclass[12pt]{article}
\pdfoutput=1

\usepackage{setspace}

\usepackage{graphicx}
\usepackage{epstopdf}
\usepackage[body={18cm, 21cm},right=1.8cm]{geometry}
\usepackage{amssymb}
\usepackage{amsmath}
\usepackage{dcolumn}

\usepackage{graphicx}
\usepackage{epstopdf}
\usepackage{amsmath}
\usepackage{amsfonts}
\usepackage{amssymb}
\usepackage[usenames]{color}
\usepackage{mathrsfs}
\usepackage{array}
\usepackage{nonfloat}

\usepackage{floatrow}
\newfloatcommand{capbtabbox}{table}[][\FBwidth]

\usepackage{caption}

\usepackage{multirow}
\usepackage{psfrag}
\pdfoutput=1

\usepackage[colorlinks,bookmarks]{hyperref}
\definecolor{linkblue}{rgb}{0,0,0.8}
\definecolor{linkgreen}{rgb}{0,0.5,0}

\hypersetup{pdfpagemode=UseNone, pdfstartview=FitH, linkcolor=linkblue, %
            citecolor=linkgreen, urlcolor=linkblue}

\newcommand\eea{\end{eqnarray}}
\newcommand\bea{\begin{eqnarray}}
\newcommand{\sfrac}[2]{{\textstyle\frac{#1}{#2}}}

\def\beq{\begin{equation}}
\def\eeq{\end{equation}}
\def\d{\partial}

\newcommand{\be}{\begin{equation}}
\newcommand{\ee}{\end{equation}}
\newcommand{\ba}{\begin{align}}
\newcommand{\ea}{\end{align}}
\newcommand{\bg}{\begin{gather}}
\newcommand{\eg}{\end{gather}}
\newcommand{\bseq}{\begin{subequations}}
\newcommand{\eseq}{\end{subequations}}

\newcommand{\knl}{k_{\rm NL}}

\def\H{{\cal H}}
\newcommand{\hinvMpc}{h\,$Mpc$^{-1}}

\newcommand{\q}{\vec{q}}

\definecolor{purple}{rgb}{0.78,0.18,0.77}

\definecolor{verde}{rgb}{0,0.5,0}
\newboolean{comment}
\setboolean{comment}{true}
\newcommand{\comment}[2]{\ifthenelse{\boolean{comment}}{\textcolor{red}{\textsf{\textbf{{#1}}}: }\textcolor{verde}{{#2}}}{}\\}

\def\epssm{\epsilon_{s<}}

\def\tin{{t_{\rm in}}}
\def\xfl{{\vec x_{\rm fl}}}
\def\xflc{{\vec x_{\rm fl,c}}}
\def\xflb{{\vec x_{\rm fl,b}}}
\def\km{{k_{\rm M}}}

\def\fnl{{f_{\rm NL}}}
\def\fnlloc{{f_{\rm NL}^{\rm loc}}}
\def\fnlequil{{f_{\rm NL}^{\rm equil}}}
\def\fnlorthog{{f_{\rm NL}^{\rm orthog}}}

\makeatletter
\newlength{\apb@width}
\newcommand{\autoparbox}[2][c]{\settowidth{\apb@width}{#2}\parbox[#1]{\apb@width}{#2}}

\makeatother

\allowdisplaybreaks

\usepackage{hyperref}
\usepackage{ifthen}
\usepackage{amsmath}
\usepackage{amssymb}
\usepackage{color}
\usepackage{graphicx}

\renewcommand{\vec} {\textbf}
\newcommand{\V} {\textbf}
\newcommand*{\df}  {\delta}
\newcommand*{\tf}  {\theta}

\newcommand*{\non}  {\nonumber}

\def\xfl{{\textbf{x}_{\rm fl}}}
\def\xflc{{\textbf{x}_{\rm fl,c}}}
\def\xflb{{\textbf{x}_{\rm fl,b}}}
\def\km{{k_{\rm M}}}

\begin{document}

\vspace{5mm}
\vspace{0.5cm}
\begin{center}

\def\thefootnote{\fnsymbol{footnote}}
{\Large \bf On the Statistics of Biased Tracers\\[0.5cm]
in the Effective Field Theory of Large Scale Structures
}
\\[0.8cm]

{\large Raul Angulo${}^{1}$, Matteo Fasiello${}^{2,3}$, Leonardo Senatore${}^{2,3}$ and Zvonimir Vlah${}^{2,3}$}
\\[0.5cm]

{\normalsize {${}^1$ \sl Centro de Estudios de Fisica del Cosmos de Aragon,\\ Plaza San Juan 1, Planta-2, 44001, Teruel, Spain
}}\\
\vspace{.3cm}

{\normalsize {${}^2$ \sl Stanford Institute for Theoretical Physics and Department of Physics, \\Stanford University, Stanford, CA 94306}}\\
\vspace{.3cm}

{\normalsize {${}^3$ \sl Kavli Institute for Particle Astrophysics and Cosmology, \\ Stanford University and SLAC, Menlo Park, CA 94025}}\\
\vspace{.3cm}

\end{center}

\vspace{.8cm}

\hrule \vspace{0.3cm}
{\small  \noindent \textbf{Abstract} \\[0.3cm]
\noindent With the completion of the Planck mission, in order to continue to 
gather cosmological information it has become crucial to understand the 
Large Scale Structures (LSS) of the universe to percent accuracy. 
The Effective Field Theory of LSS (EFTofLSS) is a novel theoretical framework 
that aims to develop an analytic understanding of LSS at long distances, 
where inhomogeneities are small. We further develop the description of biased 
tracers in the EFTofLSS to account for the effect of baryonic physics and primordial 
non-Gaussianities, finding that new bias coefficients are required. 
Then, restricting to dark matter with Gaussian initial conditions, 
we describe the prediction of the EFTofLSS for the one-loop halo-halo and 
halo-matter two-point functions, and for the tree-level halo-halo-halo, 
matter-halo-halo and matter-matter-halo three-point functions. 
Several new bias coefficients are needed in the EFTofLSS, 
even though their contribution at a given order can be degenerate 
and the same parameters contribute to multiple observables. 
We develop a method to reduce the number of biases to an 
irreducible basis, and find that, at the order at which we work, 
seven bias parameters are enough to describe this extremely 
rich set of statistics. 
We then compare with the output of an $N$-body simulation where
the normalization parameter of the linear power 
spectrum is set to $\sigma_8 = 0.9$.
For the lowest mass bin, we find percent level agreement 
up to $k\simeq 0.3\,h\,{\rm Mpc}^{-1}$ for the one-loop two-point  
functions, and up to $k\simeq 0.15\,h\,{\rm Mpc}^{-1}$ for the 
tree-level three-point  functions, with the $k$-reach decreasing with higher mass bins. 
This is consistent with the theoretical estimates, and suggests that the cosmological 
information in LSS amenable to analytical control is much more than previously believed. 
\noindent 
}

 \vspace{0.3cm}
\hrule
\def\thefootnote{\arabic{footnote}}
\setcounter{footnote}{0}

\vspace{.8cm}

\newpage
\tableofcontents




\section{Introduction}
\label{sec:intro}

With the completion of the Planck mission, most of the future cosmological information will probably need to come from Large Scale Structures (LSS). It is therefore essential to understand their behavior. In particular, it is of great interest to explore until what high $k$-value we can push our theoretical control. This is important because it gives an indication of how much information we have hope to extract about the primordial universe from LSS surveys.

The Effective Field Theory of Large Scale Structures (EFTofLSS)~\cite{Baumann:2010tm,Carrasco:2012cv,Carrasco:2013sva,Carrasco:2013mua,Pajer:2013jj,Carroll:2013oxa,Porto:2013qua,Mercolli:2013bsa,Senatore:2014via,Angulo:2014tfa,Baldauf:2014qfa,Senatore:2014eva,Senatore:2014vja,Lewandowski:2014rca,Mirbabayi:2014zca,Foreman:2015uva} 
is a theoretical framework that aims at developing a rigorous analytic understanding of  the clustering of LSS, through a perturbative expansion in the size of the fluctuations. It does so by correctly parametrizing the effect that short distance non-linearities have at large distances through an effective stress tensor where some parameters, such as the speed of sound, the viscosity, etc, need to be measured either from observations or from numerical simulations.

So far, the EFTofLSS has been compared to data only for the case of dark matter correlation functions, specifically the 2- and 3-point functions at redshift $z=0$~\cite{Carrasco:2012cv,Carrasco:2013sva,Senatore:2014eva,Angulo:2014tfa,Baldauf:2014qfa}, and the 2-pt functions has just been studied at all redshift in~\cite{Foreman:2015uva} . In these cases, the results have been extremely promising. For the power spectrum, former  analytical techniques start failing at $k\simeq 0.1\hinvMpc$~\footnote{By former analytic techniques, we mean Standard Perturbation Theory (SPT). In the literature there exist several versions of perturbations theory where a subset of the SPT diagrams is resummed, such as Renormalized Perturbation Theory~(RPT)~\cite{Crocce:2005xy}, or the Lagrangian formulation (see for example~\cite{Matsubara:2007wj,Carlson:2012bu}). As explained in~\cite{Senatore:2014eva}, when correctly implemented, these resummation techniques should give the same UV reach as SPT for the considered quantities.}, while the EFT, with the inclusion of only one single parameter, fails at $k\simeq 0.3\hinvMpc$ at one-loop order, and at $k\simeq 0.6\hinvMpc$ at two-loops. Consistently, the dark matter bispectrum at one-loop order agrees until $k\simeq 0.3\hinvMpc$ by using the same single parameter~\cite{Angulo:2014tfa}. Similarly, and consistently, the momentum power spectrum has been shown to agree with data up to $k\simeq 0.3\hinvMpc$ at one-loop order, in practice using just the same parameter~\cite{Senatore:2014via}. The slope of the power spectrum of the vorticous component of the velocity has also been correctly predicted~\cite{Carrasco:2013sva,Mercolli:2013bsa}.  A recent extension of the EFTofLSS to two species has shown that baryonic effects can be consistently included in the EFT, and this leads to the same UV reach when star formation physics is present as for dark-matter only simulations with the addition of just one parameter, the speed of sound of baryons~\cite{Lewandowski:2014rca}. 

These results are extremely promising because the available number of modes scales like the cube of the maximum wavenumber, so that an improvement of a factor of six in the $k$-reach translates in a factor of about 200 in the number of available modes. Such a gain, if confirmed for all observables in LSS, would revolutionize our expectations for what we are going to learn from LSS about the primordial universe. 

Comparably important is the fact that in the EFTofLSS it is possible to estimate the size of the terms that are not included in a calculation at a given order, so that one can forecast the UV reach of a given calculation, within order one numbers. The results obtained so far with the EFTofLSS meet this requirement.  Furthermore, since the EFTofLSS represents the description of a dynamical system, the same parameters affect many observables, so that the predictive power of the theory is still very large even in the event that one were to measure the parameters of the EFT directly from observations rather than from simulations. These are characteristics that distinguish a theory, such as the EFTofLSS, from a model.

The purpose of this work is to compare the predictions of the EFTofLSS for halos to numerical simulations. The relevant formulas for biased tracers in the EFTofLSS were worked out in~\cite{Senatore:2014eva}, importantly  generalizing and completing the treatment of~\cite{McDonald:2009dh}. The construction of~\cite{Senatore:2014eva} writes down all the bias coefficients that connect the distribution of galaxies or halos to the one of dark matter.  In particular in~\cite{Senatore:2014eva} it was pointed out that the distribution of collapsed objects at a given time does not depend on the underlying dark matter field at the same instant of time, but on the whole history of the dark matter field. This is forced by the fact that the formation time for a gravitationally collapsed object is of order of the Hubble scale, which is the same time scale over which long wavelength modes evolve.    This has the consequence that the collapsed objects depend on the underlying dark matter field through a different coefficient according to the order at which the dark matter field is evaluated. Equivalently, the collapsed objects depend also on the convective time derivatives of the dark matter field, with the peculiarity that this time derivatives are unsuppressed (as the theory is non-local in time). 

As already pointed out in~\cite{Senatore:2014eva}, at a given order in perturbation theory, the contribution of some of these terms are potentially degenerate. In Sec.~\ref{sec:equations} we will develop a method to reduce the basis of terms in~\cite{Senatore:2014eva} to a set of linearly independent ones (see also~\cite{Mirbabayi:2014zca} where an analogous irreducible basis has recently been constructed from the basis in~\cite{Senatore:2014eva} in a way that is different, and maybe more elegant, from the one we do here). This is not quite the full simplification we will find. In fact, some terms become degenerate only in some observables. For example, we will find that several cubic bias terms, which are independent at the level of fields, contribute in a degenerate way to the power spectrum. 

In this paper, we will mainly limit ourselves to study at redshift $z=0$ the halo-halo and halo-dark matter 2-point functions, $P_{hh}$ and $P_{hm}$, at one-loop order, and all the bispectra constructed from the halo field and the dark matter field, $B_{hhh},\,B_{hhm},\,B_{hmm}$,  at tree level. After this study, we will find that seven bias parameters are necessary to describe these quantities at the order we wish. We therefore expect that the UV reach will be about $k\simeq 0.3\hinvMpc$ for one loop quantities, and $k\simeq 0.1\hinvMpc$ for tree level quantities. At the order we work, we find that the terms we need to include are the same that were already present in~\cite{McDonald:2009dh}. This is just an accident of perturbation theory. 
The new terms that are identified in~\cite{Senatore:2014eva} represent a relevant physical effect that is not present in the formalism of~\cite{McDonald:2009dh}. 
They will therefore appear at an higher order in perturbation theory.  
If we were to consider the tree level trispectra ($T_{hhhh},\, T_{hhhm},\, T_{hhmm},\, T_{hmmm}$), the degeneracies leading to our current seven bias parameters get broken and we 
end up with fourteen bias parameters. It is plausible however, given the huge amount of data and highly non-trivial functional form of the trispectra,  that 
these fourteen bias parameters can be well determined by just looking at the trispectra (and maybe also at the bispectra), so that they can be used for predicting the power spectra and potentially also at some level the bispectra. 
Unfortunately we do not have at our disposal these trispecta data to compare with, but it seems however quite clear that, given the huge amount of data, it is possible to measure the required bias parameters that we mention from data, without losing much cosmological information. 
The relevant formulas and the estimates for the $k$-reach are presented in sec.~\ref{sec:equations}.

Before passing to the comparison with simulation data, we will extend the description of biased tracers in the EFofLSS of~\cite{Senatore:2014eva} to the case where baryons are present and to the case where the primordial non-Gaussianities are non-negligible. We will see that interesting new terms are allowed in these cases.

Finally, in sec.~\ref{sec:results} we compare our predictions with numerical simulations. We find that the results are quite nice, and in agreement with the theoretical expectations.

\section{Biased objects in the EFTofLSS}
\label{sec:equations}

\subsection{Overdensity of biased objects}
\label{subsec:field}

We start from the relation between halos and the underlying dark matter field. As we discussed in the introduction, 
the halo overdensities at a given location depend on the dark matter field and its derivatives, in the combinations 
allowed by the equivalence principle, through their past trajectory. Following~\cite{Senatore:2014eva}, we have for the 
overdensity of halos in Eulerian space:
\bea\label{eq:euler_bias_3}
&&\delta_h(\vec x,t)\simeq \int^t dt'\; H(t')\; \left[   \bar c_{\delta}(t,t')\; :\delta(\xfl,t'): \right.\\  \nonumber
&&\qquad+\bar c_{\delta^2}(t,t')\;  :\delta(\xfl,t')^2:  +\bar c_{s^2}(t,t')\;  :s^2(\xfl,t'):\\\nonumber
&&\qquad+\bar c_{\delta^3}(t,t')\;   :\delta(\xfl,t')^3 : +\bar c_{\delta s^2}(t,t')\;   : \delta(\xfl,t') s^2(\xfl,t'):+\bar c_{\psi}(t,t')\;   :\psi(\xfl,t'):\\ \nonumber
&&\qquad\qquad+\bar c_{\delta st}(t,t') \;   :\delta(\xfl,t') st(\xfl,t'):+\bar c_{ s^3}(t,t')\;     :s^3(\xfl,t'):\\\nonumber
&&\qquad+ \bar c_{\epsilon}(t,t')\;\epsilon(\xfl,t')\\ \nonumber
&&\qquad+\bar c_{\epsilon\delta}(t,t') \;:\epsilon(\xfl,t')\delta(\xfl,t'):+\bar c_{\epsilon  s}(t,t') \;:\epsilon s(\xfl,t'):+\bar c_{\epsilon  t }(t,t') \;:\epsilon t(\xfl,t'):\\\ \nonumber
&&\qquad+\bar c_{\epsilon^2\delta}(t,t') \;:\epsilon(\xfl,t')^2\delta(\xfl,t'):
+\bar c_{\epsilon\delta^2}(t,t') \;:\epsilon(\xfl,t')\delta(\xfl,t')^2:+\bar c_{\epsilon s^2}(t,t') \;:\epsilon(\xfl,t')s^2(\xfl,t'): \\ \nonumber
&&\qquad\qquad+\bar c_{\epsilon s \delta}(t,t') \;:\epsilon s(\xfl,t')\delta(\xfl,t'):+\bar c_{\epsilon  t \delta}(t,t') \;:\epsilon t(\xfl,t')\delta(\xfl,t'):\\ \nonumber
&&\left.\qquad+  \bar c_{\d^2\delta}(t,t')\;    \;\frac{\d^2_{x_{\rm fl}}}{\km^2}\delta(\xfl,t')+\dots\ \right] ,
\eea
Let us explain the notation. $\xfl$ is defined recursively as
\beq
\xfl(\vec x, \tau,\tau') = \vec x - \int_{\tau'}^{\tau} d\tau''\; \vec{v}(\tau'',\vec x_{\rm fl}(\vec x, \tau,\tau''))\ .
\eeq
To linear order, one can see that $\d_i v^i=-\frac{D'}{D} \delta$, where $\H=a H$, and we use the notation$'=\d/\d\tau$, 
$\tau$ being the conformal time. As usual in Standard Perturbation Theory (SPT), the velocity field can be redefined as 
$\theta\equiv\d_i\tilde v^i=-\frac{D}{D'} \d_i v^i$, such that $\theta=\delta$ at linear level. 
For the sake of simplicity, one may also redefine $\phi$ so that $\d^2\phi=\delta$, we do so henceforth. 
One can use the equations of motion to simplify the number of independent variables, as explained in~\cite{McDonald:2009dh}.
Instead of employing $v^i$ as the independent variable of choice, one can replace $\theta$ with 
\be
\eta(\vec x,t)=\theta(\vec x,t)- \delta(\vec x,t)\ ,
\ee
and $\d_j v^i$ with $t_{ij}$, which is defined as
\be
t_{ij}(\vec x,t)=\d_i v_j(\vec x,t)-\frac{1}{3}\delta_{ij} \theta(\vec x,t) -s_{ij}(\vec x,t)\ .
\ee
The quantities $\eta$ and $t_{ij}$ will start out as already second order in perturbation theory. We note here that, because of the fact that vorticity is generated only at very high order in perturbation theory~\cite{Carrasco:2013mua,Mercolli:2013bsa}, $t_{ij}$ can indeed be considered 
symmetric in $i$ and $j$ up to a high order in perturbations. The traceless tidal tensor $s_{ij}$ is defined as 
\be
s_{ij}(\vec x,t)=\d_i\d_j\phi(\vec x,t)-\frac{1}{3} \delta_{ij}\, \delta(\vec x,t) \ .
\ee
It's easy to show that, to second order in perturbations, $\eta_{(2)}=\frac{2}{7}s^{(1)}{}^2-\frac{4}{21}\delta^{(1)}{}^2$; consequently  one can introduce a variable $\psi$ that is non-zero only starting at cubic order. The expression for $\psi$ reads
\be
\psi(\vec x,t)=\eta(\vec x,t)-\frac{2}{7} s^2(\vec x,t)+\frac{4}{21}\delta(\vec x,t)^2\ .
\ee
Since calculations presented here go up to one-loop order, the quantities introduced this far will suffice for all that follows. For later convenience, we introduce non-vanishing products of the following variables:
\bea
&&s^2(\xfl,t)= s_{ij}(\xfl,t')  s^{ij}(\xfl,t')\ , \quad s^3(\xfl,t)=s_{ij}(\xfl,t')  s^{il}(\xfl,t')s_l{}^{j}(\xfl,t')\ ,\\ \nonumber
&& st(\xfl,t)=s_{ij}(\xfl,t')  t^{ij}(\xfl,t')\ , \quad \epsilon s(\xfl,t)=\epsilon_{ij}(\xfl,t')s^{ij}(\xfl,t'), \quad \epsilon t(\xfl,t)=\epsilon_{ij}(\xfl,t')t^{ij}(\xfl,t')\ ,
\eea
where indices are lowered and raised with $\delta^{ij}$. 
The trace part is absorbed into $\epsilon$, so that ~$\epsilon_{ij}$ can be treated as traceless. 
In the last line of Eq. \eqref{eq:euler_bias_3} we have the first term coming from the bias derivative expansion;  it corresponds to powers of $(k/\km)^2$.
The scale $\km$ here is the comoving wavenumber enclosing the mass of an 
object~\cite{Senatore:2014eva}. It follows from this that the derivative 
expansion terms will be of higher importance in the case of more massive halos.
Finally, the notation $:{\cal O}:$ stands for normal ordering:
\be
:{\cal O}:=\cal O-\langle\cal O\rangle\ .
\ee
The time-dependence of each order in perturbation 
theory depends on the order $(n)$ itself  and on the nature of the field (velocity, overdensity, etc.), but importantly does not depend on the wavenumber~$k$. This is due to the fact that the speed of sound for dark 
matter is only a perturbatively small effect, used as a counterterm in the perturbation-theory counting of the EFTofLSS. We can then symbolically perform the time integrals in equation~(\ref{eq:euler_bias_3}) 
to obtain~\cite{Senatore:2014eva}:
\bea\label{eq:euler_bias_4}
&&\delta_h(\vec k,t)\quad=\quad  \\ \nonumber
&&\quad  =c_{\delta,1}(t) \; \delta^{(1)}( \vec k,t)
+ c_{\delta,2}(t)\; \delta^{(2)}(\vec k,t)+ c_{\delta,3}(t)\delta^{(3)}(\vec k,t)+c_{\delta,3_{c_s}}(t)\delta^{(3)}_{c_s}(\vec k,t)\\ \nonumber
&&\qquad+ \left[c_{\delta,1}(t)-c_{\delta,2}(t)\right]\; [\d_i \delta^{(1)}\; \frac{\d^i}{\d^2}\theta^{(1)}]_{\vec k}(t)+ 
\left[c_{\delta,2}(t)-c_{\delta,3}(t)\right]\; [\d_i \delta^{(2)}\;\frac{\d^i}{\d^2}\theta^{(1)}]_{\vec k}(t)\\ \nonumber
&&\qquad+ \frac{1}{2}\left[c_{\delta,1}(t)-c_{\delta,3}(t)\right]\; [\d_i \delta^{(1)}\;\frac{\d^i}{\d^2}\theta^{(2)}]_{\vec k}(t)  \\ \nonumber
&&\qquad+\left[ \frac{1}{2} c_{\delta,1}(t)- c_{\delta,2}(t)+\frac{1}{2} c_{\delta,3}(t)\right]\;\\ \nonumber
&&\qquad \times\left[ [\d_i \delta^{(1)}\; \frac{\d_j\d^i}{\d^2}\theta^{(1)}\; \frac{\d^j}{\d^2}\theta^{(1)}]_{\vec k}(t)
+[\d_i\d_j \delta^{(1)}\;\frac{\d^i}{\d^2}\theta^{(1)}\frac{\d^j}{\d^2}\theta^{(1)}]_{\vec k}(t)\right]+ \\ \nonumber
&&\qquad+c_{\delta^2,1}(t) \; [\delta^2]^{(2)}_{\vec k}(t)+c_{\delta^2,2}(t) \; [\delta^2]^{(3)}_{\vec k}(t)-
2 \left[c_{\delta^2,1}(t)-c_{\delta^2,2}(t)\right][\delta^{(1)}\d_i \delta^{(1)}\frac{\d^i}{\d^2}\theta^{(1)}]_{\vec k} \\ \nonumber
&& \qquad
+c_{s^2,1}(t) \; [s^2]_{\vec k}^{(2)}(t)+c_{s^2,2}(t) \; [s^2]_{\vec k}^{(3)}(t)-2 \left[c_{s^2,1}(t)
-c_{s^2,2}(t)\right][s_{lm}^{(1)}\d_i (s^{lm})^{(1)}\frac{\d^i}{\d^2}\theta^{(1)}]_{\vec k}\\ \nonumber
&&\qquad +c_{st,1}(t) \; [st]^{(3)}_{\vec k}(t)+ c_{\psi,1}(t)\;  \psi^{(3)}({\vec k},t) 
+ c_{\delta^3}[\delta^3]^{(3)}_{\vec k}(t)+c_{\delta\,s^2}[\delta s^2]^{(3)}_{\vec k}(t)\\ \nonumber
&&\qquad\quad+c_{s^3}[s^3]^{(3)}_{\vec k}(t)+ [\epsilon]_{\vec k}+ c_{\delta\,\epsilon,1}[\delta \epsilon]^{(1)}_{\vec k}(t) +\ldots . \nonumber
\eea
We can reorganize this expression putting together the operators multiplying each bias coefficient:
\bea \label{eq:CoI_all}
\delta_h(\vec k,t)&=&c_{\delta,1}(t) \; \Big( \mathbb{C}^{(1)}_{{\delta,1}}(\vec k,t) + \mathbb{C}^{(2)}_{{\delta,1}}(\vec k,t) + \mathbb{C}^{(3)}_{{\delta,1}}(\vec k,t) \Big) \\ \nonumber
 &+& c_{\delta,2}(t) \; \Big( \qquad\qquad ~~~ \mathbb{C}^{(2)}_{{\delta,2}}(\vec k,t) + \mathbb{C}^{(3)}_{{\delta,2}}(\vec k,t) \Big)  \\ \nonumber
 &+& c_{\delta,3}(t) \; \Big( \qquad\qquad  \qquad \qquad ~~~+ \mathbb{C}^{(3)}_{{\delta,3}}(\vec k,t) \Big) \\ \nonumber
 &+& c_{\delta,3_{c_s}}(t) \; \mathbb{C}^{(3)}_{\delta,3_{c_s}} (\vec k,t) \\ \nonumber
 &+& c_{\delta^2,1}(t) \; \Big( \mathbb{C}^{(2)}_{\delta^2,1} (\vec k,t) +\mathbb{C}^{(3)}_{\delta^2,1} (\vec k,t) \Big) 
 + c_{\delta^2,2}(t) \; \mathbb{C}^{(3)}_{\delta^2,2} (\vec k,t)\\ \nonumber
 &+& c_{\delta^3,1}(t) \; \mathbb{C}^{(3)}_{\delta^3,1} (\vec k,t) \\ \nonumber
 &+& c_{s^2,1}(t) \; \Big( \mathbb{C}^{(2)}_{s^2,1} (\vec k,t) +\mathbb{C}^{(3)}_{s^2,1} (\vec k,t) \Big) + c_{s^2,2}(t) \; \mathbb{C}^{(3)}_{s^2,2} (\vec k,t)\\ \nonumber
 &+& c_{s^3,1}(t) \; \mathbb{C}^{(3)}_{s^3,1} (\vec k,t) \\ \nonumber
 &+& c_{st,1}(t) \; \mathbb{C}^{(3)}_{st,1} (\vec k,t) \\ \nonumber
 &+& c_{\psi,1}(t) \; \mathbb{C}^{(3)}_{\psi,1} (\vec k,t) \\ \nonumber
 &+& c_{\delta s^2,1}(t) \; \mathbb{C}^{(3)}_{\delta s^2,1} (\vec k,t) \\ \nonumber
 &+& c_{\d^2\delta ,1 }(t) \; \mathbb{C}^{(3)}_{\d^2 \delta,1} (\vec k,t) \\ \nonumber
 &+&  \mathbb{C}_{\epsilon}(\vec k,t)  \\ \nonumber
 &+&  c_{\df\epsilon,1}(t) \; \mathbb{C}_{\df\epsilon,1}^{(1)}(\vec k,t) +\ldots\ ,
\eea 
where the explicit expressions of the $\mathbb{C}_i$ operators in terms of the ones in equation~(\ref{eq:euler_bias_4}) 
are given in Appendix \ref{app:operators}, specifically in equation~(\ref{eq:C_all_explicit}). The index $n$ in 
$\mathbb{C}_i^{(n)}$ means the order in perturbation theory of the operator. Operators describing stochasticity effects are labeled 
by ``$\epsilon$", we will return to discuss them more in detail later on. 
As pointed out in~\cite{Senatore:2014eva} not all of these operators contribute in an non-degenerate functional form, and it is useful 
to check for possible degeneracies~\footnote{It is interesting to observe that one could in principle neglect the study of degeneracies 
among the operators without significantly affecting the performance of the theory against data. This is so because if we use a degenerate 
basis, we span the same functional space as a non-degenerate one. The only aspect that changes between using a linearly-dependent 
basis and an independent one is the actual value of the reduced $\chi^2$, as the number of independent degrees of freedom might be 
overcounted if one uses a degenerate basis. But, once the available number of data is much larger than the number of parameters, 
as is our case, when we compare with several 2- and 3-point functions, this is hardly a noticeable effect.}.

Note that degeneracies may occur only amongst terms of the same perturbative order. In order to get rid of these 
redundancies, we need to find the optimal operators basis that spans the full functional space we have obtained 
in equation~(\ref{eq:CoI_all}). One can see that once a product operator is introduced at a given order in perturbation theory, 
it will appear with a different bias coefficient as it is evaluated at an higher order in perturbation theory (due to the non-linear 
couplings or by being multiplied by the fluctuating fields contained in the $\xfl$ terms). However, as we move to higher orders 
in perturbation theory, new product operators are introduced for the first time. Since the new product operators might be degenerate 
with the lower ones evaluated at higher order, there will necessarily be some freedom in choosing the independent operators making 
up the basis. One convenient choice is to keep the bias operators that appear for the first time in the expansion at a given 
perturbative order only if they are not degenerate with any of the bias operators introduced earlier. In this scheme, once a 
specific independent bias operator is introduced, one keeps all of the subsequent ones from the same family that appear at 
higher perturbative orders, unless of course they are degenerate with already introduced operators (in other 
words, we keep the independent `descendants', e.g. $\mathbb{C}^{(2)}_{\delta,1}$ is a descendant of $\mathbb{C}^{(1)}_{\delta,1}$, 
and it is kept in the basis unless it is degenerate with similar descendants; $\mathbb{C}^{(2)}_{\delta,2}$ is a descendant of 
$\mathbb{C}^{(3)}_{\delta,1}$ and is kept in the basis unless is degenerate  with similar descendants, and so on. 
We dub this choice of operator basis as the `Basis of Descendants'~($BoD$) one, as, once an element is accepted 
in the basis, its descendants are by default automatically accepted~\footnote{An alternative, equally valid, choice could have been to keep the independent 
operators that appear in the local in time approximation (where one ignores the time evolution of bias coefficient) and then 
supplement them, if needed, with a properly chose minimal set of non-local in time operators necessary to complete the basis.}.

Interestingly, it turns out that up to the third perturbative order 
in the fields, once the degeneracies are dealt with, local in time operators make up the full basis (see also~\cite{Mirbabayi:2014zca}). 
This is an artifact of the lower perturbative orders and will no longer be true at higher orders.
In what follows we adopt the $BoD$ choice for constructing the operators basis, obtaining:

\bea \label{eq:CoI}
\text{(1)st order: } &&\left\lbrace \mathbb{C}^{(1)}_{\delta,1}\right\rbrace\ , \\ \nonumber
\text{(2)nd order: }&&\left\lbrace \mathbb{C}^{(2)}_{{\delta,1}},~\mathbb{C}^{(2)}_{{\delta,2}}, ~\mathbb{C}^{(2)}_{{\delta^2,1}}\right\rbrace \\ \nonumber
\text{(3)rd order: }&&\left\lbrace \mathbb{C}^{(3)}_{{\delta,1}},~\mathbb{C}^{(3)}_{{\delta,2}},~\mathbb{C}^{(3)}_{{\delta,3}}\ , 
~\mathbb{C}^{(3)}_{{\delta^2,1}}, ~\mathbb{C}^{(3)}_{{\delta^2,2}}, ~\mathbb{C}^{(3)}_{{\delta^3,1}}, \mathbb{C}^{(3)}_{\delta,3_{c_s}}
~\mathbb{C}^{(3)}_{{s^2,2}}\right\rbrace\ ,  \\ \nonumber
\text{Stochastic: }&&\left\lbrace\mathbb{C}_{\epsilon},~\mathbb{C}^{(1)}_{\df\epsilon,1}\right\rbrace\ .
\eea
In appendix~\ref{app:operators}, eq.~(\ref{eq:delta_h_CoI}), we show explicit linear combinations of the rest of the operators in terms of the \textit{BoD} basis.
In practice, it is quite straightforward to check the degeneracy of the operators. For example if we are checking the dependence of 
three second order operators $\hat{\mathit{o}}_1(\vec k_1, \vec k_2)$, $\hat{\mathit{o}}_2(\vec k_1,\vec k_2)$, 
$\hat{\mathit{o}}_3(\vec k_1,\vec k_2)$ we build a linear combination $\alpha\,\hat{\mathit{o}}_1(\vec k_1, \vec k_2)
+\beta\,\hat{\mathit{o}}_2(\vec k_1,\vec k_2)+\hat{\mathit{o}}_3(\vec k_1,\vec k_2)$ and require it to vanish for two 
independent doublets of vectors $(\vec k_1,\vec k_2)$. This determines $\alpha$ and $\beta$. We then simply ask if 
the linear combination, using the solutions for $\alpha$ and $\beta$ that we just found, vanishes for generic $(\vec k_1,\vec k_2)$. 
This procedure can then be generalized to combinations of higher order operators.   It also applies to degeneracies between bias 
operators and the operators that appear in the effective stress tensor of dark matter. 
For example, there is a degeneracy between the $\mathbb{C}^{(3)}_{\d^2 \delta ,1}$ operator, originating from the last term in 
equation~(\ref{eq:euler_bias_3}) and the operator $\mathbb{C}^{(3)}_{\delta,3_{c_s}}$ coming form the dark matter field counterterm.  
Up to cubic level, we are therefore led to write
\bea \label{eq:delta_h_CoI_t}
\delta_h(\vec k,t)&=& \tilde{c}_{\delta,1}(t) \; \Big( \mathbb{C}^{(1)}_{\delta,1}(\vec k,t)+\mathbb{C}^{(2)}_{\delta,1}(\vec k,t)+\mathbb{C}^{(3)}_{\delta,1}(\vec k,t) \Big)\\ \nonumber
&+& \tilde{c}_{\delta,2(2)}(t) \; \mathbb{C}^{(2)}_{\delta,2}(\vec k,t) + \tilde{c}_{\delta,2(3)}(t) \; \mathbb{C}^{(3)}_{\delta,2}(\vec k,t)\\ \nonumber
&+& \tilde{c}_{\delta,3}(t) \; \mathbb{C}^{(3)}_{\delta,3}(\vec k,t)
+ \tilde{c}_{\delta,3_{c_s}}(t) \; \mathbb{C}^{(3)}_{\delta,3_{c_s}}(\vec k,t)\\ \nonumber
&+& \tilde{c}_{\delta^2,1(2)}(t) \; \mathbb{C}^{(2)}_{\delta^2,1}(\vec k,t) + \tilde{c}_{\delta^2,1(3)}(t) \; \mathbb{C}^{(3)}_{\delta^2,1}(\vec k,t)\\ \nonumber
&+& \tilde{c}_{\delta^2,2}(t) \; \mathbb{C}^{(3)}_{\delta^2,2}(\vec k,t) + \tilde{c}_{s^2,2}(t) \; \mathbb{C}^{(3)}_{s^2,2}(\vec k,t)\\ \nonumber
&+& \tilde{c}_{\delta^3,1}(t) \; \mathbb{C}^{(3)}_{\delta^3,1}(\vec k,t)\\ \nonumber
&+& \tilde{c}_{\epsilon,1}(t) \; \mathbb{C}^{(3)}_{\epsilon,1}(\vec k,t) + \ldots
\eea
where the new bias coefficients $\tilde{c}_i$ are given in terms of the old ones from equation~(\ref{eq:CoI_all}) in~(\ref{eq:c_tildas}) of Appendix~\ref{app:operators}.

In order to simplify the organization of the computations, it is usually useful to write the perturbative expansion of the dark matter overdensity field in Eulerian framework:
\bea \label{eq:CoI_all_dm}
\delta(\vec k,t)&=& \delta^{(1)} (\vec k,t) + \delta^{(2)} (\vec k,t) + \delta^{(3)} (\vec k,t) + \delta^{(3)}_{c_s} (\vec k,t) + \ldots\ ,
\eea
where we have 
\begin{align} \label{eq:CoI_all_kern}
\delta^{(1)} (\vec k,t)&= \int \frac{d^3q_1}{(2\pi)^3}\;F^{(1)}_s (\vec q_1)\,\delta_D^{(3)}(\vec k- \vec q_1)\,\delta^{(1)} (\vec q_1,t)\ , \\ \nonumber
\delta^{(2)} (\vec k,t)&= \int \frac{d^3q_1}{(2\pi)^3}\;\frac{d^3q_2}{(2\pi)^3}\;\; 
F^{(2)}_s (\vec q_1, \vec q_2)\,\delta_D^{(3)}(\vec k-\vec q_1-\vec q_2)\,\delta^{(1)} (\vec q_1,t)\delta^{(1)} (\vec q_2,t)\ , \\ \nonumber
\delta^{(3)} (\vec k,t)&= \int \frac{d^3q_1}{(2\pi)^3}\;\frac{d^3q_2}{(2\pi)^3}\;\frac{d^3q_3}{(2\pi)^3}\;\; 
F^{(3)}_s (\vec q_1,\vec q_2,\vec q_3)\,\delta_D^{(3)}(\vec k-\vec q_1-\vec q_2-\vec q_3)\,\delta^{(1)} (\vec q_1,t)\delta^{(1)} (\vec q_2,t)\delta^{(1)} (\vec q_3,t)\ ,
\end{align}
and $F^{(n)}_s$ are standard Eulerian perturbation theory kernels (see e.g.~\cite{Bernardeau:2001qr} for definitions).
By analogy, for the overdensity field of biased tracers it is useful to rewrite (\ref{eq:CoI_all}) as 
\bea \label{eq:CoI_all_hl}
\delta_h(\vec k,t)&=& \delta^{(1)}_h (\vec k,t) + \delta^{(2)}_h (\vec k,t) + \delta^{(3)}_h (\vec k,t) + \delta^{(3)}_{h, c_s} (\vec k,t) + \ldots + (\epsilon\text{-terms})
\eea
where we have introduced the new kernels for biased tracers 
\begin{align} \label{eq:CoI_all_hl_kern}
\delta^{(1)}_h (\vec k,t)&= \int \frac{d^3q_1}{(2\pi)^3}\;K^{(1)}_s (\vec q_1)\,\delta_D^{(3)}(\vec k-\vec q_1)\, \delta^{(1)} (\vec q_1,t)\ , \\ \nonumber
\delta^{(2)}_h (\vec k,t)&= \int \frac{d^3q_1}{(2\pi)^3}\;\frac{d^3q_2}{(2\pi)^3}\;\; 
K^{(2)}_s (\vec q_1,\vec q_2)\,\delta_D^{(3)}(\vec k-\vec q_1-\vec q_2)\,\delta^{(1)} (q_1,t)\delta^{(1)} (\vec q_2,t)\ , \\ \nonumber
\delta^{(3)}_h (\vec k,t)&= \int \frac{d^3q_1}{(2\pi)^3}\;\frac{d^3q_2}{(2\pi)^3}\;\frac{d^3q_3}{(2\pi)^3}\;\; 
K^{(3)}_s (\vec q_1,\vec q_2,\vec q_3)\,\delta_D^{(3)}(\vec k-\vec q_1-\vec q_2-\vec q_3)\,\delta^{(1)} (\vec q_1,t)\delta^{(1)} (\vec q_2,t)\delta^{(1)} (\vec q_3,t)\ .
\end{align}
The newly introduced kernels $K^{(i)}_s$ can be expressed in the \textit{BoD} basis as 
\bea \label{eq:CoI}
 K^{(1)}_s (\vec q_1,t) &=& \tilde{c}_{\delta,1}(t) \; \widehat{c}^{(1)}_{\delta,1}(\vec q_1)\ , \\ \nonumber
 K^{(2)}_s (\vec q_1,\vec q_2,t) &=& \tilde{c}_{\delta,1}(t) \; \widehat{c}^{(2)}_{{\delta,1}}( \vec q_1,\vec q_2)+\tilde{c}_{\delta,2(2)}(t) \; \widehat{c}^{(2)}_{{\delta,2}}(\vec q_1,\vec q_2)
                                      +\tilde{c}_{\delta^2,1(2)}(t) \; \widehat{c}^{(2)}_{{\delta^2,1}}(\vec q_1,\vec q_2)\ , \\ \nonumber
 K^{(3)}_s (\vec q_1,\vec q_2, \vec q_3,t) &=& \tilde{c}_{\delta,1}(t) \; \widehat{c}^{(3)}_{{\delta,1}}(\vec q_1,\vec q_2,\vec q_3) 
                                      +\tilde{c}_{\delta,2(3)}(t) \; \widehat{c}^{(3)}_{{\delta,2}}(\vec q_1,\vec q_2,\vec q_3)
                                            + \tilde{c}_{\delta,3}(t) \; \widehat{c}^{(3)}_{{\delta,3}}(\vec q_1,\vec q_2,\vec q_3) \\ \nonumber
                                       &&+\tilde{c}_{\delta^2,1(3)}(t) \; \widehat{c}^{(3)}_{{\delta^2,1}}(\vec q_1,\vec q_2,\vec q_3)       
                                            + \tilde{c}_{\delta^2,2}(t) \;\widehat{c}^{(3)}_{{\delta^2,2}}(\vec q_1,\vec q_2,\vec q_3) \\ \nonumber
                                       &&+ \tilde{c}_{\delta^3,1}(t) \; \widehat{c}^{(3)}_{{\delta^3,1}}(\vec q_1,\vec q_2,\vec q_3)
                                            + \tilde{c}_{s^2,2}(t) \; \widehat{c}^{(3)}_{{s^2,2}}(\vec q_1,\vec q_2,\vec q_3)\ .
\eea
We see that the kernels $K^{(i)}_s$ are linear combinations of bias parameters $\tilde c_{\dots}(t)$, that are in turn 
linear combinations of the bias $c_{\dots}(t)$ in eq.~(\ref{eq:CoI_all}), as well as subkernels $\hat c^{(n)}_{\dots}(q_1,q_2,\ldots)$, 
that are defined in Appendix \ref{app:operators} (see  eq.~(\ref{eq:c_all_explicit}) and eq.~(\ref{eq:c_tildas})).

\subsection{Construction of observables: one-loop power spectrum and tree-level bispectrum}
\label{subsec:obs}

We are interested in computing the set of observables which can later be compared to 
$N$-body simulations data. In doing so, we will also  extract the values for all the bias parameters. We compute the halo-matter 
cross power spectrum $P_{hm}$, the halo-halo auto power spectrum $P_{hh}$ at one-loop order 
as well as the matter-matter-halo bispectrum $B_{mmh}$, matter-halo-halo bispectrum $B_{mhh}$ 
and halo-halo-halo bispectrum $B_{hhh}$ at tree level.

For the halo-matter cross power spectrum $P_{hm}$ at one-loop order we have:
\begin{align}
\label{eq:Phm}
 P_{hm}(k,t) =&~\langle \delta^{(1)} (\vec k,t) \delta^{(1)}_h (\vec k,t) \rangle ' + \langle \delta^{(2)} (\vec k,t) \delta^{(2)}_h (\vec k,t) \rangle ' 
                    +  \langle \delta^{(1)} (\vec k,t) \delta^{(3)}_h (\vec k,t) \rangle ' \non \\
                    &~~~~~~~~+ \langle \delta^{(3)} (\vec k,t) \delta^{(1)}_h (\vec k,t) \rangle ' +\langle \delta^{(1)} (\vec k,t) \delta^{(3)}_{h,c_s} (\vec k,t) \rangle ' 
                    + \langle \delta^{(3)}_{c_s} (\vec k,t) \delta^{(1)}_h (\vec k,t) \rangle ' \non \\
                  =&~\tilde{c}_{\delta,1}(t)\; P_{11}(k;t,t) \non \\
                  &+2\int \frac{d^3q}{(2\pi)^3} \;
                            F^{(2)}_s(\vec{k}-\vec{q},\vec{q})\,K^{(2)}_s(\vec{k}-\q,\q)\,P_{11}(q;t,t)\,P_{11}( |\vec{k}-\vec{q}|;t,t) \non \\
                  &+ 3\; P_{11}(k;t,t)\, \int \frac{d^3q}{(2\pi)^3} \;
                  \Big( \tilde{c}_{\delta,1}(t)\,F^{(3)}_s(\vec{k},-\q,\q)+K^{(3)}_s(\vec{k},-\q,\q)\Big) P_{11}(q;t,t) \non \\
                  &+( \tilde{c}_{\delta,3_{c_s}}(t)+\tilde{c}_{\delta,1} (t) )\left(-(2\pi)c^2_{s(1)}(t)\right)\frac{k^2}{k^2_{\text{NL}}}P_{11}(k;t,t)\ ,
\end{align}
where the primed brackets $\langle \rangle^{\prime}$ means that we are dropped a momentum conserving Dirac $\delta$-function 
from the expectation value. Notice that since the kernels $K^{(n)}_s$ contain unknown bias parameters, therefore they need to be 
expanded. For example, for the convolution term in the second line in the last equality in eq.~(\ref{eq:Phm}), we have: 
\begin{align}
2\int \frac{d^3q}{(2\pi)^3} \;F^{(2)}_s(\vec{k}-\vec{q},\vec{q})&K^{(2)}_s(\vec{k}-\q,\q)P_{11}(q;t,t)P_{11}( |\vec{k}-\vec{q}|;t,t) =  \non \\
      \tilde{c}_{\delta,1}(t)\; 2 &\int \frac{d^3q}{(2\pi)^3} \;
                     F^{(2)}_s(\vec{k}-\vec{q},\vec{q})\widehat{c}^{(2)}_{{\delta,1}}(\vec{k}-\q,\q)P_{11}(q;t,t)P_{11}( |\vec{k}-\vec{q}|;t,t) \non \\
   +\tilde{c}_{\delta,2(2)}(t)\; 2 &\int \frac{d^3q}{(2\pi)^3} \;
                    F^{(2)}_s(\vec{k}-\vec{q},\vec{q})\widehat{c}^{(2)}_{{\delta,2}}(\vec{k}-\q,\q)P_{11}(q;t,t)P_{11}( |\vec{k}-\vec{q}|;t,t) \non \\
   +\tilde{c}_{\delta^2,1(2)}(t)\; 2 &\int \frac{d^3q}{(2\pi)^3} \;
                    F^{(2)}_s(\vec{k}-\vec{q},\vec{q})\widehat{c}^{(2)}_{{\delta^2,1}}(\vec{k}-\q,\q)P_{11}(q;t,t)P_{11}( |\vec{k}-\vec{q}|;t,t),
\end{align}
A similar expression for the second term is given in eq.~(\ref{eq:P13_K}) and eq ~(\ref{eq:kkhaloterm}) in Appendix~\ref{app:operators}.
Similarly, for the halo-halo auto power spectrum $P_{hh}$ we have:
\begin{align}
 P_{hh}(k,t) =&~\langle \delta^{(1)}_h (\vec k,t) \delta^{(1)}_h (\vec k,t) \rangle ' + \langle \delta^{(2)}_h (\vec k,t) \delta^{(2)}_h (\vec k,t) \rangle ' 
                   + 2 \langle \delta^{(1)}_h (\vec k,t) \delta^{(3)}_h (\vec k,t) \rangle ' \non \\ 
                   &~~~~+ 2 \langle \delta^{(1)} (\vec k,t) \delta^{(3)}_{h,c_s} (\vec k,t) \rangle '
                   + c_{\epsilon}^2 \langle \left[\epsilon\right]_{\vec k} \left[\epsilon\right]_{\vec k} \rangle '  \non \\
                 =&~\tilde{c}_{\delta,1}^2(t) P_{11}(k;t,t) \non \\
                 &+2\int \frac{d^3q}{(2\pi)^3} \;
                            \left[K^{(2)}_s(\vec{k}-\q,\q)\right]^2P_{11}(q;t,t)P_{11}( |\vec{k}-\vec{q}|;t,t) \non \\
                  &+ 6\; \tilde{c}_{\delta,1}\; P_{11}(\vec{k};t,t) \int \frac{d^3q}{(2\pi)^3} \;K^{(3)}_s(\vec{k},-\q,\q)P_{11}(k;t,t) \non \\
                  &+2\; \tilde{c}_{\delta,1}\tilde{c}_{\delta,3_{c_s}}(t) \left(-(2\pi)c^2_{s(1)}(t)\right) \frac{k^2}{k^2_{\text{NL}}}P_{11}(k;t,t) 
                    + {\rm \tilde{C}onst}_{\epsilon,P} \non \\
                  &+  \tilde{c}_{\epsilon \delta,1}^2 {\rm \tilde{C}onst}_{\epsilon,P} \int \frac{d^3q}{(2\pi)^3}  P_{11}(q;t,t).   
\label{eq:Phh}
\end{align}
Notice that the term containing the $K^{(3)}_s$ kernel contributes similarly to  both  
$P_{hm}$ and $P_{hh}$ and no new terms need be computed. 
This is not the case for the $K^{(2)}_s$ kernel where the halo auto power spectrum $P_{hh}$ 
brings new contributions relative to the ones in the cross power spectrum $P_{hm}$. 
An expression for the second term of the final expression in (\ref{eq:Phh}) is given in eq.~(\ref{eq:kkhaloterm}) in Appendix~\ref{app:operators}.
Before proceeding further, let us consider the contributions to the power spectrum arising from auto correlations 
of the stochastic field $\epsilon$ as well as potential cross correlations of matter fields and the stochastic field $\epsilon$.
In case of the halo-matter cross power spectrum $P_{hm}$, as shown in~\cite{Senatore:2014eva}, the leading stochastic contribution 
has a scale dependence:
\begin{align}
\int^t dt'\; H(t')\; \bar c_{\epsilon}(t,t')\; \left\langle \epsilon(\vec{k},t') \left[ \partial^2\Delta \tau \right]_{\vec k}(t) \right\rangle'_{\text{EFT-one-loop}}
= \frac{\gamma^{1/2}}{( {k_s}^3k_{\rm NL}^3)^{1/2}}\frac{k^2}{k^2_{NL}}\ ,
\label{20}
\end{align}
where $\gamma$ is expected to be an order one number,
and $k_s$ is the scale connected to the constant stochastic contribution which
we discuss further below.
We can see that this term is highly suppressed relative to the rest of the terms at one-loop order 
in the regime of validity of one-loop results, so we will neglect it here. 
Doing so may no longer be correct at higher loops order, where this and higher stochastic terms should be appropriately  taken into account.
The same term also appears in the case of auto power spectrum $P_{hh}$ because of the correlation of the stochastic bias 
with the stochastic dark matter stress tensor, but it is again negligible due to similar arguments  (see~\cite{Senatore:2014eva} for details).

Last,  we consider the auto correlations of the halo stochastic terms which contain contributions of the form 
$\text{Const}_\epsilon \sim  \left\langle \left[ \epsilon \right]_k \left[ \epsilon \right]_k 
\right\rangle' \sim   ((2\pi)/k_s)^{3} $.  
This defines a new scale $k_s$ which corresponds to the mean distance between 
the biased objects of a given type (e.g. halos of specific mass).
In contrast to the aforementioned terms (which scale as $k^2$), 
this one generates a constant contribution at all scales. We may obtain an estimate for $ ((2\pi)/k_s)^{3}$ reasoning 
along the following lines: a constant term well described by a Poisson-type shot noise contribution will scale like $1/\bar{n}$ where 
$\bar{n}$ is the number density of halos in the specific mass bin. As mentioned above, $2\pi/k_s$ then represent the mean distance among halos of the same kind.
There exists higher derivative stochastic biases, such as for example 
$ \langle \left[ \frac{\d^2}{k_M^2}\epsilon \right]_k \left[ \epsilon \right]_k\rangle\sim  \frac{k^2}{k_M^2}\frac{(2\pi)^3}{k_s^{3}} $. 
Notice that the scale suppressing the higher derivative corrections is expected not to be the inverse distance of the objects $k_s$,  
but the inverse size of the objects $k_M$, or the non-linear scale $\knl$.
Plugging in the numbers for our simulation (see sec.~\ref{sec:results}) 
we arrive at a value of the order $  ((2\pi)/k_s)^{3} \sim 10^3-10^4 (\text{Mpc}/h)^3$ depending on the bin. Correspondingly, the value of the 
contribution in eq.~(\ref{20}) will be of the order $\sim (10^1-10^2) (\text{Mpc}/h)^3\times (k/k_{\rm NL})^2$ which shows why we have 
treated this contribution as negligible. We have  verified that these estimates are consistent with our findings in sec.~\ref{sec:results}, 
table~\ref{tb:reach}. 

The bispectrum is given by correlating three fields, each one being either the matter or the halos overdensity. For the three combinations of interest we get:
\begin{align}
 B_{mmh}(\vec k_1,\vec k_2,\vec k_3,t)
                 =&~ 2\;\tilde{c}_{\delta,1} (t) \Big( F^{(2)}_s(\vec{k}_2,\vec{k}_3) P_{11}(k_2;t,t) P_{11}(k_3;t,t) + (k_2 \rightarrow k_1) \Big) \non \\
                   &\qquad\qquad\qquad + 2\; K^{(2)}_s(\vec{k}_1,\vec{k}_2,t) P_{11}(k_1;t,t) P_{11}(k_2;t,t) +\ldots \non \\
                =&~ 2\; \tilde{c}_{\delta,1} (t) \Big( F^{(2)}_s(\vec{k}_2,\vec{k}_3) P_{11}(k_2;t,t) P_{11}(k_3;t,t) + (k_2 \rightarrow k_1) \non \\
                  &\qquad\qquad\qquad\qquad\qquad\qquad +  \widehat{c}^{(2)}_{\delta,1,s} (\vec{k}_1,\vec{k}_2) P_{11}(k_1;t,t) P_{11}(k_2;t,t)  \Big) \non \\
                  &~ + 2\,\tilde{c}_{\delta,2(2)}(t) \; \widehat{c}^{(2)}_{{\delta,2,s}}(\vec{k}_1,\vec{k}_2)  P_{11}(k_1;t,t) P_{11}(k_2;t,t) \non \\
                  &~ + 2\,\tilde{c}_{\delta^2,1(2)}(t) \; \widehat{c}^{(2)}_{{\delta^2,1,s}}(\vec{k}_1,\vec{k}_2)  P_{11}(k_1;t,t) P_{11}(k_2;t,t) \ , \non \\
                  &&\non\\
B_{mhh}(\vec k_1,\vec k_2,\vec k_3,t)                  =&~ 2\;\tilde{c}_{\delta,1}^2(t) F^{(2)}_s(\vec{k}_2,\vec{k}_3) P_{11}(k_2;t,t) P_{11}(k_3;t,t) \non \\
                   &\qquad\qquad + \tilde{c}_{\delta,1}(t) \Big( 2\; K^{(2)}_s(\vec{k}_1,\vec{k}_2) P_{11}(k_1;t,t) P_{11}(k_2;t,t) + (k_2 \rightarrow k_3) \Big)  +\ldots \non \\
                 =&~ 2\;\tilde{c}_{\delta,1}^2(t) \Big( F^{(2)}_s(\vec{k}_2,\vec{k}_3) P_{11}(k_2;t,t) P_{11}(k_3;t,t) \non \\
                  &\qquad\qquad\qquad\qquad +  \widehat{c}^{(2)}_{\delta,1,s} (\vec{k}_1,\vec{k}_2) P_{11}(k_1;t,t) P_{11}(k_2;t,t) + (k_2 \rightarrow k_3) \Big) \non \\
                  &~ + 2\,\tilde{c}_{\delta,1}(t)\tilde{c}_{\delta,2(2)}(t) \; 
                  \Big( \widehat{c}^{(2)}_{{\delta,2,s}}(\vec{k}_1,\vec{k}_2)  P_{11}(k_1;t,t) P_{11}(k_2;t,t) + (k_2 \rightarrow k_3) \Big) \non \\
                  &~ + 2\,\tilde{c}_{\delta,1}(t)\tilde{c}_{\delta^2,1(2)}(t) \; 
                  \Big( \widehat{c}^{(2)}_{{\delta^2,1,s}}(\vec{k}_1,\vec{k}_2)  P_{11}(k_1;t,t) P_{11}(k_2;t,t) + (k_2 \rightarrow k_3) \Big) \non\ , \\
                  &~ +{ 2\,\tilde{c}_{\epsilon\delta,1}(t) {\rm \tilde{C}onst}_{\epsilon,P}  \; P_{11}(k_1;t,t) }\non\ , \\
                                    &&\non\\
B_{hhh}(\vec k_1,\vec k_2,\vec k_3,t)                  =&~ 2\; K^{(2)}_s(\vec{k}_1,\vec{k}_2) P_{11}(k_1;t,t) P_{11}(k_2;t,t) + 2 \text{ permutations} \non \\
                 =&~ 2\;\tilde{c}_{\delta,1}^3(t) \Big(\widehat{c}^{(2)}_{\delta,1,s} (\vec{k}_1,\vec{k}_2) P_{11}(k_1;t,t) P_{11}(k_2;t,t) + 2 \text{ permutations} \Big) \non \\
                  &~ + 2\,\tilde{c}_{\delta,1}^2(t)\tilde{c}_{\delta,2(2)}(t) \; 
                  \Big( \widehat{c}^{(2)}_{{\delta,2,s}}(\vec{k}_1,\vec{k}_2)  P_{11}(k_1;t,t) P_{11}(k_2;t,t) + 2 \text{ permutations} \Big) \non \\
                  &~ + 2\,\tilde{c}_{\delta,1}^2(t)\tilde{c}_{\delta^2,1(2)}(t) \; 
                  \Big( \widehat{c}^{(2)}_{{\delta^2,1,s}}(\vec{k}_1,\vec{k}_2)  P_{11}(k_1;t,t) P_{11}(k_2;t,t) + 2 \text{ permutations} \Big)\non\\
                  &~ +{ 2\,\tilde{c}_{\delta,1}(t)\tilde{c}_{\epsilon\delta,1}(t) 
                  {\rm \tilde{C}onst}_{\epsilon,P}  \Big( P_{11}(k_1;t,t)+ P_{11}(k_2;t,t)+P_{11}(k_3;t,t) \Big) }\non \\
                  &~+ {{\rm \tilde{C}onst}_{\epsilon,B} }\ .
\label{eq:Bisspec}
\end{align}
The bispectrum of the halo-halo-halo density has a stochasticity contribution $ {\rm \tilde{C}onst}_{\epsilon,B}$. 
This quantity can be related to ${\rm \tilde{C}onst}_{\epsilon,P} $. As usual with quantities that are renormalziaed 
in perturbation theory, this matching can be done only after the UV behavior is taken into account and renormalization 
performed (for an early discussion of these subtleties in this context, see~\cite{Carrasco:2012cv}). 
We shall do this in the following subsection.

\subsection{UV dependence and renormalization of bias coefficients}
\label{subsec:obs}

In order to correctly take into account the effect that short distance physics has at long distances, the bias parameters need to be renormalized.
This is arrived at by readjusting the bias parameter so as to cancel the contribution from perturbation theory that is obtained from a regime 
where it cannot be trusted, and readjusting it so that the effect of short distance physics at long distance is correctly reproduced~\cite{McDonald:2009dh}. 
The results for dark matter power spectra~\cite{Carrasco:2013sva,Senatore:2014via} and bispectra~\cite{Angulo:2014tfa,Baldauf:2014qfa} 
seem to show that there is perturbative control up to the relatively UV scale of $k\simeq 0.6\hinvMpc$. We will therefore take the point of 
view that modes with wavenumber higher than that are not under perturbative control. At about  $k\simeq 0.6\hinvMpc$, the matter power 
spectrum is well described by a scaling behavior with slope $n\sim -2.1$. By approximating our universe with a scaling one with the same slope, 
we can estimate which ones, among the various contributions, are UV sensitive~\cite{Carrasco:2013sva}. 
This can be done by Taylor expanding in $q\gg k$ the integrands in the former expressions, and
identifying the ones that are UV divergent if the power spectrum has a slope $n\sim -2.1$.

Let us find these contributions starting from the terms in eq.~(\ref{eq:P13_K}). 
We can ignore the loop terms that appear already in the calculation of the dark matter correlation functions, 
as they are renormalized by the counterterms present in the equations for dark matter. 
We therefore can focus on the other terms. The only divergent terms are the leading ones in the Taylor expansion and lead
to the following expression:

\begin{align}
\label{eq:P13_K_UV1}
\bigg(3\; P_{11}&(k;t,t) \int \frac{d^3q}{(2\pi)^3} \; K^{(3)}_s(\vec{k},-\q,\q) P_{11}(q;t,t) \bigg)_{UV}= \non \\
& \sigma^2(t) \left(- \frac{13}{21} \tilde{c}_{\delta, 1} - \frac{34}{21} \tilde{c}_{\delta, 2} + 
 \frac{47}{21} \tilde{c}_{\delta, 3} + 2 \tilde{c}_{\delta^2, 1} + \frac{26}{21} \tilde{c}_{\delta^2, 2} + \frac{136}{63} \tilde{c}_{s^2, 2}
 + 3 \tilde{c}_{\delta^3, 1}\right) P_{11}(k;t,t)
\end{align}
where 
\begin{align}
\sigma^2(t) = \langle \delta(\V x,t)^2 \rangle = \int \frac{d^3q}{(2\pi)^3} \;P_{11}(q;t,t).
\end{align}
To appropriately parametrize the UV-sensitivity this expression, we need to readjust one of our bias parameters to change this contribution and make it correct. 
This is possible precisely because the form of the UV divergent terms here is the one of the linear bias. We may then renormalize the linear bias coefficient so as to absorb it. 
In practice, one may imagine $\tilde{c}_{\delta,1}$ is a bare parameter, the sum of a finite part and a counterterm:
\begin{align}
\label{eq:c1_conter}
\tilde{c}_{\delta,1}(t) = \tilde{c}_{\delta,1,\text{ finite}}(t)  + \tilde{c}_{\delta,1,\text{ counter}}(t)\ ,
\end{align}
where $\tilde{c}_{\delta,1,\text{ counter}}(t)$ is chosen to cancel the contribution 
from (\ref{eq:P13_K_UV1}), while $\tilde{c}_{\delta,1,\text{ finite}}(t)$ will give the correct contribution of short distance physics at large distances.

After renormalization we have:
\bea\label{eq:P13_K_UV2}
&&\left[3\; P_{11}(k;t,t) \int \frac{d^3q}{(2\pi)^3} \; K^{(3)}_s(\vec{k},-\q,\q) P_{11}(q;t,t) \right]_{\rm finite} = \non \\
                  && 3\; \tilde{c}_{\delta,1} P_{11}(k;t,t)  \int \frac{d^3q}{(2\pi)^3} \;\left(\widehat{c}^{(3)}_{{\delta,1,s}}(\vec{k},-\q,\q) 
                                                                       + \sfrac{13}{63}\right)  P_{11}(q;t,t) \non \\
&&               + 3\; (\tilde{c}_{\delta,3}+15\tilde{c}_{s^2,2}) P_{11}(k;t,t) 
               \int \frac{d^3q}{(2\pi)^3} \;\left( \widehat{c}^{(3)}_{{\delta,3,s}}(\vec{k},-\q,\q) - \sfrac{47}{63}\right)  P_{11}(q;t,t) 
\eea
Here with the subscript $_{\rm finite}$ we mean that we have removed the UV dependent part. 
This `finite' term is however meaningful only when added to the counterterm. 
In other words, only loop terms plus counterterms give sensible results that are independent of the UV physics and the renormalization procedure.
From~(\ref{eq:P13_K_UV2}) we see that renormalization brings to light one additional degeneracy between the bias parameters 
$\tilde{c}_{\delta,3}$ and $\tilde{c}_{s^2,2}$, which appears both in cross and and auto power spectrum. 
Similar behavior appears in convolution-type contributions of the halo-halo power spectrum in 
equation~(\ref{eq:Phh}).  As an example, we can look at the term proportional to the 
$\tilde{c}^2_{\delta^2,1\,(2)}$ coefficient, which is of the form
\bea
\label{eq:const_conter}
\tilde{c}^2_{\delta^2,1\,(2)}\left( \int \frac{d^3 q}{(2\pi)^3} P_{11}(q;t,t) P_{11}(\vec{k}-\vec{q};,t,t)\right)_{UV} = \tilde{c}^2_{\delta^2,1\,(2)}\, \Sigma(t)^2\,,
\eea
where
\bea
\Sigma(t)^2=  \int \frac{d^3 q}{(2\pi)^3} \big[ P_{11}(q;t,t)\big]^2 \,.
\eea
In the UV these contributions will be analytic functions of $k$ and as such they can be reabsorbed in the definition 
of the stochastic coefficient ${\rm\tilde{C}onst}_{\epsilon, P}$ and the higher derivative ones (at the order at which 
we are working, there is no need of the higher derivative ones). Similarly to eq.~(\ref{eq:c1_conter}), the latter will then be 
\bea
{\rm {\tilde C}onst}_{\epsilon, P} =  {\rm {\tilde C}onst}_{\epsilon, P\,\, {\rm finite}} + {\rm {\tilde C}onst}_{\epsilon, P\,\, {\rm counter}} ~,
\eea 
where 
$ {\rm {\tilde C}onst}_{\epsilon, P\,\, {\rm counter}}$ is chosen to cancel the contribution from eq.~(\ref{eq:const_conter}).
Note that, by symmetry reasons, $P_{hm}$ does not have the $k^0$ stochastic ${\rm \tilde{C}onst_{\epsilon, P}}$ contribution. 
Indeed all the convolution-type divergencies cancel exactly so the final result is finite.

Note that only the finite part of ${\rm {\tilde C}onst}_{\epsilon, P}$ corresponds to the physical contribution and 
is thus connected to the stochastic contribution in the bispectrum, ${\rm {\tilde C}onst}_{\epsilon, B}$. 
If, as it is expected, we assume a Poissonian nature for the statistics corresponding to the mean halo number density, 
we have ${\rm Const}_{\epsilon,B} = {\rm Const}_{\epsilon,P\,\,{\rm finite}}^2$ , which is the well known relation 
of the shot noise contribution in the power spectrum and bispectrum~(see e.g.~\cite{Peebles1980, Smith:2007sb, Schmittfull:2014tca}).
Equipped with this relation, one is able to reduce the number of stochastic bias parameters down to one.

At this stage we can give the final set of bias parameters that describe all the introduced statistics. We perform the following replacements:
\begin{align}
\label{eq:renscheme}
&\tilde{c}_{\delta,1}\rightarrow b_{\delta,1}
- \sigma^2(t) \left(- \frac{13}{21} \tilde{c}_{\delta, 1} - \frac{34}{21} \tilde{c}_{\delta, 2} + 
 \frac{47}{21} \tilde{c}_{\delta, 3} + 2 \tilde{c}_{\delta^2, 1} + \frac{26}{21} \tilde{c}_{\delta^2, 2} + \frac{136}{63} \tilde{c}_{s^2, 2}
 + 3 \tilde{c}_{\delta^3, 1}\right)
, \\ 
&\tilde{c}_{\delta,2\,(2)}\rightarrow b_{\delta,2}\ ,\non\\ 
&\tilde{c}_{\delta,3}+15\tilde{c}_{s^2,2}\rightarrow b_{\delta,3}\ ,\non\\ 
&\tilde{c}^2_{\delta^2,1\,(2)}\rightarrow b_{\delta^2}\ ,\non\\ 
&\tilde{c}_{\delta,3_{c_s}} \rightarrow b_{c_s}\ , \non\\
& \tilde{c}_{\epsilon \delta,1} \rightarrow b_{\epsilon \delta,1}\ , \non\\
&{\rm \tilde{C}onst}_{\epsilon,P}\rightarrow {\rm Const}_{\epsilon}   \Big( 1 - \tilde{c}_{\epsilon \delta,1}^2 \sigma^2(t) \Big) \non\\ 
&\qquad\qquad\qquad   -\Sigma(t)^2 \left( \tilde{c}_{\delta,1}^2 + \tilde{c}_{\delta,2\,(2)}^2 + \tilde{c}^2_{\delta^2,1\,(2)} - 2 \; \tilde{c}_{\delta,1}\tilde{c}_{\delta,2\,(2)}  
- 2 \; \tilde{c}_{\delta,1}\tilde{c}_{\delta^2,1\,(2)} + 2 \; \tilde{c}_{\delta,2\,(2)}\tilde{c}_{\delta^2,1\,(2)} \right) 
\non\ .
\end{align}
where the bias coefficients $b_{\delta,1},~b_{\delta,2},~b_{\delta,3},~b_{\delta^2},~b_{\delta^2}$ and ${\rm Const}_{\epsilon} $ 
are finite contributions and we have used the relation ${\rm Const}_{\epsilon,B} = {\rm Const}_{\epsilon}^2$~\footnote{Notice that strictly speaking, in the renormalization of $b_{\delta,1}$ and ${\rm \tilde{C}onst}$ one could introduce additional contributions, such as those arising from operators such as $\delta \epsilon^{n\geq2},\, s^2 \epsilon,\, \delta^2\epsilon,\, \ldots $. However, all of these terms do not contribute to the bispectrum at the order we are working on, and are trivially reabsorbable by a shift in ${\rm \tilde{C}onst}$ and $b_{\delta,1}$. We therefore omit to write them explicitly.}. 
The cross power spectra of biased tracers and dark matter is therefore given by:
\begin{align}
 P_{hm}(k,t) =& b_{\delta,1}(t)\left( P_{11}(k;t,t) 
  +2\int \frac{d^3q}{(2\pi)^3} \;F^{(2)}_s(\vec{k}-\vec{q},\vec{q})\, \widehat{c}^{(2)}_{{\delta,1,s}}(\vec{k}-\q,\q)\, P_{11}(q;t,t)P_{11}( |\vec{k}-\vec{q}|;t,t) \right. \non\\
  &\qquad\qquad\qquad~~~~\left.+3 \; P_{11}(k;t,t) \int \frac{d^3q}{(2\pi)^3} \;\left(F^{(3)}_{s}(\vec{k},-\q,\q)+
  \widehat{c}^{(3)}_{{\delta,1,s}}(\vec{k},-\q,\q) + \sfrac{13}{63}\right)  P_{11}(q;t,t) \right) \non \\
  &+b_{\delta,2}(t) \; 2 \int \frac{d^3q}{(2\pi)^3} \;
                    F^{(2)}_s(\vec{k}-\vec{q},\vec{q})\left( F^{(2)}_s(\vec{k}-\vec{q},\vec{q}) - \widehat{c}^{(2)}_{{\delta,1,s}}(\vec{k}-\q,\q) \right)
                    P_{11}(q;t,t)P_{11}( |\vec{k}-\vec{q}|;t,t) \non \\
  &+b_{\delta, 3}(t)\; 3\; P_{11}(k;t,t) \int \frac{d^3q}{(2\pi)^3} \;\left( \widehat{c}^{(3)}_{{\delta,3,s}}(\vec{k},-\q,\q) - \sfrac{47}{63}\right)  P_{11}(q;t,t) \non \\
  &+b_{\delta^2}(t)\; 2 \int \frac{d^3q}{(2\pi)^3}F^{(2)}_s(\vec{k}-\vec{q},\vec{q})P_{11}(q;t,t)P_{11}( |\vec{k}-\vec{q}|;t,t)\non \\
  &+\left(b_{c_s}(t)-2(2\pi)c^2_{s(1)}(t)b_{\delta,1} (t) \right)\frac{k^2}{k^2_{\text{NL}}}P_{11}(k;t,t)\,.
\label{eq:Phm_fin}
\end{align}
Note that the following relation holds: 
\begin{align}
\int \frac{d^3q}{(2\pi)^3} \;F^{(3)}_{s}(\vec{k},-\q,\q) P_{11}(q;t,t) = 
\int \frac{d^3q}{(2\pi)^3} \;\left( \widehat{c}^{(3)}_{{\delta,1,s}}(\vec{k},-\q,\q) 
          + \widehat{c}^{(3)}_{{\delta,3,s}}(\vec{k},-\q,\q)  - \sfrac{34}{63} \right)  P_{11}(q;t,t) \non
\end{align}
and for $\lbrace b_{\delta,1} \rightarrow 1, ~  b_{\delta,2} \rightarrow b_{\delta,1}, ~ b_{\delta,3} \rightarrow b_{\delta,1}, ~ \
b_{\delta^2} \rightarrow 0, ~ b_{c_s}\rightarrow 0 \rbrace $ our formula reduces to that of the standard dark matter case (below at (\ref{eq:PS_eft})), as it should.

Similarly, for the auto power spectra of biased tracers we have:
\begin{align}
 P_{hh}(k,t) =& b^2_{\delta,1}(t)\left( P_{11}(k;t,t) 
  +2\int \frac{d^3q}{(2\pi)^3} \;\left[\widehat{c}^{(2)}_{{\delta,1,s}}(\vec{k}-\q,\q)\right]^2P_{11}(q;t,t)P_{11}( |\vec{k}-\vec{q}|;t,t) \right. \non\\
  &\qquad\qquad\qquad\qquad\qquad\qquad \left.+6 \; P_{11}(k;t,t) \int \frac{d^3q}{(2\pi)^3} \;\left(\widehat{c}^{(3)}_{{\delta,1,s}}(\vec{k},-\q,\q) 
  + \sfrac{13}{63}\right)  P_{11}(q;t,t) \right) \non \\
  &+b_{\delta,1}(t) b_{\delta, 3}(t)\; 6\; P_{11}(\vec{k};t,t) \int \frac{d^3q}{(2\pi)^3} \;\left( \widehat{c}^{(3)}_{{\delta,3,s}}(\vec{k},-\q,\q) - \sfrac{47}{63}\right)  P_{11}(q;t,t) \non \\
  &+b^2_{\delta,2}(t) \; 2 \int \frac{d^3q}{(2\pi)^3} \;\left[F^{(2)}_s(\vec{k}-\vec{q},\vec{q}) - \widehat{c}^{(2)}_{{\delta,1,s}}(\vec{k}-\q,\q)\right]^2
                                                       P_{11}(q;t,t)P_{11}( |\vec{k}-\vec{q}|;t,t) \non \\
  &+b^2_{\delta^2}(t) \; 2 \int \frac{d^3q}{(2\pi)^3} \;P_{11}(q;t,t)P_{11}( |\vec{k}-\vec{q}|;t,t)\non \\
  &+b_{\delta,1}(t)b_{\delta, 2}(t)\; 4 \int \frac{d^3q}{(2\pi)^3} \;\widehat{c}^{(2)}_{{\delta,1,s}}(\vec{k}-\q,\q) \left(F^{(2)}_s(\vec{k}-\vec{q},\vec{q})\right.  \non \\
    &\qquad\qquad\qquad\qquad\qquad\qquad\qquad\qquad \left. - \widehat{c}^{(2)}_{{\delta,1,s}}(\vec{k}-\q,\q)\right) P_{11}(q;t,t)P_{11}( |\vec{k}-\vec{q}|;t,t)\non \\
  &+b_{\delta,1}(t)b_{\delta^2}(t)\; 4 \int \frac{d^3q}{(2\pi)^3} \;\widehat{c}^{(2)}_{{\delta,1,s}}(\vec{k}-\q,\q) P_{11}(q;t,t)P_{11}( |\vec{k}-\vec{q}|;t,t)\non \\
  &+b_{\delta,2}(t)b_{\delta^2}(t)\; 4 \int \frac{d^3q}{(2\pi)^3} \;
                                   \left(F^{(2)}_s(\vec{k}-\vec{q},\vec{q}) - \widehat{c}^{(2)}_{{\delta,1,s}}(\vec{k}-\q,\q)\right) P_{11}(q;t,t)P_{11}( |\vec{k}-\vec{q}|;t,t)\non \\
  &+b_{\delta,1}(t)b_{\delta,3_{c_s}}(t)\; 2\; \frac{k^2}{k^2_{\text{NL}}}P_{11}(k;t,t) + \text{Const}_{\epsilon} \non \\
& -\Sigma(t)^2 \left( b_{\delta,1}^2 + b_{\delta,2}^2 + b^2_{\delta^2} - 2 \; b_{\delta,1}b_{\delta,2}  
- 2 \; b_{\delta,1}b_{\delta^2} + 2 \; b_{\delta,2}b_{\delta^2} \right)   \, .
\label{eq:Phh_fin}
\end{align}
Again, as a check, taking bias parameters $\lbrace b_{\delta,1} \rightarrow 1, ~  b_{\delta,2} \rightarrow b_{\delta,1}, ~ b_{\delta,3} \rightarrow b_{\delta,1}, ~ \
b_{\delta^2} \rightarrow 0, ~ b_{c_s}\rightarrow 0 \rbrace $ we reproduce the dark matter power spectrum result (\ref{eq:PS_eft}) below.

Let us consider the effects of renormalisation on the bispectrum expressions in eq.~(\ref{eq:Bisspec}). 
Here too, in analogous way to eq.~(\ref{eq:renscheme}), we employ renormalized coefficients. 
Notice however that we have to include only the finite part of the coefficients (schematically shown 
in eq.~(\ref{eq:c1_conter})). We do so because the bispectrum terms are in no need of being renormalized 
at tree level. The `counter' part of eq. (\ref{eq:c1_conter}) is present in order to correct for the incorrect 
contribution from loop diagrams. Since we do not include loop diagrams in the bispectrum, we should not 
include the `counter' part of the counterterm, as at this order, there is nothing to correct for. This is the usual 
fact that tree-level expressions do not need to be renormalized. In summary, and schematically, at tree level 
for the bispectrum we  perform the following replacement (for $\tilde{c}_{\delta,1}$ and correspondingly for the rest):
 $ \tilde{c}_{\delta,1} \rightarrow \tilde{c}_{\delta,1,\text{ finite}} = b_{\delta_1} $.

\subsection{IR-resummation}
\label{subsec:irresum}

As detailed in~\cite{Porto:2013qua,Senatore:2014eva}, in the EFTofLSS, after the renormalization is 
performed, the loop expansion, complemented by the insertion of the relevant higher order bias coefficients,
corresponds to an expansion in the parameters that regulate the dark matter expansion, those are $\epsilon_{\delta<}$ 
and $\epsilon_{s>}$~\cite{Porto:2013qua,Senatore:2014via}. These quantities are defined according to 
\bea
&&\epsilon_{s >} =k^2  \int_k^\infty {d^3k' \over (2 \pi)^3}  {P_{11}(k') \over k'^2}\ , \\ \nonumber
&&\epsilon_{\delta <} = \int^k_0 {d^3k' \over (2 \pi)^3} P_{11}(k')\ .
\eea
where $P_{11}(k)$ stands for the dark matter power spectrum. For a given wavenumber $k$, $\epsilon_{s >} $ 
will represent the effect of the displacement due to short wavelength modes, while $\epsilon_{\delta <}$ 
describes the effect of tidal forces due to long wavelength modes.  Both these quantities scale proportionally to 
$k/\knl$. The bias derivative expansion corresponds to one in powers of $(k/\km)^2$, 
where $\km$ is the comoving wavenumber enclosing the mass of an object~\cite{Senatore:2014eva}. 
Within an Eulerian treatment we ought to expand also in the displacement due to long wavelength modes 
$\epsilon_{s<}= (k\, \delta s_<)^2$, defined as 
\bea
\epsilon_{s_<} &=&k^2  \int_0^k {d^3k' \over (2 \pi)^3}  {P_{11}(k') \over k'^2}\ .
\eea
As explained in~\cite{Senatore:2014via}, in the $k$-range of interest $\epsilon_{s<}$ is of order one; 
as a consequence one cannot Taylor expand in this parameter. Fortunately, for equal-time, dark-matter correlators 
the only effect of $\epsilon_{s<}$ is to affect the Baryon Acoustic Oscillation (BAO) peak. In Fourier space, 
this shows up in about $2\%$ residual oscillations that are not correctly reproduced if one Taylor expands 
in~$\epsilon_{s<}$. In this work, we aim at exploring the UV reach of the EFTofLSS for biased tracers, 
and therefore we perform the IR-resummation that corresponds to treating $\epsilon_{s<}$ 
non-perturbatively only for the power spectrum. In this case, since there is no velocity bias that is not higher 
derivative~\cite{Senatore:2014eva,Mirbabayi:2014zca}, the formulas for IR-resummation simply extend the 
ones for dark matter~\cite{Senatore:2014via,Senatore:2014eva}.
\be\label{eq:Lagrangian_bias_5} 
\left.P_{\delta_a\delta_b}(k;t_1,t_2)\right|_N=\sum_{j=0}^N \int \frac{d^3k'}{(2\pi)^3}\; 
M_{||_{N-j}}( k, k';t_1,t_2)\; P_{\delta_a\delta_b,\,j}(k';t_1,t_2)\ .
\ee
where $M_{||_{N-j}}( k, k';t_1,t_2)$ is defined as the double Fourier transform of $ P_{{\rm int}||_{N-j}}(r|q;t_1,t_2)$
\bea
&&M_{||_{N-j}}( k, k';t_1,t_2)=\frac{1}{4\pi}\int d^3 r\; d^3 q \; P_{{\rm int}||_{N-j}}(r|q;t_1,t_2) 
\; e^{i \vec k \cdot\vec r} \;e^{-i \vec k' \cdot\vec q}\\ \nonumber
&&\qquad\qquad\qquad  
\qquad=\int d^3 q \; F_{||_{N-j}}(\vec q,-\vec k;t_1,t_2) \; e^{i (\vec k-\vec k') \cdot\vec q}\ ,
\eea
and $P_{\delta_a\delta_b,\,j}$ stands for the $j$-th order term for the power spectrum in the Eulerian calculation. 
We refer the interested reader to~\cite{Senatore:2014eva} for further details on the notation.

\subsection{Estimated theory errors}
\label{subsec:theoryerrors}

An important property of the EFTofLSS, and of every theory that is not a model, 
is that it allows for a reliable estimate of the error at every given perturbative order.  We can readily do this by assuming a simplified scaling universe 
behavior corresponding to a slope $n\simeq -1.7$ for the linear power spectrum.  There are several contributions that come at the next order both in the power spectra and in the bispectra, which scale differently accordingly to size of the bias parameters and of the stochastic terms. This means that, when this is relevant, for some quantities we quote estimates for the contribution from more than one diagram, and we take the largest  among those as our theoretical error. 
For the matter power spectrum we thus obtain
\bea
&&\Delta P_{mm} \sim (2 \pi)^2\, \left(\frac{k}{k_{\rm NL}} \right)^{2.6} P_{11}(k)\ , 
\eea
which comes two-loops of dark matter self-interactions. For the halo-matter power spectrum, we take
\bea
&& \Delta P_{hm} \sim (2 \pi)^2\,{\rm Max} \left[ b_{\delta,1}, b_{\delta^2,1} \right]\, \left(\frac{k}{k_{\rm NL}} \right)^{2.6} P_{11}(k) \ ,\\ \nonumber
\label{eq:Power_Errors_mm_hm}
\eea
which again come from a two-loop diagram using dark matter self-interactions. The first term comes from using the linear bias, while the second comes from the quadratic one. For the halo-halo power spectrum, we have similarly
\bea
&& \Delta P_{hh} \sim (2 \pi)^2\,{\rm Max} \left[ b_{\delta,1}^2 , b_{\delta^2,1}^2 \right]\, \left(\frac{k}{k_{\rm NL}} \right)^{2.6} P_{11}(k) \ ,\\ \nonumber
\label{eq:Power_Errors_mm_hm}
\eea
depending if we use the linear or the quadratic bias. 

Similarly for the tree-level halo-halo-halo,  and halo-matter-matter bispectra, we derive the following error estimates, associated to dark matter loops: 
\bea
&&\Delta B_{mmm}\sim (2\pi)\, \left(\frac{k}{k_{\rm NL}} \right)^{1.3} B_{{\rm tree\,level}} \ , \\
&&
\Delta B_{hmm}\sim   (2\pi)\, {\rm Max}\left[b_{\delta,1},b_{\delta^2,1}\right]\left(\frac{k}{k_{\rm NL}} \right)^{1.3} B_{\rm tree\,level}\ ,
\eea
where $k$ is the largest wavenumber of a given triangle. 
For tree-level halo-halo-matter, the contribution from the stochastic bias in $\epsilon\,\delta$ can become relevant. We therefore write
\bea
\Delta B_{hhm}\sim{\rm Max}\left[ (2\pi)\, {\rm Max}\left[b_{\delta,1}^2,b_{\delta^2,1}^2\right] \left(\frac{k}{k_{\rm NL}} \right)^{1.3} B_{{\rm tree\,level}} ,\; (2\pi)\;b_{\epsilon \delta,1}\left(\frac{2\pi}{k_S}\right)^3 \left(\frac{k_{\rm dm}}{k_{\rm NL}} \right)^{1.3} P_{11}(k_{\rm dm}) \right]  \ ,
\eea
where $k_{\rm dm}$ is the wavenumber associated to the dark matter field. Finally, for the tree-level halo-halo-halo bispectrum, we take
\bea
&&\Delta B_{hhh}\sim  {\rm Max}\left[{\rm Max}[(2\pi) b_{\delta,1}{}^3,\, b_{\delta^2,1}{}^3] \left(\frac{k}{k_{\rm NL}} \right)^{1.3} B_{{\rm tree\,level}} ,\; (2\pi)\,b_{\delta,1}b_{\epsilon \delta,1}\left( \frac{2\pi}{k_S} \right)^3
\left(\frac{k}{k_{\rm NL}} \right)^{1.3} P_{11}(k)\right] . \nonumber \\ 
\label{eq:Bispec_Errors}
\eea
Strictly speaking, the relation given just above for the bispectum error is obtained for the 
equilateral triangles configuration where all three momentum vectors have the same amplitude. 
Arbitrary triangle configurations would make the relation more cumbersome, but without 
significantly changing the result. Similarly, for the porpuse of estimates, we work in the local in time approximation, and therefore we do not distinguish between $b_{\delta,1},b_{\delta,2},b_{\delta,3},\ldots$ and we do not take into account of possible degeneracies. Importantly, we notice that tracers with high values of their 
bias coefficients will tend to be less well behaved in perturbation theory, 
in a way that can still be predicted in the EFTofLSS. These estimates will be later used to estimate the $k$-reach of our calculations with the EFTofLSS.

\section{Baryons and Primordial non-Gaussianities}

The results presented in the former section would allow us to proceed to compare the EFTofLSS with data of biased tracers. 
We delay this  to first extend the formalism to account for two interesting effects, even though we will not study them 
in the comparison with data in this paper. The derivation of bias results in the EFTofLSS, as described in sec.~\ref{sec:equations}, 
has been relying on a few essential assumptions: the validity of general relativity, the presence of only one species in the 
universe (a dark matter component), and the absence of primordial non-Gaussianities. The fact that general relativity 
holds was assumed when, for example, we did not allow the halos in Eulerian space to depend directly on the gravitational 
potential $\Phi$ or its first derivative $\d\Phi$. The presence of only one species was assumed in the fact that we wrote 
only one matter component, and that the halos were not allowed to depend on the velocity of the particles, apart for the 
streaming terms. Finally, the assumption of Gaussian initial conditions was implied in the fact that the only effect of long 
modes on the formation of halos was simply due to evolution effects, such as the fact that long tidal forces or velocity 
gradients might enhance or delay the collapse of a certain short region into halos. The initial conditions of short modes 
that end up collapsing were not affected by the long modes; this is where the hypothesis of gaussian initial conditions, 
which indeed implies the fact that modes of different wavelengths do not affect each other, has entered. 

In this section, we will generalize the effective description of biased tracers  presented  in sec.~\ref{sec:equations} to the 
case in which we include the presence of baryons and of primordial non-Gaussianities. We will leave the inclusion of 
deviations from general relativity to a later time. We will simply provide the relevant equations, 
and we will compare to data in subsequent studies.

\subsection{Including Baryons}

The inclusion of baryon effects in the EFTofLSS was recently performed in~\cite{Lewandowski:2014rca}. 
There, it was shown that baryons can be described at large distances, even in the presence of star formation events, 
as a second species on top of the dark matter, where the effect of short distance non-linearities and of star formation 
physics at long distances is encoded in an effective stress tensor and an effective force. The effective force, absent in 
the case of only one species, takes into account the momentum exchange at short distances between dark matter and 
baryons through gravity. By comparison with measurements of the power spectrum from simulations, it was shown that 
the EFTofLSS can reproduce baryon effects at percent level with a one-loop calculation up to the relatively high 
wavenumber $k\simeq 0.6\hinvMpc$, consistent with the estimates of the theoretical errors. 
We now proceed to the theory for biased tracers.

The generalization of the formulas of sec.~\ref{sec:equations} is rather straightforward. We start from Eulerian space, 
where the derivation follows in analogy the writing of the effective stress tensor and force terms in~\cite{Lewandowski:2014rca}. 
In the EFTofLSS with two species, the degrees of freedom that we are able to use are the gravitational field, the density and 
velocity gradients of the two species, and, finally, the relative velocity of the two species in the center of mass frame.

The contribution of long wavelength perturbations of dark matter and of baryons is expected to be weighted by the 
relative abundances of baryons and dark matter: $w_b=\frac{\Omega_b}{\Omega_b+\Omega_c},\ w_c
=\frac{\Omega_c}{\Omega_b+\Omega_c}$. This is important because, as shown explicitly in~\cite{Lewandowski:2014rca}, 
the factor of $w_b$ suppresses perturbatively the contribution of baryons. When introducing baryons, there is therefore a 
sort of doubling of the bias parameters that represent the effect of matter fluctuations, even though those associated to 
baryons are perturbatively less important. 

As anticipated, in the presence of two species, the bias can depend directly on the relative velocity, 
which is not zero in the center of mass frame~\cite{Lewandowski:2014rca,Mirbabayi:2014zca}. In fact, in the center of mass frame we have
\bea
&&v_{c,{\rm CM}}= v_c-\left(w_c v_c+w_b v_b\right)=w_b (v_c- v_b)\ ,\quad v_{b,{\rm CM}}=v_b-\left(w_c v_c+w_b v_b\right)=w_c (v_b- v_c)\ .
\eea 
The halo fields should therefore be allowed to depend on them. 
Due to the units of the velocity, the only combination that involves no derivatives on the velocity field directly 
(as these terms are allowed in the case of single species as well) 
are quadratic and effectively proportional to $w_b$, such as 
$w_\sigma\, v^i_{\rm \sigma, CM} \d_i\delta_{\sigma'}/H$, where $\sigma,\sigma'$ run over $c,b$.

Summarizing, if in the only-dark-matter case we write, as pointed out in earlier sections from~\cite{Senatore:2014eva},
\bea\label{eq:euler_bias_2}
&&\delta_h(\vec x,t)\simeq \int^t dt'\; H(t')\; \left[   \bar c_{\d^2\phi}(t,t')\; \frac{\d^2\phi(\xfl,t')}{H(t')^2} \right.\\  \nonumber
&&\quad+ \bar c_{\d_i v^i}(t,t') \;  \frac{\d_i v^i(\xfl,t')}{H(t')}
+\bar c_{\d_i \d_j \phi \d^i \d^j \phi}(t,t') \;\frac{\d_i\d_j \phi(\xfl, t')}{H(t')^2}\frac{ \d^i \d^j \phi(\xfl,t')}{H(t')^2} + \ldots\\\nonumber
&&\quad+ \bar c_{\epsilon}(t,t')\;\epsilon(\xfl,t')+\bar c_{\epsilon\d^2\phi}(t,t') \;\epsilon(\xfl,t')\frac{\d^2\phi(\xfl,t')}{H(t')^2}+ \ldots \\ \nonumber
&&\left.\quad+  \bar c_{\d^4\phi}(t,t')   \;\frac{\d^2_{x_{\rm fl}}}{\km^2}\frac{\d^2\phi(\xfl,t')}{H(t')^2}+\dots\ \right] ,
\eea
in the presence of baryons we have the following generalization 
\bea\label{eq:euler_bias_3_baryons}
&&\delta_h(\vec x,t)\simeq \int^t dt'\; H(t')\; \left[   \bar c_{\d^2\phi}(t,t')\; \frac{\d^2\phi(\xfl,t')}{H(t')^2}+ \bar c_{\delta_b}(t,t')\, w_b \, \delta_b(\xfl_b) \right. \\  \nonumber
&&\quad+ \bar c_{\d_i v_c^i}(t,t') \; w_c\, \frac{\d_i v^i_c(\xflc,t')}{H(t')}+ \bar c_{\d_i v_b^i}(t,t') \; w_b\, \frac{\d_i v^i_b(\xflb,t')}{H(t')}\\ \nonumber
&&\quad+\bar c_{\d_i \d_j \phi \d^i \d^j \phi}(t,t') \;\frac{\d_i\d_j \phi(\xfl, t')}{H(t')^2}\frac{ \d^i \d^j \phi(\xfl,t')}{H(t')^2} + \ldots\\\nonumber
&&\quad+ \bar c_{\epsilon_c}(t,t')\;w_c\,\epsilon_c(\xflc,t')+ \bar c_{\epsilon_b}(t,t')\;w_b\,\epsilon_b(\xflb,t')\\ \nonumber
&&\quad+\bar c_{\epsilon_c\d^2\phi}(t,t') \;w_c\,\epsilon_c(\xflc,t')\frac{\d^2\phi(\xfl,t')}{H(t')^2}
+\bar c_{\epsilon_b\d^2\phi}(t,t') \;w_b\,\epsilon_b(\xflb,t')\frac{\d^2\phi(\xfl,t')}{H(t')^2} \ldots \\ \nonumber
&&\left.\quad+  \bar c_{\d^4\phi}(t,t')   \;\frac{\d^2_{x_{\rm fl}}}{\km^2}\frac{\d^2\phi(\xfl,t')}{H(t')^2}+
\sum_{\sigma,\sigma'=b,c} w_\sigma\, v^i_{\sigma, \rm CM} (\xfl_\sigma,t') \frac{\d_i\delta_{\sigma'}(\xfl_{\sigma',t'})}{H}+
\ldots \right] \ ,
\eea
where
\beq
\xflb(\vec x, \tau,\tau') = \vec x - \int_{\tau'}^{\tau} d\tau''\; 
\vec{v}_b(\tau'',\vec x_{\rm fl}(\vec x, \tau,\tau''))\ ,\quad \xflc(\vec x, \tau,\tau') = \vec x - \int_{\tau'}^{\tau} d\tau''\; \vec{v}_c(\tau'',\vec x_{\rm fl}(\vec x, \tau,\tau''))\ .
\eeq
and where the $\xfl$ term in $\frac{\d^2\phi(\xfl,t')}{H(t')^2} $ should be meant as by using the equations of motion 
\be
\frac{\d^2\phi(\xfl,t)}{H(t)^2} =w_c\,\delta_c(\xflc,t)+w_b\,\delta_b(\xflb,t)\ .
\ee
Notice that we have inserted a factor of $w_b$ in front of the baryonic fluctuations. 
This is because the effect of baryons on halos is expected to be suppressed by $w_b$. 
The suppression in $w_b$ is probably absent in the case the collapsed objects are galaxies, 
so if one were to use these formulas for galaxies, one should allow for these terms to be 
non parametrically suppressed by $w_b$.

At this point, we proceed in the analogous way as we did for the case of dark-matter-only~\cite{Senatore:2014eva}.
We Taylor expand in all perturbations, and we perform symbolically all the time integrals 
to obtain independent bias coefficients for each order in perturbation theory; one then 
finally explores the degeneracy between the functional terms, along the same lines we elaborated upon in sec.~\ref{sec:equations}.

In the case of two species, the expression for the biased tracers written in Eulerian space, as the ones above, 
expand not only in $\epsilon_{\delta<},\epsilon_{s>}$ and $\epsilon_{s<}$, but also in the parameters associated 
to the relative displacements~\cite{Lewandowski:2014rca}:
\bea\label{eq:epsrelparameters}
\epsilon_{s < }^{\rm rel}(k) = k^2  \int_0^k \frac{ d^3 k' }{(2\pi)^3} \frac{\tilde P_{11}(k')}{k'{}^2} 
\ ,\qquad \epsilon_{s > }^{\rm rel}(k) =  k^2 \int_k^\infty \frac{ d^3 k' }{(2\pi)^3} \frac{\tilde P_{11}(k')}{k'{}^2} \ .
\eea
where  $\tilde P_{11}(k)$ is the power spectrum of the log-derivative with respect to the scale factor $a$ of the 
difference in the dark matter and baryon overdensities: $\d(\delta_c-\delta_b)/\d\log a$. 
The parameters $\epsilon_{s<}$ and $\epsilon_{s < }^{\rm rel}$ are order one in the current universe for the 
$k$'s of interest ($\epsilon_{s < }^{\rm rel}$ is order one only at high redshifts), and so cannot be treated perturbatively. 
Fortunately, they are IR dominated, and can therefore be resummed. Formulas for the IR-resummation in the case of two 
species were provided in~\cite{Senatore:2014via} for the baryon and dark matter fields. As the usual IR-resummation for 
single species extends trivially from overdensities to tracers~\cite{Senatore:2014via}, so it does for the case of two 
species~\cite{Lewandowski:2014rca}. For the overdensity power spectrum, we have
\be \label{resumeq}
P^\alpha ( k ; t ) |_N = \sum_{j=0}^N \int dk' \, \hat{M}^\alpha_{||_{N-j}} ( k , k' ; t, t ) \,  P^\alpha_{ j} ( k'; t) \ ,
\ee
where $\alpha = hh$ or $hm$.    
The quantity $P^\alpha_{ j} ( k'; t) $ is the Eulerian result at $j$-th order obtained by expanding in 
$\epsilon_{\delta<},\,\epsilon_{s>},\, \epsilon_{s<},\,\epsilon_{s < }^{\rm rel},\,\epsilon_{s > }^{\rm rel}$ treated as equal, 
$P^\alpha ( k ; t ) |_N $ is the IR-resummed result up to $N$-th order by expanding only in $\epsilon_{\delta<},\,\epsilon_{s>}$ 
and $\epsilon_{s > }^{\rm rel}$, and treating them as equal. The matrix $\hat{M}^\alpha_{||_{N-j}}$ is defined in eq.~(4.5) 
of~\cite{Lewandowski:2014rca}. It is important to keep in mind that this formula is obtained by treating at non-linear level  
the displacements of the dark matter and baryon particles, and not of the adiabatic and isocurvature modes. 
It is in this basis that the IR-resummation takes a particularly simple form.
Notice finally that for this formulas to be consistent, it is crucial that there is no linear, 
non-derivatively suppressed, velocity bias, as it is indeed the case.

\subsection{Including primordial non-Gaussianities}

In the case when the primordial fluctuations are non-Gaussian, the formulas of sec.~\ref{sec:equations} are incomplete. 
In fact, in the derivation of those formulas, it was assumed that the {\it only} way the long modes affect the number density 
of collapsed objects is by affecting the evolution of the short modes that eventually collapse. However, in the presence of 
non-Gaussianities that correlate modes of different wavelengths in the initial conditions, the presence of long wavelength 
fluctuations can in principle affect the actual distribution of the short modes before horizon re-entry, when all evolution 
effects are still null. The explanation of this effect is quite intuitive: suppose that the presence of a non-Gaussian correlation 
is such that, in the presence of a positive primordial fluctuation, the variance of the short fluctuations is enhanced. Then, it 
is quite intuitive that it would be particularly easy for collapse to happen in that region. This effect is particularly interesting 
because the initial conditions are, contrary to the evolution effects, non-local in space, and can therefore have a different 
$k$-dependence than the one obtained from standard biases~\footnote{The fact that we need a way to generate non-local 
in space initial conditions is what is named as the horizon problem in cosmology and leads to the need of inflation. 
It is quite interesting to see the resurgence of such an historical problem in this unusual setting.}.

The description  of bias in the EFTofLSS is based on an hierarchy of scales between the long modes $k_L$ that affects the 
collapse of short modes with wavenumber $k_S$ of the order of the one whose wavelength encloses a region with the same 
mass $M$ as the collapsed object. It is therefore an expansion in $k_L/k_S\ll 1$. Since there is no hierarchy in time, 
the EFTofLSS for biased tracers is non-local in time~\cite{Senatore:2014eva,Carrasco:2013mua,Carroll:2013oxa}, 
so there is no rigorously justified analogous expansion in $\omega_L/\omega_S$. The limit $k_L/k_S\ll 1$ implies that the 
functional dependence of the  number density of collapsed objects on the long-wavelength ones will probe the squeezed limit of non-Gaussian correlation functions.

In summary, two ingredients are necessary to include the effect of primordial non-Gaussian correlations into the bias. 
The first is that these correlations are present in the initial conditions, and not developed by the time evolution when modes are sub-Hubble. 
Second, the effect can be well described by the squeezed limit, $k_L\ll k_S$, of correlation functions.

Let us follow through these considerations and derive an equation for biased tracers. We will focus our discussion on the case where there is a non-trivial three-point function, the so-called $f_{\rm NL}$ case. The study can be straightforwardly extended to the case of primordial non-Gaussianities for higher $N$-point functions. In the squeezed limit $k_L\ll k_S$, we can write any three-point function as the one induced by the field~\footnote{This formula does not account for anisotropic squeezed limits that are generated in some inflationary models, which however can be included using the same formalism as developed here (see  \cite{Assassi:2015jqa,Assassi:2015fma, Lewandowski:2015ziq} for subsequent works).
}
\be\label{eq:squeezed1}
\zeta(\vec k_S)\simeq \zeta_g(\vec k_S)+f_{\rm NL} \left(\frac{k_L}{k_S}\right)^{\alpha} \zeta_g(\vec k_S-\vec k_L) \zeta_g(\vec k_L) \ ,
\ee
where $\zeta$ is the curvature perturbation that is constant on super-Hubble scales~\footnote{For the proof of the constancy of $\zeta$ at quantum level see~\cite{Senatore:2012ya}.}. $\zeta_g$ is an auxiliary gaussian variable. $\alpha$ is some power that depends on the particular inflationary model. For example, in the local case, $\fnlloc$, $\alpha=0$, for the equilateral, $\fnlequil$~\cite{Creminelli:2005hu}, and orthogonal, $\fnlorthog$~\cite{Senatore:2009gt}, cases, $\alpha=2$~\footnote{Concerning the squeezed limit on the orthogonal case, there is some confusion in the literature due to the fact that in ref.~\cite{Senatore:2009gt} there are two templates for the three point function that is induced by $\fnlorthog$. The one in the main text has an incorrect squeezed limit, but it is used for CMB analysis where the signal in the squeezed limit is limited. The one in the appendix  has the correct squeezed limit, leading to $\alpha=2$, and should be used for the bias.}, quasi-single field inflation~\cite{Chen:2009we,Chen:2009zp}, gives $0\leq\alpha\leq 3/2$. Strongly coupled conformal sectors can give rise to $0\leq\alpha\leq 2$~\cite{Green:2013rd}. If precision requires, additional subleading terms in the squeeze limit can be added in (\ref{eq:squeezed1}) in a similar fashion.

Well after horizon re-rentry, but still early enough to neglect all gravitational non-linearities, the primordial density fluctuation can therefore be written as 
\be
\delta^{(1)}(\vec k_S,t_{\rm in })\simeq \delta_g(\vec k_S)+f_{\rm NL}  \tilde\phi(\vec k_L, t_{\rm in})\delta_g(\vec k_S-\vec k_L,t_{\rm in }) \ ,
\ee
where
\be\label{eq:tildephi}
\tilde\phi(\vec k_L, t_{\rm in})= \frac{3}{2} \frac{H_0^2 \, \Omega_m} {D(t_{\rm in})}\frac{1}{k_S^2\, T(k)} \left(\frac{k_L}{k_S}\right)^{\alpha} \delta_g(\vec k_L,t_{\rm in})
\ee
and where $T(k)$ is the transfer function. The super-script ${}^{(1)}$ signals that this is the initial density field that will then be non-linearly evolved up to the present time. It is non-Gaussian simply because of the primordial non-Gaussianities (see for example~\cite{Baldauf:2010vn}). 

In general, the number of collapsed objects at a given location, $n_h(\vec x,t)$, will be a function of the evolution of the short fluctuations evaluated on the path that led them to the final point. The short fluctuations will be well described by the set of all their correlation functions. Symbolically, 
\be
n_h(\vec x,t)=\int^tdt'\; K(t,t')\; f_h\left(\langle\delta_s^2 \rangle_{l}(\xfl(t,t'),t'),\langle\delta_s\d_iv_s^i\rangle_{l}(\xfl(t,t'),t'),\langle\delta_s^3\rangle_{l}(\xfl(t,t'),t'),\ldots\right)\ .
\ee
The subscript $\langle\rangle_l$ means that the average is taken over the short modes by keeping fixed the long modes.
Obviously, it is impossible to compute perturbatively this expression. But, as for the case of no non-Gaussianities, we can still Taylor expand in the long wavelength fluctuations. Due to the primordial non-Gaussianities, the dependence on the long modes is not just the one induced by the gravitational dynamics. For example, the two-point function of short modes contains terms that scale as: 
\be
\langle\delta_s^2\rangle_l(\vec x_{\rm in},t_{\rm in})\supset \langle\delta_s^2\rangle_0 (t_{\rm in}) \fnl   \tilde\phi(\vec x_{\rm in}, t_{\rm in})\ .
\ee
This term allows us to deduce that the overdensity of collapsed objects will depend on terms of the form 
\bea
\delta_h(\vec x,t)\supset \int dt^{\prime} K(t,t') \bar{b}(t') \fnl \tilde\phi(\xfl(t,t_{\rm in})) \,\,=\,\, \fnl \, b^{ \tilde{\phi}}_{\,1}\,\,   \tilde\phi(\xfl(t,t_{\rm in}))  \qquad \qquad \qquad \qquad \qquad \qquad  \nonumber \\
\simeq   b^{ \tilde{\phi}}_{\,1}\,\fnl\, \Bigg[\tilde{\phi}(\vec{x}, t_{\rm in})-  \d_i \tilde{\phi}(\vec{x}, t_{\rm in}) \frac{\d^i}{\d^2}\theta^{(1)}(\vec{x},t)  +\frac{1}{2} \d_i\d_j \tilde{\phi}(\vec{x}, t_{\rm in}) \frac{\d^i}{\d^2}\theta^{(1)}(\vec{x},t) \frac{\d^j}{\d^2}\theta^{(1)}(\vec{x},t)  \nonumber \\ \qquad -\frac{1}{2}  \d_i \tilde{\phi}(\vec{x}, t_{\rm in}) \frac{\d^i}{\d^2}\theta^{(2)}(\vec{x},t)
+\frac{1}{4} \beta(t) \d_i \tilde{\phi}(\vec{x}, t_{\rm in}) \frac{\d}{\d_i}\left(\frac{\d^j}{\d^2}\theta^{(1)}(\vec{x},t)\right)^2  +\dots  \Bigg]        \, \nonumber 
\eea
where in the last line we have expanded in $\vec{x}_{\rm fl}$ to cubic order.

As said, the correlation functions of the short modes cannot be computed explicitly. However, the correlation functions that are present in the initial conditions will be enhanced by those that come from gravitational interactions.
 Concerning the effect of the long modes, this means that on top of depending on  $\tilde\phi(\xfl(t,t_{\rm in}),\tin)$, the collapsed objects will depend on the field that were allowed to be present in the absence of primordial non-Gaussianities, evaluated on their past trajectoreis, plus all products of these fields evaluated on their past trajectories with powers and derivatives of~$\tilde\phi(\xfl(t,t_{\rm in}),\tin)$. The peculiarity of these terms is associated to the presence of the transfer function in (\ref{eq:tildephi}), as well as potential non-analytic powers of $k_L$: this non-analytic behavior is non-degenerate with terms that we could have written as arising from local evolution. Indeed, local dynamics is mirrored in analyticity in Fourier space, and therefore any non-analytic behavior must come from the initial conditions.

In the presence of primordial non-Gaussianities, we can therefore write the following additional terms on top of the ones in (\ref{eq:euler_bias_4}):

\bea
\label{eq:euler_bias_3_fnl}
&&\delta_h(\vec x,t)\simeq\fnl \; \tilde\phi(\xfl(t,t_{\rm in}),\tin)\;\int^t dt'\; H(t')\; \left[  \bar c^{\,\, \tilde{\phi}}(t,t') +  \bar c^{\,\, \tilde{\phi}}_{\d^2\phi}(t,t')\; \frac{\d^2\phi(\xfl,t')}{H(t')^2} \right.\\  \nonumber
&&\quad\quad\quad\quad\quad\quad\quad\quad+ \bar c^{\,\, \tilde{\phi}}_{\d_i v^i}(t,t') \;  \frac{\d_i v^i(\xfl,t')}{H(t')}+\bar c^{\,\, \tilde{\phi}}_{\d_i \d_j \phi \d^i \d^j \phi}(t,t') \;\frac{\d_i\d_j \phi(\xfl, t')}{H(t')^2}\frac{ \d^i \d^j \phi(\xfl,t')}{H(t')^2} + \ldots\\\nonumber
&&\quad\quad\quad\quad\quad\quad\quad\quad+ \bar c^{\,\, \tilde{\phi}}_{\epsilon}(t,t')\;\epsilon(\xfl,t')+\bar c^{\,\, \tilde{\phi}}_{\epsilon\d^2\phi}(t,t') \;\epsilon(\xfl,t')\frac{\d^2\phi(\xfl,t')}{H(t')^2}+ \ldots \\ \nonumber
&&\left.\quad\quad\quad\quad\quad\quad\quad\quad+  \bar c^{\,\, \tilde{\phi}}_{\d^4\phi}(t,t')   \;\frac{\d^2_{x_{\rm fl}}}{\km^2}\frac{\d^2\phi(\xfl,t')}{H(t')^2}+\dots\ \right] \\ \nonumber
&&\qquad\qquad+\fnl^2 \; \tilde\phi(\xfl(t,t_{\rm in}),\tin)^2 \int^t dt'\; H(t')\; \left[  \bar c^{{\,\, \tilde{\phi}}^2}(t,t') +  \bar c^{{\,\, \tilde{\phi}}^2}_{\d^2\phi}(t,t')\; \frac{\d^2\phi(\xfl,t')}{H(t')^2}+\ldots\right] +\ldots\, .
\eea
As in the former sections, the time integral can be symbolically done, to lead to the following new bias coefficients:
\bea\label{eq:euler_bias_fnl}
\delta_h(k,t)=&&\fnl 
\left( \,  c^{\,\, \tilde{\phi}}_{, 1}(t) \;[\tilde{\phi}]_k(t)
- c^{\,\, \tilde{\phi}}_{,1}(t) \;[\d_i \tilde{\phi} \frac{\d^i}{\d^2}\theta^{(1)}]_k(t) + \dots \right. \\
&&\left. \qquad + c^{\,\, \tilde{\phi}}_{\delta,1}(t) \;[\tilde{\phi}\, \delta^{(1)}]_k(t)- c^{\,\, \tilde{\phi}}_{\delta,1}(t) \;[\d_i \tilde{\phi} \frac{\d^i}{\d^2}\theta^{(1)}\, \delta^{(1)}]_k(t)+ c^{\,\, \tilde{\phi}}_{\delta,2}(t)\; [\tilde{\phi}\, \delta^{(2)}]_k(t)+\dots \right. \non \\ \nonumber
&&\left. \qquad \, + \left[c^{\,\, \tilde{\phi}}_{\delta,1}(t)-c^{\,\, \tilde{\phi}}_{\delta,2}(t)\right]\; [\tilde{\phi}\, \d_i \delta^{(1)}\; \frac{\d^i}{\d^2}\theta^{(1)}]_k(t)+ \dots +c^{\,\, \tilde{\phi}}_{\delta^2,1}(t) \; [\tilde{\phi}\, \delta^2]^{(3)}_k(t)+\dots    \right. \\ \nonumber
&&\left. \,  \qquad+c^{\,\, \tilde{\phi}}_{s^2,1}(t) \; [\tilde{\phi}\,s^2]_k^{(3)}(t)+\dots  +c^{\,\, \tilde{\phi}}_{\epsilon,1}(t)  [\tilde{\phi}\,\epsilon^{(1)}]_k(t) 
- c^{\,\, \tilde{\phi}}_{\epsilon,1}(t) \;[\d_i \tilde{\phi} \frac{\d^i}{\d^2}\theta^{(1)}\, \epsilon^{(1)}]_k(t) \right.  \\ \nonumber
&& \left. \,  \qquad  +c^{\,\, \tilde{\phi}}_{\epsilon,2}(t)  [\tilde{\phi}\,\epsilon^{(2)}]_k(t)+ \dots  \right)  \\ \nonumber
&&+ \fnl^2 \left(
 c^{\,\, \tilde{\phi}^2}_{,1}(t) \;[\tilde{\phi}^2]_k(t) + \dots
+  c^{\,\, \tilde{\phi}^2}_{\delta,1}(t) \;[\tilde{\phi}^2 \, \delta^{(1)}]_k(t)+ \dots   c^{\,\, \tilde{\phi}^2}_{\epsilon,1}(t)  [\tilde{\phi}^2\,\epsilon^{(1)}]_k(t) + \dots    \right)
\eea
where the dots stand for terms higher than cubic order and we have expanded the $\vec{x}_{\rm fl}$ dependence.

In the local in time approximation, the linear bias coefficients due to primordial non-Gaussianities have been known for some time. The linear scale-dependent bias induced by $\fnlloc$ was first discovered  in~\cite{Dalal:2007cu}, even though the presence of the transfer function factor was introduced only in~\cite{Slosar:2008hx}. It has since been studied extensively in the literature. The presence of the transfer function factor that makes the functional form of the linear bias distinguishable from the standard one was pointed out, for the equilateral and orthogonal cases, in~\cite{Schmidt:2010gw,Matsubara:2012nc}. We point out here the generalization to the non-linear biases, to the non-locality in time, to higher derivative terms as well as, next, to the IR-resummation.

In fact, so far this formulas are written in Eulerian space, which expands in $\epssm$. This parameter is order one in the current universe, and so needs to be resummed. However, the same considerations that apply  in the absence of primordial non-Gaussianities apply here: only long wavelength displacements need to be resummed, and, as explained in~\cite{Senatore:2014eva}, the IR-resummation for biased tracers is the same as for dark matter as developed in~\cite{Senatore:2014via}~\footnote{The reader concerned by the fact that the variable $\tilde\phi$ appears only evaluated at $\tin$, might notice that this can be thought of as a non-local in time bias with support as a $\delta$-function at an early times. Therefore, the same discussion as in~\cite{Senatore:2014eva} applies here.}.

\section{Results and Comparison to $N$-body simulations}
\label{sec:results}

In this section we compare our results for halo cross and auto power spectrum and bispectrum of dark matter halos with Gaussian initial conditions, against $N$-body simulations. We fit for seven independent bias parameters; six enter in the halo-halo auto power 
spectrum, five enter in the halo-matter cross power spectrum, four enter in the halo-halo-halo auto bispectrum, 
and three in the halo-halo-matter and halo-matter-matter cross bispectra. 
We fit these parameters by minimizing  the joined $\chi^2$ function.  
We include below the results and corresponding plots obtained using these values.

\subsection{$N$-body simulations}
\label{subsec:sims}

We compare our theoretical results with the Millennium-XXL (MXXL) $N$-body simulation~\cite{Angulo2012ep}; 
it evolved $6720^3$ particles inside a comoving cubical region whose side measures $L_{box} = 3000\text{Mpc}/h$. 
The evolution onset corresponds to $z= 63$ and lasts until the present day. The MXXL simulation was performed 
with the following settings for cosmological parameters: a matter density of $\Omega_m = 0.25$ in units of the 
critical density, a cosmological constant with $\Omega_\Lambda = 0.25$, a Hubble constant $h = 0.73$, a 
spectral index $n_s = 1$ and a normalization parameter $\sigma_8 = 0.9$ for the primordial linear density 
power spectrum. The initial conditions were set using the linear theory power spectrum obtained from the 
Boltzmann code CAMB~\cite{Lewis:1999bs}. 

In this work we focus on the auto and cross power spectrum and bispectrum of the halo and matter density 
contrast at redshift $z = 0$. For the dark matter component, as described in~\cite{Angulo:2014tfa}, we first 
construct a density field by assigning simulation particles onto a $1024^3$ cubic grid using a ``Cloud-in-Cells'' deposit 
scheme. 
This is followed by Fast Fourier transforming the given density field and correcting for the effects of the assignment 
scheme by dividing each Fourier mode by the Fourier transform of a cubical top-hat window function. 

\begin{figure*}[t!]
   \begin{center}
   \hspace*{-0.5cm}
   \includegraphics[scale=0.55]{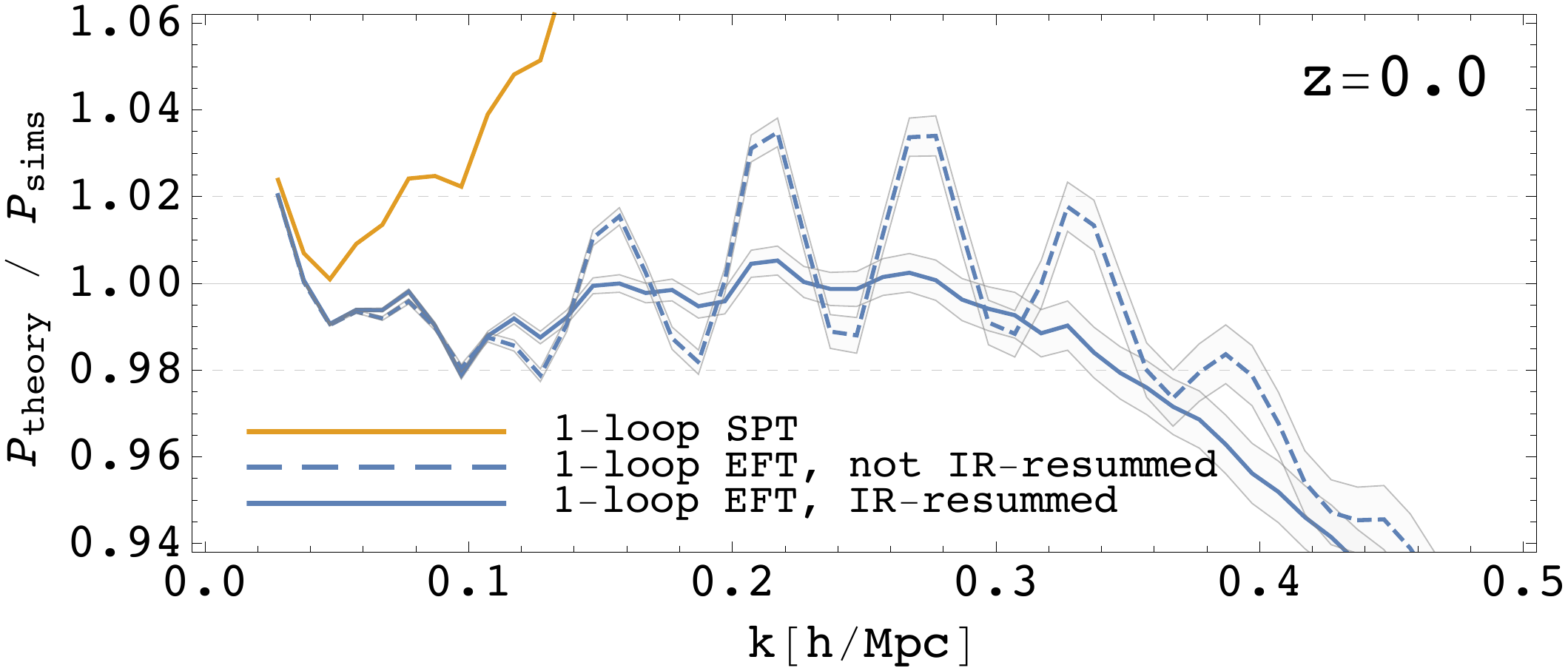}
   \end{center}
   \vspace*{-0.5cm}
   \caption{\small Dark matter power spectrum at redshift $z=0$ divided by the nonlinear simulation data.
   The one-loop EFTofLSS result is shown after (solid blue line) and before (dashed blue line) performing the IR resummation procedure.  
   For comparison, we also show the one-loop SPT result (solid orange line). 
   The parameter $c^2_{s(1)}$ is determined by fitting the IR-resummed EFT predictions over the range $0.15\hinvMpc < k < 0.3\hinvMpc$.   
  The  thin gray dashed lines signal the $2\%$ systematics error assumed for the simulation data.
   }
   \label{fig:Pmm}
\end{figure*} 

Dark matter halos are identified using the ``Friends-of-Friends'' algorithm. 
Linking length is set to 0.2 times the mean particle separation and only halos with more than 20 particles are used in the analysis. 
Since we are interested in how clustering of halos depends on the halo mass, the dark matter halo sample is divided into four 
subsamples (bin0, 1, 2 and 3) according to their mass. Mass bins are defined as in~\cite{Vlah:2013lia, Okumura:2012xh}. 
Because of the fact that the scaling of measurement of the bispectrum goes as the sixth power of the number of grid cells 
per dimension, cross and auto bispectrum are measured up to modes $k < 0.15\hinvMpc$. 

The cross and auto power spectra are computed by spherically averaging the amplitude of Fourier modes in 
a range of radius $\Delta k$. All the bispectra $B(k_1, k_2, \theta)$, as detailed in~\cite{Angulo:2014tfa} for the dark 
matter case, are obtained by performing a nested loop over all grid points ($1024^6$) and averaging the quantity 
$\df(k_1)\df(k_2)\df(k_3)$ over triangles whose sides satisfy:
\begin{eqnarray}
k^i_1=k_1\pm\frac{1}{2}\Delta k, \quad  
k^i_2=k_2\pm\frac{1}{2}\Delta k, \quad  \tf^i=\tf\pm\frac{1}{2}\Delta \tf \,\, ,
\end{eqnarray}
and where 
$\tf^i=\cos^{-1}\left(\hat{k}_1\cdot\hat{k}_2\right)$.  For the purposes of this work we have set 
\begin{eqnarray}
\Delta k = 2 \pi /L_{box} = 0.0021 \text{Mpc}/h, \quad  \delta \tf = \pi /20\,\, .
 \end{eqnarray}

\noindent Finally, we review the estimation of the errors in the measurement of the power spectra and bispectra. 
For the cross and auto power spectrum we have for the following:
\begin{align}
\left[\Delta P^{(xy)}(k)\right]^2 = \frac{2}{n} P^{(xy)}_{NL}(k)^2,
\label{eq:Power_Errors}
\end{align} 
where $(xy)$ indices may stand for both, halo-matter $(hm)$ or halo-halo $(hh)$ power spectrum, and $n$ is the number of modes contributing to the given $k-$bin.
Similarly, for the auto and cross bispectra one has 
(see e.g.~\cite{Feldman1994, Fry1993, Scoccimarro:1997, Scoccimarro:2004}):
\begin{align}
\left[\Delta B^{(hxy)}(k_1, k_2, \theta)\right]^2 = s_{123} \frac{L_{\text{box}}}{n_{\text{triangles}}} 
P^{(hh)}_{NL}(k_1)P^{(xx)}_{NL}(k_2)P^{(yy)}_{NL}(k_3)\,\,,
\label{eq:BErrror}
\end{align} 
where $x$ and $y$ encompass the halo $h$ or matter $m$ case. Note that above  $s_{123}$ is equal to  $6, 2, 1$ for equilateral, 
isosceles and general triangles respectively  and that the relation $k_3^2=k_1^2+k_2^2-2k_1k_2\cos(\theta)$ is satisfied.
The quantity $n_{\text{triangles}}$ stands for the number of triangles contributing to a given configuration, which, in the case at hand, 
is directly counted inside the bispectrum code. 

Even though derived in the Gaussian limit, these estimates give a better approximation for the scales we shall be concerned 
with if we keep $P_{NL}$ in lieu of $P_{11}$. In addition to the error estimates presented above, we add a 2$\%$ 
error for each bispectrum data point, in order to account for possible systematic errors in the simulation.

\subsection{Determining $c^2_{s(1)}$ from the matter power spectrum}

\begin{figure*}[t!]
   \begin{center}
   \hspace*{-0.5cm}
   \includegraphics[scale=1.0]{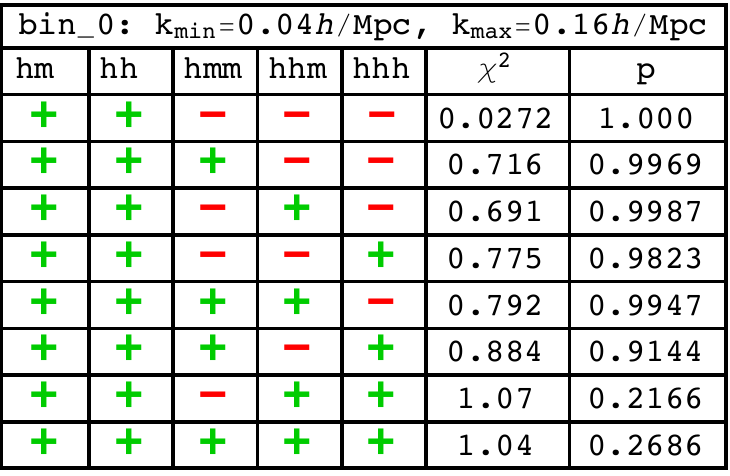} \hspace*{0.5cm}
   \includegraphics[scale=1.0]{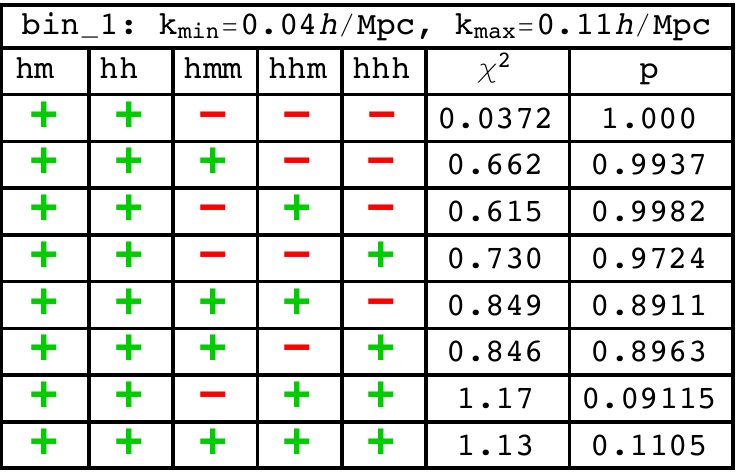} \\ \vspace*{0.5cm} \hspace*{-0.55cm}
   \includegraphics[scale=1.0]{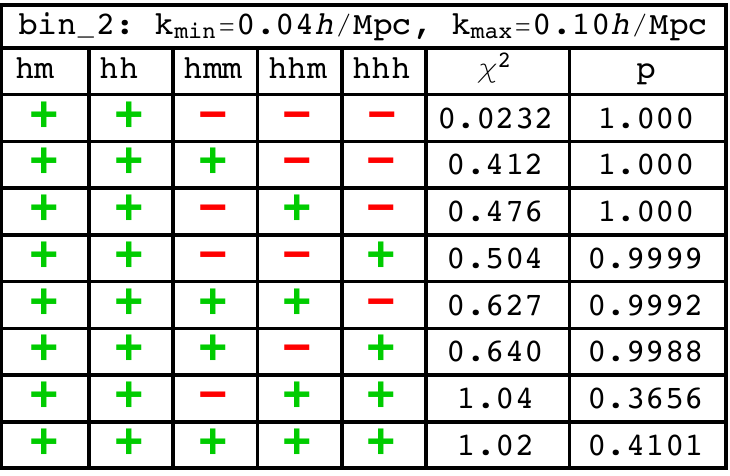} \hspace*{0.5cm}
   \end{center}
   \vspace*{-0.5cm}
   \caption{\small $\chi^2$ and $p$-values are given for best fit bias parameters procedure for the three 
   mass bins. In each table we indicate the bispectrum fitting $k$-range (for the power spectrum the $k$-range is 
   always $0.04-0.3\hinvMpc$). Green plus (red minus) signs are indicating which statistics is included (excluded)
   from the fit. We see that adding the tree level bispectrum significantly improves the constraining power of the 
   fitting procedure.
   }
   \label{fig:chi-p_4tables}
\end{figure*} 

Before we proceed with determining all the bias coefficients, the EFT $c_{s(1)}$ 
dark matter parameter needs to be evaluated. This coefficient enters the EFT calculation of the halo-matter
power spectrum equation~(\ref{eq:Phm_fin}) and is determined by fitting the one-loop EFTofLSS dark matter power 
spectrum to nonlinear data from $N$-body simulations. 
As shown in~\cite{Carrasco:2012cv}, the one-loop EFTofLSS power spectrum takes the form
\begin{align}
P_{\text{EFT-1-loop}} = P_{11}+P_{\text{1-loop}}-2(2\pi)c^2_{s(1)}\frac{k^2}{k^2_{\text{NL}}}P_{11}\,\,,
\label{eq:PS_eft}
\end{align}
and is expected to be in good agreement with data up to a scale of  $k\sim0.3\hinvMpc$. 
Beyond this scale, it would not be consistent to neglect the two-loop contribution~\cite{Carrasco:2013sva}. 
By fitting equation (\ref{eq:PS_eft}) to the power spectrum measured from the $N$-body simulations 
(see sec.~\ref{subsec:sims}) in the $k$-range $[0.15-0.3] 
\hinvMpc$ one finds:
\begin{align}
c^2_{s(1)} = (2.31\pm 0.02)\times\frac{1}{2\pi}\left(\frac{k_{\text{NL}}}{h\,\text{Mpc}^{-1}}\right)^{2}.
\end{align}
In Fig.~\ref{fig:Pmm} is shown the resulting comparison of the one-loop EFTofLSS power spectrum divided by the 
nonlinear spectrum $P_{NL}$. As detailed in~\cite{Angulo:2014tfa}, we also place uniform 2$\%$ errorbars on 
the data to account for possible systematic errors in the simulation and in the comparison procedure between the simulation and the EFT. The residual oscillations of about $2\%$ in 
size around the nonlinear data originate from bulk flow effects and can be appropriately taken into account by 
resumming the IR modes (see~\cite{Senatore:2014vja}).

\begin{table*}[t!]
\caption{
Table of estimated $k_{\rm max}$ values for the power spectra and bispectra. 
Power spectra values correspond to the range where theoretical 
error estimates obtained from higher perturbative order are estimated to be about 3\%.
Bispectra $k_{\rm max}$ ranges correspond to the scale where theoretical errors 
obtained from higher perturbative order are expected to reach the same value as the variance 
of simulation data given by equation~(\ref{eq:BErrror}).}

\centering 
\setlength{\tabcolsep}{8pt}
\renewcommand{\arraystretch}{1.0}
\begin{tabular}{c|cccc}
\hline\hline
  $k_{\rm max}\left[h/\text{Mpc}\right]$ & bin0 & bin1 & bin2 \\ [0.5ex] 
\hline
${mm}$   & $  0.22-0.31$ & $ 0.22-0.31$ & $ 0.22-0.31$ \\
${hm}$    & $  0.24-0.35$ & $ 0.22-0.35$ & $ 0.22-0.50$ \\
${hh}$     & $  0.19-0.32$ & $ 0.17-0.30$ & $ 0.20-0.35$ \\
${mmm}$& $  0.14-0.22$ & $ 0.14-0.22$ & $ 0.14-0.22$ \\
${hmm}$ & $  0.13-0.22$ & $ 0.13-0.22$ & $ 0.13-0.21$ \\
${hhm}$  & $  0.13-0.22$ & $ 0.13-0.22$ & $ 0.14-0.23$ \\
${hhh}$   & $  0.13-0.21$ & $ 0.13-0.21$ & $ 0.11-0.23$ \\
\hline
\end{tabular}
\label{tb:reach}
\end{table*}

\subsection{Procedure for comparing to data}
\label{subsec:fitting}

We briefly describe below the method that allows us to compare to simulation  data and at the same time to determine the bias parameters from the same data, in particular from 
power spectrum and bispectrum measurements. We employ a total of five observables: halo-matter (cross) 
$P^{(hm)}$ and halo-halo (auto) power spectra $P^{(hh)}$, as well as halo-halo-halo $B^{(hhh)}$, 
halo-halo-matter $B^{(hhm)}$ and halo-matter-matter $B^{(hmm)}$ bispectra. The EFT predictions for these 
observables involve all together seven independent bias parameters: $b_{\delta,1}$, $b_{\delta,2}$, $b_{\delta,3}$, 
$b_{\delta^2}$, $b_{c_s}$, $b_{\delta\,\epsilon}$ and a constant contribution which we label ${\rm Const}_\epsilon$. 
These bias coefficients correspond to our choice of the basis ($BoD$). Note that only five of these ($b_{\delta,1}$, $b_{\delta,2}$, 
$b_{\delta^2}$, $b_{\delta\,\epsilon}$ and ${\rm Const}_\epsilon$) enter the bispectrum results at tree level, and $b_{\delta\,\epsilon}$
parameter does not enter the one loop power spectra.

\begin{table*}[t!]
\caption{Best fit bias parameters table. After renormalization seven bias parameters remain to be 
evaluated. We use $N$-body simulations to evaluate the best fit values for these parameters on five 
observables: halo-matter cross power spectrum $P_{hm}$, halo-halo auto power spectrum $P_{hh}$, the
two cross halo-matter-matter and halo-halo-matter bispectra $B_{hmm}$, $B_{hhm}$ and finally the 
auto halo-halo-halo bispectra $B_{hhh}$. Parameter values are given for three mass bins (see~\cite{Vlah:2013lia, Okumura:2012xh} for the mass ranges), ${\rm bin0}$ being the lightest and bin2 the most massive.}
\centering 
\setlength{\tabcolsep}{8pt}
\renewcommand{\arraystretch}{1.0}
\begin{tabular}{c|cccccc}
\hline\hline
  & bin0 & bin1 & bin2 \\ [0.5ex] 
\hline
$b_{\df,1}$  & $  1.00\pm0.01$ & $1.32\pm0.01$ & $1.89\pm0.02$ \\
$b_{\df,2}$  & $  0.23\pm0.01$ & $0.52\pm0.01$ & $-0.51\pm0.04$  \\
$b_{\df,3}$  & $  0.48\pm0.12$ & $0.66\pm0.13$ & $1.48\pm0.28$  \\
$b_{\df^2}$ & $  0.28\pm0.01$ & $0.30\pm0.01$ & $1.88\pm0.03$  \\
$b_{c_s}$     & $  0.72\pm0.16$ & $0.27\pm0.17$ & $-0.66\pm0.37$  \\
$b_{\df\epsilon}$ &  $  0.31\pm0.08$ & $0.76\pm0.17$ & $1.38\pm0.30$  \\
${\rm Const}_\epsilon$  &  $5697\pm108$ & $10821\pm169$ & $28856\pm570$  \\
\hline
\end{tabular}
\label{tb:bias}
\end{table*}

Previous results on the dark matter one-loop power spectrum in the EFTofLSS are believed to 
be valid up to scales $k\sim0.3\hinvMpc$~\cite{Carrasco:2012cv}. Considering that dark matter fields are the building blocks 
of halos observables, we should thus not assume for halo cross or auto power spectrum that the $k$-range is any larger. 
In order to reduce the cosmic variance effect on the two point statistics, instead of using the halo-halo $P^{(hh)}$ and 
halo-matter $P^{(hm)}$ power spectra directly, we divide these by the dark matter predictions so as to obtain the 
ratios: $r^{(hh)}=P^{(hh)}/P^{(mm)}$, $r^{(hm)}=P^{(hm)}/P^{(mm)}$ (see~\cite{Seljak:2008xr, Seljak:2009af}). 
In this way one is able to reduce the error in simulations due to sampling variance at low $k$'s, where the linear approximation becomes exact.

We complement the above by adding uniform $2\%$ errorbars on the quantities $r^{(hh)}$ and $r^{(hm)}$ obtained 
from the simulations data to account for possible systematics. We also note again here that all the final bias
components in eq.~(\ref{eq:Phm_fin}) and in eq.~(\ref{eq:Phh_fin})  (i.e. the functional forms multiplying the different bias terms) take into account 
the appropriate resummation of IR modes, as shown in~\cite{Senatore:2014vja} and displayed for the dark matter case in Fig.~\ref{fig:Pmm}.

\begin{figure*}[t!]
   \begin{center}
   \hspace*{-0.5cm}
   \includegraphics[scale=0.51]{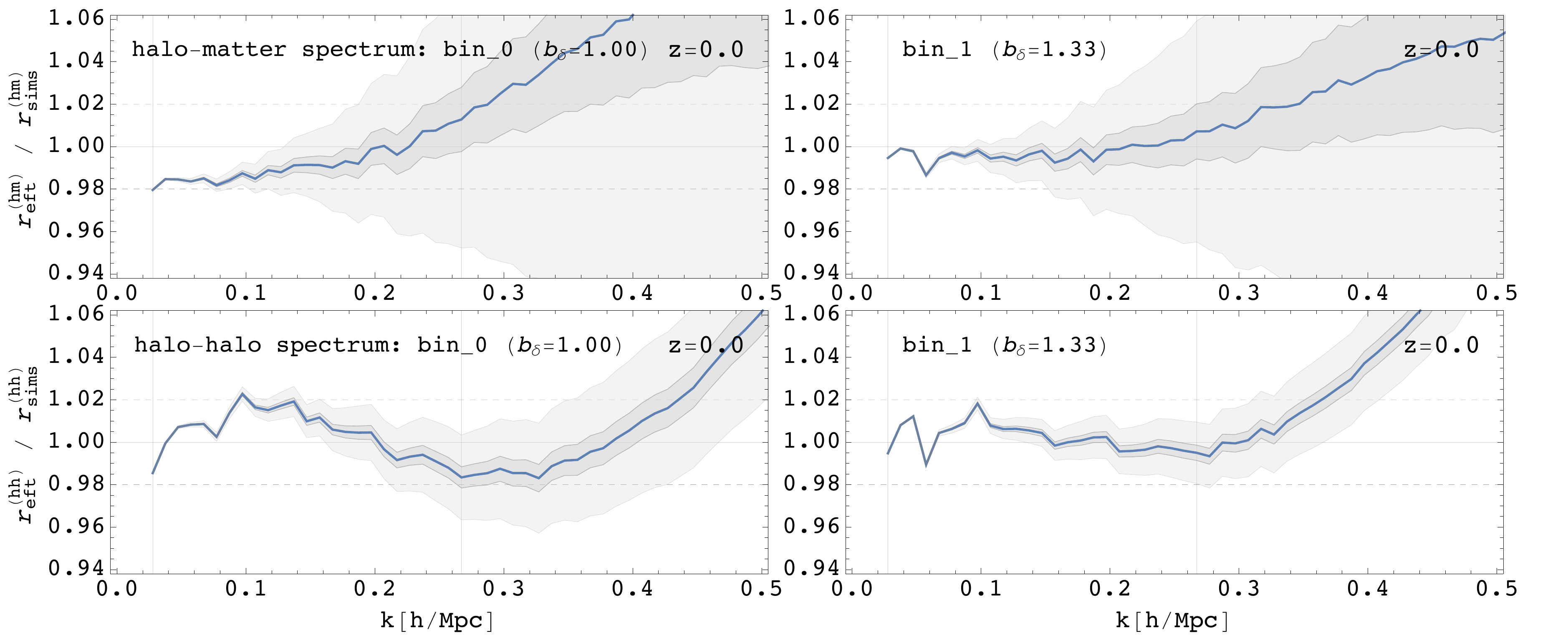}
   \end{center}
   \vspace*{-0.5cm}
   \caption{\small Halo-matter (upper panels) and halo-halo (lower panels) power spectrum ratios
   are shown for the mass bin0 ($b_{\df}=1.0$) on the left, and for bin1 ($b_{\df}=1.33$) on the right. The solid blue line is our theoretical prediction  
   containing seven bias parameters (see table~\ref{tb:bias}), divided by the nonlinear simulation 
   data measurements. The thin gray lines correspond to the theoretical error estimates stemming 
   from our neglecting two-loop corrections. 
   The horizontal thin dashed gray lines signal the $2\%$ error  that we input to account for the systematics in simulations.
  Consistently, we see that the theory stops matching the data when the theoretical error becomes sizable.   }
   \label{fig:PSbin0}
\end{figure*}

We construct the halo-halo $\chi^2_{hh}$ and halo-matter  $\chi^2_{hm}$ function, which depend polynomially 
on six bias parameters ($b_{\delta\,\epsilon}$ enters only in the bispectra):  
\begin{align}
 \chi^{2}_{r_{xy}} = \sum_{k_{i}\le k_{{\text{max}},r(k)}}
 \frac{\left[r^{(xy)}(k_i)-r^{(xy)}_{\text{NL}}(k_{i})\right]^{2}}{\Delta r^{(xy)}_{\text{NL}}(k_{i})^{2}}\, , 
 \label{eq: chi2_00}
\end{align}
where the $(xy)$ indices may represent both, halo-matter $(hm)$ cross-correlation or halo-halo $(hh)$ auto-correlation. 
The theoretical prediction $r^{(xy)}$ is evaluated at the $k_i$ points of simulation measurements denoted as $r^{(xy)}_{\text{NL}}$.
The error on the spectrum ratios $\Delta r^{(xy)}_{\text{NL}}$, as already noted above, consists of two components: 
the leading error estimate is derived from the equation~(\ref{eq:Power_Errors}), to which a $2\%$ is added to account for possible unkown systematics in simulations and in the comparison between the EFTofLSS and simulations. 
Finally, $k_{{\text{max}},r(k)}$ is the largest wavenumber in the power spectrum ratio considered in the sum.
 
In a similar fashion, one may construct the $\chi^2$ function for the three bispectrum observables:  
\begin{align}
 \chi^{2}_{B_{hxy}} = \sum_{k_{1i,2j,3l}\le k_{{\text{max}},B(k)}}
 \frac{\left[B^{hxy}(k_{1i},k_{2j},k_{3l}) -B^{hxy}_{\text{NL}}(k_{1i},k_{2j},k_{3l})\right]^{2}}{\Delta B^{hxy}_{\text{NL}}(k_{1i},k_{2j},k_{3l})^{2}}, 
 \label{eq: chi2_Bk}
\end{align}
where $x$ and $y$ can represent halo $h$ or matter $m$ . 
The theoretical bispectrum predictions $B^{hxy}$ are expressed according to equation~(\ref{eq:Bisspec}) and evaluated at the $k_i$ 
points of the simulation measurements $B^{hxy}_{\text{NL}}$. Again, the errors of the bispectra $\Delta B^{hxy}_{\text{NL}}$ 
also have two components: the first is given by the simulation error estimate in equation~(\ref{eq:Bispec_Errors}), the second is the $2\%$ 
covering for possible unidentified systematics. The quantity $k_{{\text{max}},B(k)}$ is the largest wavenumber allowed into the sum \footnote{Incidentally, we remind the reader that the $B^{hmm}$ bispectrum depend only on three bias 
parameters $b_{\delta,1}$, $b_{\delta,2}$, and $b_{\df^2}$, which is, at tree level, equivalent to the local in time approximation
(see e.g.~\cite{Chan:2012jj, Baldauf:2012hs, Saito:2014qha,Schmittfull:2014tca} ), but this degeneracy is expected to break down at higher orders. 
On the other hand, $B^{hhh}$ and $B^{hhm}$ include two shot noise contributions that were not considered earlier in the literature, even in the local-in-time approximation: 
${\rm Const}_\epsilon$ and $b_{\epsilon\delta}$. Furthermore, we remind the reader that all former analytic techniques, when applied to halos, are harmed by the fact that the analytic description of dark matter is incorrect as soon as loop corrections are considered.}.

Assuming independence for both the two and three point statistics, i.e. neglecting the correlation between any of the 
$\chi^2$ functions, the total $\chi^2_{\text{tot}}$ is a sum of all five components:
\begin{align}
\chi^2_{\text{tot}} = \chi^{2}_{r_{hm}} +\chi^{2}_{r_{hh}} +\chi^{2}_{B_{hmm}}+\chi^{2}_{B_{hhm}}+\chi^{2}_{B_{hhh}}.
\end{align}
Since $\chi^2_{\text{tot}}$ is a polynomial of sixth order in the bias parameters, it has well defined minima (assuming reasonable 
constraints on parameters). As a consequence, minimization can be achieved by applying any of the standard minimization tools (as found e.g. in the Mathematica package).
 
This is a good point to discuss our choice of the maximum wavenumber over which, for each given observable, we perform the fit to data. We choose this interval with care in order to avoid as much as possible any issue with overfitting. First of all, we can estimate the $k$-reach of each observable by using the estimates in sec.~\ref{subsec:theoryerrors}. The results are reported in table~\ref{tb:reach}~\footnote{We notice in passing that this error bar is determined by when the higher order terms are estimated to be a 3\% fraction of the non-linear results, after subtracting for the shot noise, which dominates the signal at relatively high~$k$'s.}. We therefore choose the following $k_{\rm max}$ for the various observables and bins over which we optimize the fit between theory and simulations. For the power spectra, for each bin the $k$-interval is shown in Fig.s~\ref{fig:PSbin0} and~\ref{fig:PSbin123} below as delimited by two vertical lines between approximately $0.03\hinvMpc\leq k_{\rm fit}\leq 0.26\hinvMpc$. Instead, for the bispectra, we choose the $k_{\rm max}$ depending on the bin: $k_{\rm max,B}=0.16,0.11,0.10\hinvMpc$, for bin0,1, and 2 respectively. It is important to notice two aspects of these choices. First, the risk of overfitting by choosing too a large $k_{\rm max}$ in the bispectrum is very limited once the available number of data is large enough. This is due to the complexity of the functional form of the bispectrum~\footnote{ Unfortunately, the $k$-reach of bin2 is so low that we  have only about 32 bispectra data. This is dangerously small a number, and with a dangerously large error due to the large cosmic variance at low $k$'s, to sufficiently determine all our bias coefficients with confidence. This means that there is a risk that the results for bin2 are overfitted, and they should be therefore taken with great caution. This issue should not apply to bin0 and bin1, where the bispectra data points we use are respectively 112 and 41.}. For the power spectra, where there appear three bias coefficeints that do not appear in the bispectra, the risk of overfitting is more realistic, and we therefore limit the $k$-region of the fit to a region which is consistent with our expectations from the theoretical errors. Second aspect to notice is that in order to establish the estimates for the $k$-reach in table~\ref{tb:reach}, one needs to have the numerical values of the biases, which can be determined only after fitting. We have therefore implemented the fitting procedure iteratively, obtaining a self-consitent solution.

\subsection{Results and Plots}
\label{subsec:fits}

\begin{figure*}[t!]
   \begin{center}
   \hspace*{-0.5cm}
   \includegraphics[scale=0.65]{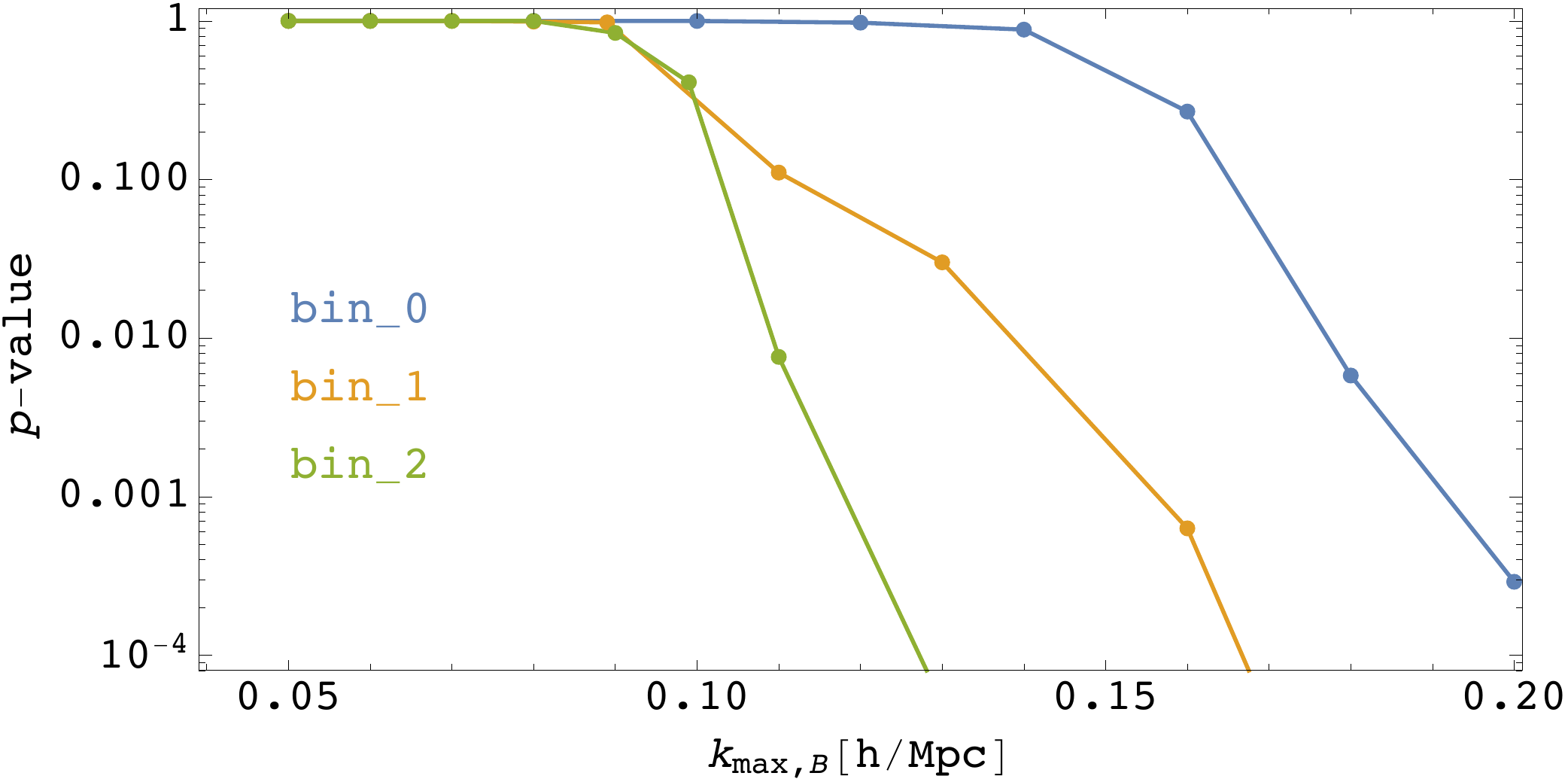}
   \end{center}
   \vspace*{-0.5cm}
   \caption{\small $p$-values corresponding to three halo mass bins, shown as a function of the maximum 
   triangle side length $k_{{\rm max},B}$ for the bispectra (and fixed $k_{{\rm max},P}=0.3\hinvMpc$). 
   We see the characteristic sharp drop in the $p$-value after the maximal scale $k_{{\rm max},B}$.
   As expected for higher mass bins results fail earlier, around $k_{{\rm max},B}=0.08$ for bin0 and then at slightly 
   higher scales for middle mass halos (around  scales of $k_{{\rm max},B}=0.09$) for bin1 and bin2). For lithest 
   mass bin0 expansion preforms up to scales $\sim k_{{\rm max},B}=0.15$) 
   }
   \label{fig:chi-p_plot}
\end{figure*} 

In this part we present bias parameters measured from the simulated data of halo-matter, halo-halo power spectra as well as 
halo-halo-halo, halo-halo-matter and halo-matter-matter bispectra.

In Fig.~\ref{fig:chi-p_4tables} we show the $p$-values and minimal $\chi^2$ for individual and joined best fits. 
We start from joint fits just for the two point functions, we do so for the three different halo mass ranges. 
We find that using seven free parameters gives  good $p$-value up to the theoretically expected $k$-ranges for both, power spectra and bispecta statistics. 
Adding the three bispectrum observables results in more constraining power and gives a $p$-value of
$30\%$ for bin0 ($b_{\delta,1}\sim 1.0$) with a maximal scale of $k_{{\rm max},B}=0.16\hinvMpc$. 

Increasing the halo mass (i.e. higher $b_{\delta,1}$) whilst keeping the $p$-value in the range $15-40\%$, causes the
$k_{{\rm max},B}$ scale to drop to the value $0.11\hinvMpc$ and $0.10\hinvMpc$ for bin1 and bin2 respectively.  We also note that, having the same value of $k_{{\rm max},B}$ for bin1 and $2$ is due to the discreteness of our 
simulation data points, and that a finer $k$ binning would probably slightly increase the range $k_{{\rm max},B}$ of bin1 
or/and decrease it for bin2 (as suggested by the p-value in table~\ref{fig:chi-p_4tables}).

In table~\ref{tb:bias} we show the best fit values of the bias parameters for the three halo mass bins. 
The parameters are obtained by means of a global fit to all five observables in Fig.~\ref{fig:chi-p_4tables} (last line in the tables) 
with the corresponding $k_{{\rm max},B}$ and with $k_{{\rm max},P}=0.3\hinvMpc$. The error bars of the best fit parameters given in the table are obtained by estimating the 50\% drop in the of  $p$-value relative to the one given by the best fit parameters.

In Fig.~\ref{fig:PSbin0} we show the results for the halo-matter and halo-halo power spectrum ratios for mass bin0 (on the left), and bin1 (on the right), 
using the parameter values given in table~\ref{tb:bias}. We show the (ratio of) power spectra ratios relative to the 
results obtained from $N$-body simulation measurements. As a shaded region in the plot we also display the estimate for the theoretical 
error due to our neglecting two-loop effects, multiple and divided by 2 to account for our uncertainties. As always, the $2\%$ error 
corresponding to the potential systematics in the simulations in included. Notice that in the cross correlation between matter and halos, the deviation of the prediction of the EFTofLSS from the non-linear data happens approximately when the theoretical error becomes relevant.  This is an important confirmation of the correctness of the EFTofLSS, whose predictions fail when they are expected to fail. The failure of the EFTofLSS happens at the $k$-values as estimated from the theoretical error also for the halo-halo power spectrum,  even though this is not visible in the plot, because the signal is dominated by shot noise, while the theoretical error is computed relative to the subleading non-shot noise signal. This is why the theoretical error appears small.  Similar plots corresponding to the higher 
halo mass bin2 are in Fig.~\ref{fig:PSbin123} in Appendix~\ref{app:bbins}. 
The plots of all the bispectra (cross and auto): halo-matter-matter, halo-halo-matter, halo-halo-halo, for all the 
mass bin0 can be found in the Appendix Fig.  \ref{fig:BiMMHbin0}, \ref{fig:BiMHHbin0}, \ref{fig:BiHHHbin0}.

In Fig.~\ref{fig:chi-p_plot} we show the $p$-values corresponding to three halo mass bins, as a function 
of the maximum triangle side length $k_{{\rm max},B}$ for the bispectra (and fixed $k_{{\rm max},P}=0.26\hinvMpc$). 
We see the characteristic sharp drop in the $p$-value after the maximal scale $k_{{\rm max},B}$. Within these scales 
our EFT results fit the data well, and then fail after crossing of $k_{{\rm max},B}$  scales.
Higher mass (higher bias $b_{\df,1}$) objects have typically smaller $k_{{\rm max},B}$ range. The sharp drop in the $p$-value as a function of $k_{{\rm max},B}$ is indication of the power of the bispectra in constraining the parameters of the theory.

Finally we look at the results when we exclude bias parameters $b_{\epsilon\,\delta}$ or $b_{\df^2}$ from our theory predictions.
We find that removing the $\epsilon\delta$ operator induces a sharp drop in the $p$-value. This is not the case if one alternatively removes the operator in $\delta^2$. This suggests that the functional form of some of these parameters is similar, given the error bars of our data. We leave a study of the similarity of the contribution of the various operators to future work. We notice that once the number of data points is much larger than the number of parameters, it is not a big issue to have degenerate, or quasi degenerate, parameters, unless one is using large prefactors to force unnatural cancellations between the contributions, as it is not the case here.

\subsection{Configuration-space results for the cross and auto power spectra}

\begin{figure*}[t!]
   \begin{center}
   \hspace*{-0.5cm}
   \includegraphics[scale=0.7]{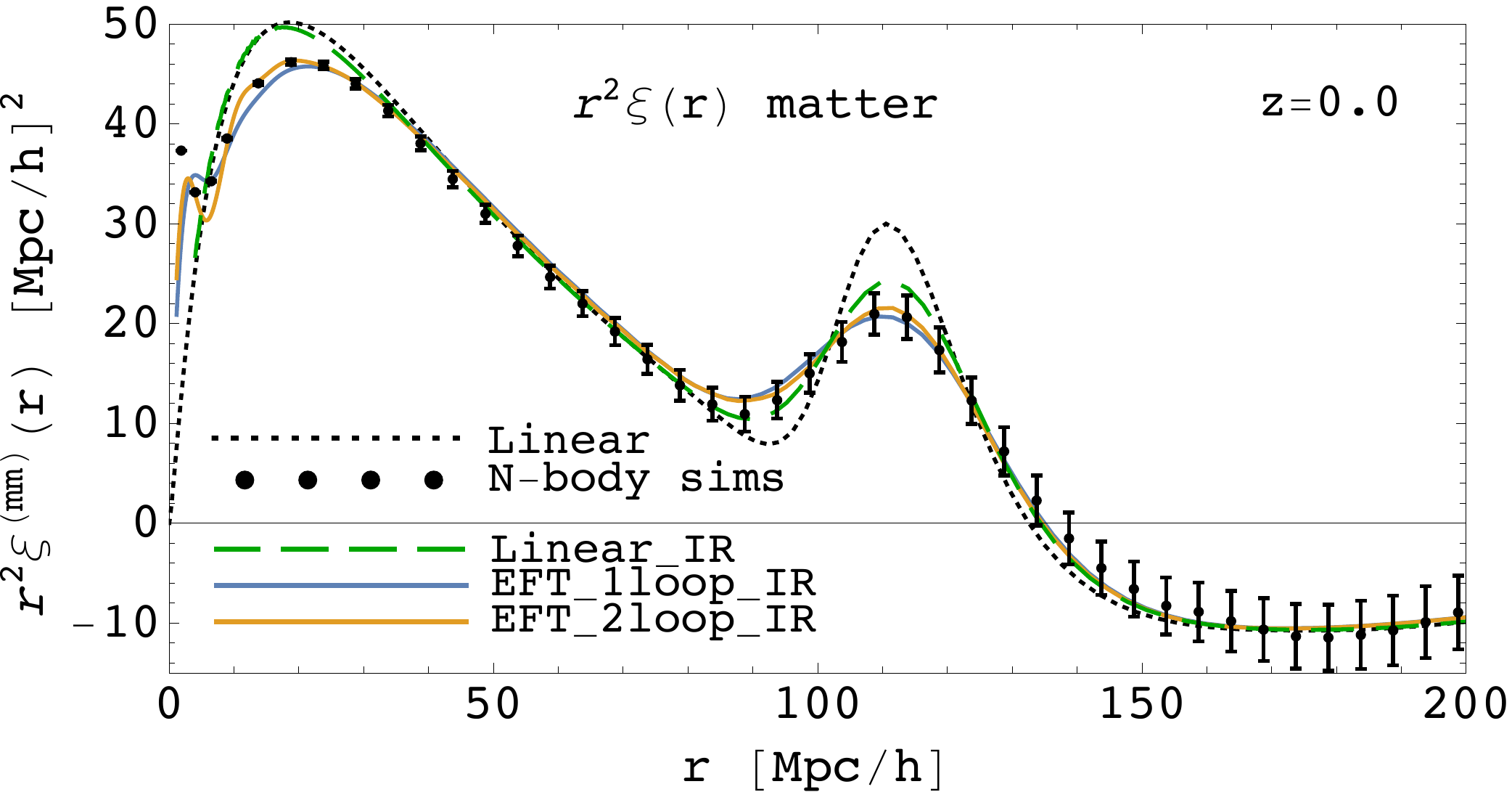}
   \end{center}
   \vspace*{-0.5cm}
   \caption{\small Dark matter correlation function. The subscript ${}_{\rm IR}$ means that we have performed the IR resummation. Error bars from simulations represent the estimated $2\sigma$ (see~\cite{Sanchez:2008iw}). Error bars are correlated though we do not quantify their correlation.
   }
   \label{fig:corfncDM}
\end{figure*} 

\begin{figure*}[t!]
   \begin{center}
   \hspace*{-0.5cm}
   \includegraphics[scale=0.43]{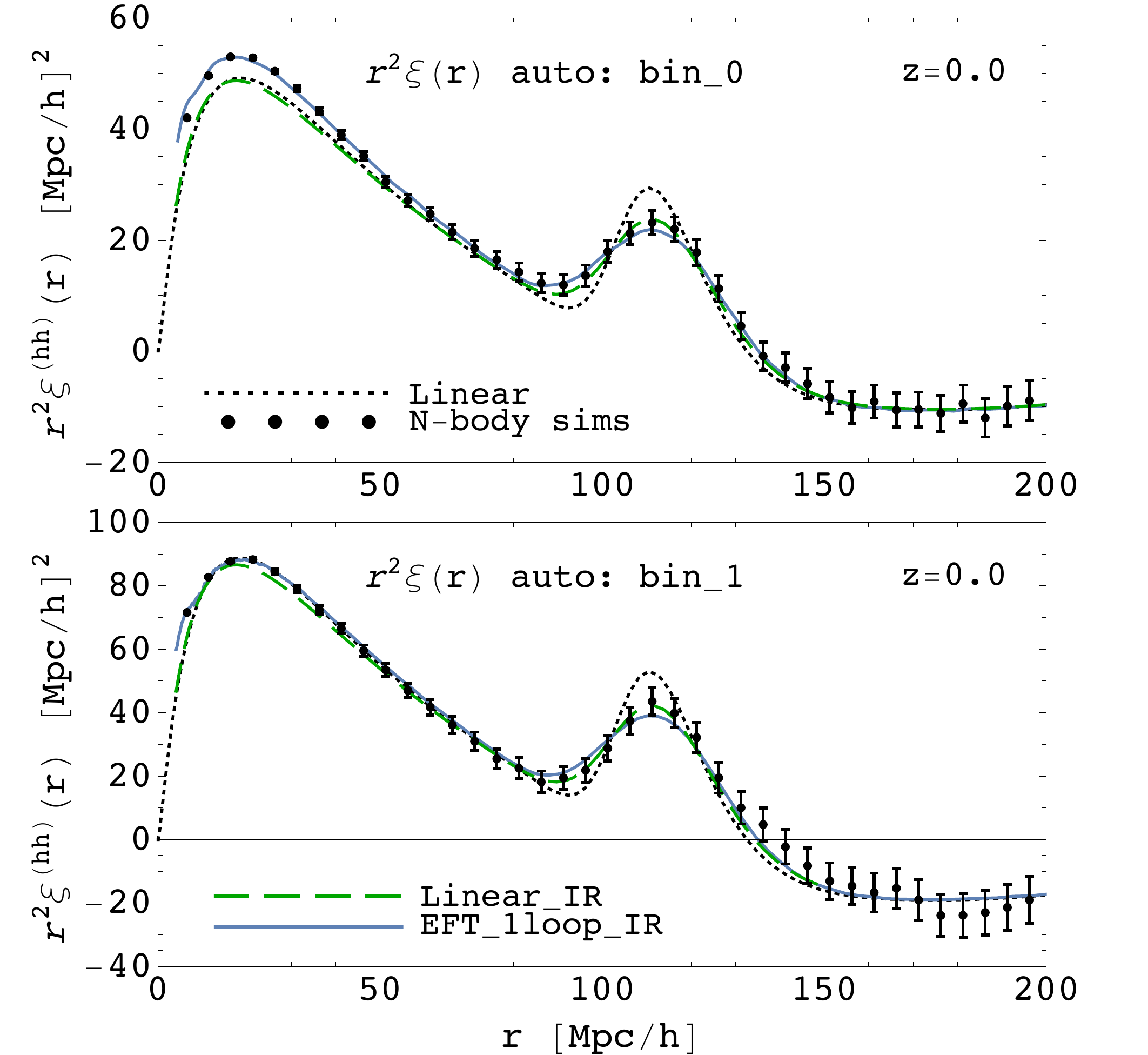} \hspace*{-0.8cm}
   \includegraphics[scale=0.43]{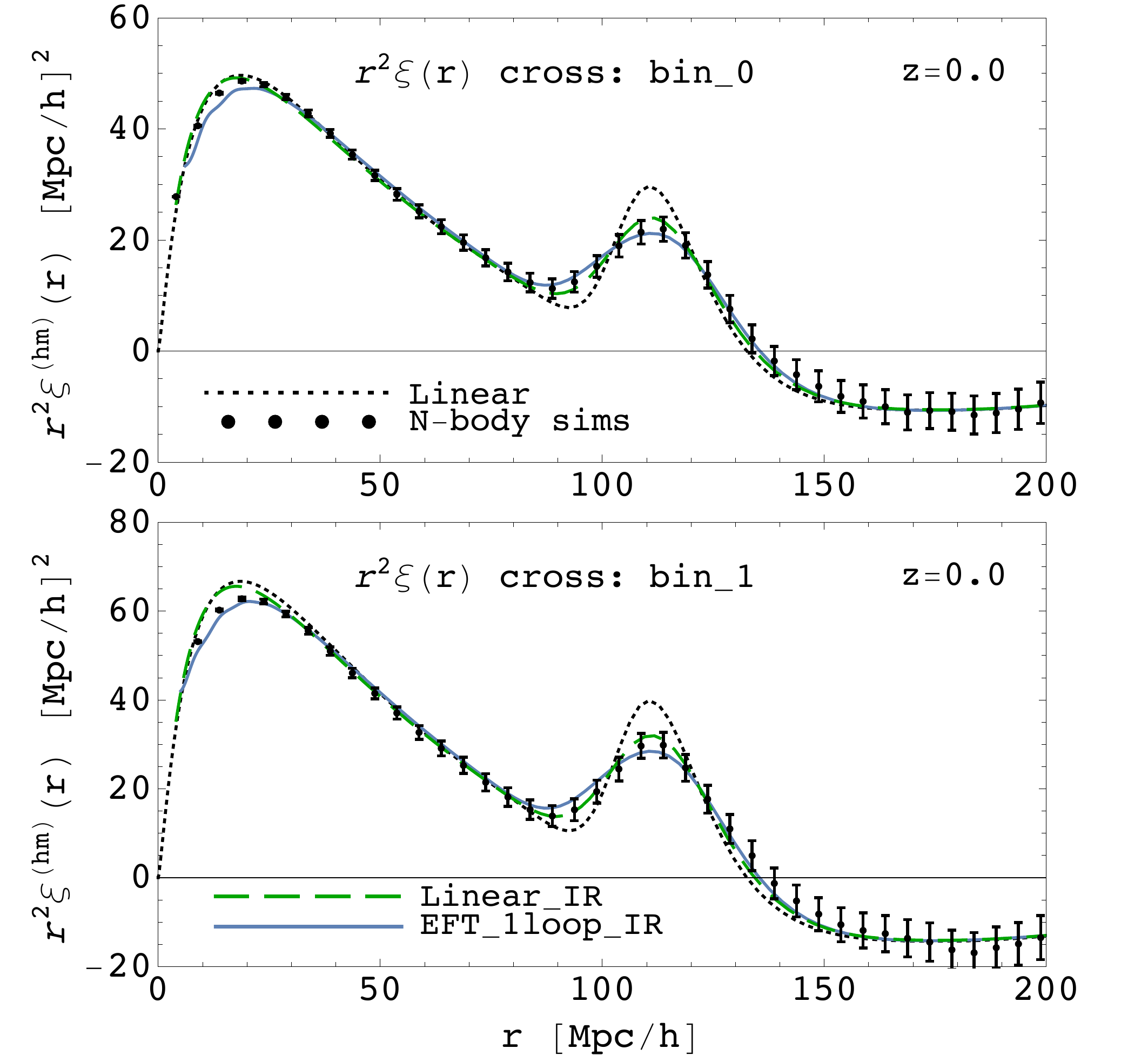}
   \end{center}
   \vspace*{-0.5cm}
   \caption{\small Halo-matter and halo-halo correlation functions. Error bars from simulations represent the estimated $2\sigma$ region (see~\cite{Sanchez:2008iw}). Error bars are correlated though we do not quantify their correlation. The green curve is obtained by using only the $b_{\delta_1}$ bias term and linear theory, with tree-level IR-resummation, while the blue curve represent the EFTofLSS prediction at one-loop, with one-loop IR-resummation. }
   \label{fig:corfnc}
\end{figure*} 

For completeness,we show the results in configuration space.
Correlation functions can be obtained by simple Fourier transform
\begin{align}
\xi^{xy}(r)=\int{\frac{d^3k}{(2\pi)^3}~P^{xy}(k)e^{-i\vec {k} \cdot \vec {r}}}
 = \int \frac{k^2dk}{2\pi^2}~P^{xy}(k) j_0(kr)~,
\end{align}
where $j_0$ is the spherical Bessel function. The only subtle point in this procedure is that the EFTofLSS in Fourier space fails at high $k$'s beyond the estimated theoretical reach. When it fails, it does so with large deviations from the non-linear power spectrum. In order for the errors at high wavenumber not to erroneously affect the correlation functions at large distances, we extrapolate the power spectra in the EFT to high $k$ beyond their reach, about $k\simeq 0.3\hinvMpc$ for one loop and $k\simeq 0.6\hinvMpc$ for two loops, by using some well behaved function. Results are insensitive to the procedure when the extrapolation function is reasonable. For example, we extrapolated the power spectrum by continuously connecting the EFT result with a rescaled linear power spectrum or by applying a Gaussian damping $e^{-\frac{k^2}{20 (h /{\rm Mpc})^{2}}}$, without finding appreciable changes in the plots we show below.

In Fig.~\ref{fig:corfncDM} we show the dark matter cross correlation. In Fig.~\ref{fig:corfnc} we show the cross and auto correlation functions corresponding to the 
halo-halo and halo-matter power spectra. For both figures, the $1\sigma$ error bars on the simulation data are given by the square root of the diagonal elements of the respective
covariance matrices calculated in the Gaussian limit following~\cite{Sanchez:2008iw}. Focussing on the region of the BAO peak, we can see that IR-resummation of the contribution from the displacements is crucial to reach agreement with the non-linear power spectrum (see for example~\cite{Senatore:2014via}). Subsequent non-linear corrections at one loop and two loops (for dark matter) seem to provide a relatively small effect in that region, shifting the curve within the simulation error bars. At short distances, the theory curves begin to fail, as expected by the fact that the power spectrum in the EFT stops being correct at high wavenumbers. Given that at low wavenumbers the EFT power spectra do not present any apparent residual, we indeed find that at long distances, and in particular around the BAO peak, the results are compatible with the error bars from simulations.

\section{Conclusions}
\label{sec:conclusion}

Understanding the behavior of the Large Scale Structures (LSS) of the Universe has become a vital prerequisite if we wish to continue to gather information about the primordial universe. The EFTofLSS provides a novel technique to analytically describe the LSS at large distances, where perturbations are small. So far, the EFTofLSS has been compared with simulation data for statistics involving the dark matter field, and, as we have described in the Introduction, the success has been remarkable. In this paper we have first extended the theoretical description of biased tracers in the EFTofLSS by including the effect of baryonic physics and primordial non-Gaussianity. We have then compared the predictions for biased tracers of the  EFTofLSS against dark-matter-only simulations with Gaussian initial conditions. We have focused on five different statistics: the halo-halo power spectrum, the halo-matter cross correlation at one-loop level, and the halo-halo-halo, halo-halo-matter, and halo-matter-matter bispectra at tree level. We have found that the predictions of the EFTofLSS for biased tracers at this order depend on seven bias parameters, after taking account of degeneracies. Four of these parameters enter in all of the five statistics we consider.  According to each mass bins, we have estimated the theoretical error associated to the next order terms, allowing us to estimate the $k$-reach of each computations, and finding that lower mass bins should have a larger $k$-reach than the higher ones.  We have restricted our study to three mass bins, and found that, at the order at which we compute, the two-point functions from EFTofLSS match the numerical data to percent level up to $k\simeq0.3\hinvMpc$, while the bispectra match until $k\simeq0.16\hinvMpc$, $k\simeq0.11\hinvMpc$ and $k\simeq0.10\hinvMpc$ for bin0, bin1 and bin2 respectively. These results are consistent with the theoretical expectations~\footnote{Notice that our simulations have $\sigma_8=0.9$, implying that the universe we compare with is quite more non-linear than the one we currently live in. This suggests that the $k$-reach of our calculations will be slightly larger in more realistic cosmologies.}. We also shown results for the same two-point functions as just described, but in configuration space, finding consistent results.

We believe that our findings, if confirmed by other studies and comparisons, open the way to understanding in an analytical way the statistics of biased tracers up to quite larger wavenumbers than pervasively believed, and to complement the role of simulations at smaller distances where physics is non-linear. Hopefully, this complementary approach will allow us to extract much cosmological information from on-going and next generation LSS surveys.

\section*{Acknowledgments}

We would like to thank Simon Foreman for sharing some of his Mathematica codes. M.F. is supported by NSF grant PHY-1068380.  L.S. is supported by DOE Early Career Award DE-FG02-12ER41854 and by NSF grant PHY-1068380. Z.V. is supported in part by the U.S. Department of Energy contract to SLAC no. DE-AC02-76SF00515.


\appendix
\section*{Appendix}

\section{$\mathbb{C}_i$ operators}
\label{app:operators}

In this appendix we provide a direct connection between the operators in equation~(\ref{eq:euler_bias_4}) and the ones identified by $\mathbb{C}^{(N)}_i$
in eq. (\ref{eq:CoI_all}): 
\bea \label{eq:C_all_explicit}
&&\text{1st order:}  \\ \nonumber
&&\mathbb{C}^{(1)}_{\delta,1} (\vec k,t) = \delta^{(1)} (\vec k,t)\, ,   \\ \nonumber
&& {\mathbb{C}_{\epsilon}(\vec k,t)= [\epsilon]_{\vec k}  }  \, , \\ \nonumber
&& {\mathbb{C}^{(1)}_{\epsilon\delta,1}(\vec k,t)= [\epsilon\delta]^{(1)}_{\vec k} } \, ,\\ \nonumber
&& \\ \nonumber
&&\text{2nd order:}\\ \nonumber
&&\mathbb{C}^{(2)}_{\delta,1} (\vec k,t) = [\d_i \delta^{(1)}\; \frac{\d^i}{\d^2}\theta^{(1)}]_{\vec k}(t)\,  , \\ \nonumber
&&\mathbb{C}^{(2)}_{\delta,2} (\vec k,t) = \delta^{(2)}(\vec k,t) - [\d_i \delta^{(1)}\; \frac{\d^i}{\d^2}\theta^{(1)}]_{\vec k}(t) \, ,\\ \nonumber
&&\mathbb{C}^{(2)}_{\delta^2,1} (\vec k,t) =    [\delta^2]_{\vec k}^{(2)}(t) \, ,\\ \nonumber
&&\mathbb{C}^{(2)}_{s^2,1}(\vec k,t) =   [s^2]_{\vec k}^{(2)}(t)  \,  ,\\ \nonumber
&&\mathbb{C}^{(2)}_{\epsilon,2} (\vec k,t) = [\epsilon^{(2)}]_{\vec k} \, , \\ \nonumber
&& \\ \nonumber
&&\text{3rd order:}\\ \nonumber
&&\mathbb{C}^{(3)}_{\delta,1} (\vec k,t) = \frac{1}{2}  [\d_i \delta^{(1)}\; \frac{\d^i}{\d^2}\theta^{(2)}]_{\vec k}(t) 
+\frac{1}{2}  \left( [\d_i \delta^{(1)}\; \frac{\d_j\d^i}{\d^2}\theta^{(1)}\; \frac{\d^j}{\d^2}\theta^{(1)}]_{\vec k}(t) \right.  \\ \nonumber
&&\qquad\qquad\qquad\qquad\qquad\qquad\qquad\qquad\qquad\qquad \left.
+[\d_i\d_j \delta^{(1)}\;\frac{\d^i}{\d^2}\theta^{(1)}\frac{\d^j}{\d^2}\theta^{(1)}]_{\vec k}(t)\right)  \, ,\\  \nonumber
&&\mathbb{C}^{(3)}_{\delta,2} (\vec k,t) =  [\d_i \delta^{(2)}\; \frac{\d^i}{\d^2}\theta^{(1)}]_{\vec k}(t)  
- \left( [\d_i \delta^{(1)}\; \frac{\d_j\d^i}{\d^2}\theta^{(1)}\; \frac{\d^j}{\d^2}\theta^{(1)}]_{\vec k}(t) \right. \\  \nonumber
&&\qquad\qquad\qquad\qquad\qquad\qquad\qquad\qquad\qquad\qquad \left.
+[\d_i\d_j \delta^{(1)}\;\frac{\d^i}{\d^2}\theta^{(1)}\frac{\d^j}{\d^2}\theta^{(1)}]_{\vec k}(t)\right)   \, ,\\ \nonumber
&&\mathbb{C}^{(3)}_{\delta,3} (\vec k,t) =  \delta^{(3)}(\vec k,t) - [\d_i \delta^{(2)}\; \frac{\d^i}{\d^2}\theta^{(1)}]_{\vec k}(t) 
+ \frac{1}{2} \left( [\d_i \delta^{(1)}\; \frac{\d_j\d^i}{\d^2}\theta^{(1)}\; \frac{\d^j}{\d^2}\theta^{(1)}]_{\vec k}(t) \right. \\  \nonumber
&&\qquad\qquad\qquad\qquad\qquad\qquad\qquad\qquad\qquad\qquad \left.
+[\d_i\d_j \delta^{(1)}\;\frac{\d^i}{\d^2}\theta^{(1)}\frac{\d^j}{\d^2}\theta^{(1)}]_{\vec k}(t)\right)   \, , \\ \nonumber
&&\mathbb{C}^{(3)}_{\delta^2,1} (\vec k,t) = -2 [\delta^{(1)} \d_i \delta^{(1)} \frac{\d^i}{\d^2} \theta^{(1)}]_{\vec k}(t)  \, ,\\ \nonumber 
&&\mathbb{C}^{(3)}_{\delta^2,2} (\vec k,t) = [\delta^2]^{(3)}_{\vec k}(t)+2  [\delta^{(1)} \d_i \delta^{(1)} \frac{\d^i}{\d^2} \theta^{(1)}]_{\vec k}(t)     \, ,\\ \nonumber 
&&\mathbb{C}^{(3)}_{\delta^3,1} (\vec k,t) = [\delta^3]^{(3)}_{\vec k}(t) \,  ,\\ \nonumber 
&&\mathbb{C}^{(3)}_{s^2,1} (\vec k,t) = -2 [s_{lm}^{(1)}\d_i (s^{lm})^{(1)}\frac{\d^i}{\d^2}\theta^{(1)}]_{\vec k}(t) \, , \\ \nonumber 
&&\mathbb{C}^{(3)}_{s^2,2} (\vec k,t) = [s^2]^{(3)}_{\vec k}(t)+2  [s_{lm}^{(1)}\d_i (s^{lm})^{(1)}\frac{\d^i}{\d^2}\theta^{(1)}]_{\vec k}(t)    \,  ,\\ \nonumber 
&&\mathbb{C}^{(3)}_{s^3,1} (\vec k,t) = [s^3]^{(3)}_{\vec k}(t)   \,  , \\ \nonumber 
&&\mathbb{C}^{(3)}_{st,1} (\vec k,t) = [st]^{(3)}_{\vec k}(t) \, ,\\ \nonumber 
&&\mathbb{C}^{(3)}_{\psi,1} (\vec k,t) = \psi^{(3)}(\vec k,t)   \,  ,\\ \nonumber 
&&\mathbb{C}^{(3)}_{\delta s^2,1} (\vec k,t) =   [\delta s^2]^{(3)}_{\vec k}(t)   \ .  \nonumber
\eea 
Below we give the explicit form of the $\mathbb{C}_i$ operators introduced in the equation~(\ref{eq:CoI_all}) in term of linear densities and kernels:
\begin{align} \label{eq:C_all_explicit_2}
\mathbb{C}^{(1)}_i (\vec k,t)&= \int \frac{d^3q_1}{(2\pi)^3\;)}\widehat{c}^{(1)}_{s,i} (q_1)\delta_D^{(3)}(\vec k-\vec q_1)\delta^{(1)} (\vec q_1,t), \\ \nonumber
\mathbb{C}^{(2)}_i (\vec k,t)&= \int \frac{d^3q_1}{(2\pi)^3}\;\frac{d^3q_2}{(2\pi)^3}\;\; 
                                           \widehat{c}^{(2)}_{s,i} (q_1,q_2)\delta_D^{(3)}(\vec k-\vec q_1-\vec q_2)\delta^{(1)} (\vec q_1,t)\delta^{(1)} (\vec q_2,t), \\ \nonumber
\mathbb{C}^{(3)}_i (\vec k,t)&= \int \frac{d^3q_1}{(2\pi)^3}\;\frac{d^3q_2}{(2\pi)^3}\;\frac{d^3q_3}{(2\pi)^3}\;\; 
                                           \widehat{c}^{(3)}_{s,i} (\vec q_1,\vec q_2,\vec q_3)
                                           \delta_D^{(3)}(\vec k-\vec q_1-\vec q_2-\vec q_3)\delta^{(1)} (\vec q_1,t)\delta^{(1)} (\vec q_2,t)\delta^{(1)} (\vec q_3,t), \\ \nonumber
\mathbb{C}^{(3)}_{\delta,3_{c_s}} (\vec k,t) &= \delta^{(3)}_{c_s}(\vec k,t)\, ,
\end{align}
where $n$-th kernels $\widehat{c}^{(n)}_{s,i}$ are symmetrized in the momentum variables. The unsymmetrized  version of the kernels is given by:
\bea \label{eq:c_all_explicit}
\text{1st order:} \\ \nonumber
\widehat{c}^{(1)}_{\delta,1}(\vec q_1) &= & 1, \\ \nonumber
\\ \nonumber
\text{2nd order:}\\ \nonumber
\widehat{c}^{(2)}_{\delta,1}(\vec q_1,\vec q_2) &=& \sfrac{\V q_1\cdot\V q_2 }{q_1^2}, \\ \nonumber
\widehat{c}^{(2)}_{\delta,2}(\vec q_1,\vec q_2) &=& F^{(2)}\left(\V q_1, \V q_2\right) -  \sfrac{\V q_1\cdot\V q_2 }{q_1^2},\\ \nonumber
\widehat{c}^{(2)}_{\delta^2,1} (\vec q_1,\vec q_2) &=& 1,\\ \nonumber
\widehat{c}^{(2)}_{s^2,1} (\vec q_1,\vec q_2) &=& \sfrac{(\V q_1\cdot\V q_2)^2}{q_1^2 q_2^2} - \sfrac{1}{3},\\ \nonumber
\nonumber\\
\text{3rd order:}\nonumber \\ \nonumber 
\widehat{c}^{(3)}_{\delta,1}(\vec q_1,\vec q_2,\vec q_3) &=& \sfrac{1}{2} \left(\sfrac{\left(\V q_1 \cdot \V q_2+\V q_1 \cdot \V q_3\right)}{q_2^2+q_3^2
                                                                     +2 \V q_2 \cdot \V q_3} G^{(2)}\left(\V q_2, \V q_3\right)
                                                                     +\sfrac{\V q_1 \cdot \V q_2 \left(\V q_1 \cdot \V q_3
                                                                     + \V q_2\cdot \V q_3\right)}{q_2^2 q_3^2}\right), \\ \nonumber
\widehat{c}^{(3)}_{\delta,2}(\vec q_1,\vec q_2,\vec q_3) &=& \sfrac{\left(\V q_1\cdot \V q_3+\V q_2 \cdot \V q_3\right) }{q_2^2 q_3^2}
                                                                \left(F^{(2)}\left(\V q_1, \V q_2\right) q_2^2-\V q_1 \cdot \V q_2 \right), \\ \nonumber
\widehat{c}^{(3)}_{\delta,3}(\vec q_1,\vec q_2,\vec q_3) &=& F^{(3)}\left(\V q_1, \V q_2,  \V q_3\right)+\sfrac{\left(\V q_1+ \V q_2\right)\cdot \V q_3}{2q_2^2 q_3^2}
                                                                 \left(\V q_1\cdot \V q_2-2  F^{(2)}\left(\V q_1, \V q_2\right) q_2^2\right)\\ \non
                                                                 &&\qquad\qquad\qquad\qquad\qquad
                                                                 -\sfrac{\V q_1\cdot \left(\V q_2+ \V q_3\right)}{2(q_2^2+q_3^2 + 2 \V q_2\cdot \V q_3)} G^{(2)}\left(\V q_2, \V q_3\right) \\ \non
\widehat{c}^{(3)}_{\delta^2,1} (\vec q_1,\vec q_2,\vec q_3) &=& -2\sfrac{\V q_2 \cdot \V q_3}{q_3^2} \\ \nonumber 
\widehat{c}^{(3)}_{\delta^2,2} (\vec q_1,\vec q_2,\vec q_3) &=& 2 F^{(2)}\left(\V q_1, \V q_2\right) + 2\sfrac{\V q_2 \cdot \V q_3}{q_3^2}  \\ \nonumber 
\widehat{c}^{(3)}_{\delta^3,1} (\vec q_1,\vec q_2,\vec q_3) &=& 1 \\ \nonumber 
\widehat{c}^{(3)}_{s^2,1} (\vec q_1,\vec q_2,\vec q_3) &=& -2\sfrac{\V q_2 \cdot \V q_3}{q_3^2} 
                                                                  \left( \sfrac{\left(\V q_1 \cdot \V q_2\right)^2}{q_1^2 q_2^2} -\sfrac{1}{3} \right) \\ \nonumber 
\widehat{c}^{(3)}_{s^2,2} (\vec q_1,\vec q_2,\vec q_3) &=& 2F^{(2)}\left(\V q_1, \V q_2\right)\left( \sfrac{\left((\V q_1 + \V q_2)\cdot \V q_3\right)^2}
                                                                  {(\V q_1 + \V q_2)^2 q_3^2} -\sfrac{1}{3} \right)+2\sfrac{\V q_2 \cdot \V q_3}{q_3^2} 
                                                                  \left( \sfrac{\left(\V q_1 \cdot \V q_2\right)^2}{q_1^2 q_2^2} -\sfrac{1}{3} \right)\\ \nonumber 
\widehat{c}^{(3)}_{s^3,1} (\vec q_1,\vec q_2,\vec q_3) &=& \sfrac{9 \V q_1\cdot \V q_2 \V q_1\cdot \V q_3 \V q_2\cdot \V q_3-3 \left(\V q_1\cdot \V q_3\right)^2 q_2^2-3 
                                                                \left(\V q_1\cdot \V q_2\right)^2 q_3^2 
                                                                + q_1^2 \left( -3 \left(\V q_2\cdot \V q_3\right)^2+2 q_2^2 q_3^2\right)}{9 q_1^2 q_2^2 q_3^2} \\ \nonumber 
\widehat{c}^{(3)}_{st,1} (\vec q_1,\vec q_2,\vec q_3) &=& \left( G^{(2)}\left(\V q_1, \V q_2\right) - F^{(2)}\left(\V q_1, \V q_2\right) \right) 
                                                                 \left( \sfrac{\left(\V q_1 \cdot \V q_2\right)^2}{q_1^2 q_2^2} -\sfrac{1}{3} \right)\\ \nonumber 
\widehat{c}^{(3)}_{\psi,1} (\vec q_1,\vec q_2,\vec q_3) &=&   G^{(3)}\left(\V q_1, \V q_2, \V q_3\right) - F^{(3)}\left(\V q_1, \V q_2, \V q_3\right)  \\ \nonumber 
                                                                  &&+ 2 F^{(2)}\left(\V q_1, \V q_2\right) \left( F^{(2)}\left(\V q_1 + \V q_2, \V q_3\right) 
                                                                  - G^{(2)}\left(\V q_1+ \V q_2, \V q_3\right)  \right) \\ \nonumber 
\widehat{c}^{(3)}_{\delta s^2,1} (\vec q_1,\vec q_2,\vec q_3) &=& \sfrac{\left(\V q_1 \cdot \V q_2\right)^2}{q_1^2 q_2^2} -\sfrac{1}{3}  \\ \nonumber 
\eea

There are degeneracies amongst the set of operators in equation~(\ref{eq:C_all_explicit}).
Using the $BoD$ basis introduced in the section~\ref{subsec:field} we are able to identify  degenerate operators and express them as linear combinations of the basis element: 
\bea \label{eq:C_all_explicit_4}
\text{2nd order operators:}\\ \nonumber
\mathbb{C}^{(2)}_{s^2,1} &=&  \sfrac{7}{2}\mathbb{C}^{(2)}_{\delta,2}- \sfrac{17}{6}\mathbb{C}^{(2)}_{\delta^2,1}, \\ \nonumber
&&\\ \nonumber
\text{3rd order operators:}\\ \nonumber
\mathbb{C}^{(3)}_{s^2,1} &=&  -\sfrac{7}{2}\mathbb{C}^{(3)}_{\delta,2} - \sfrac{17}{6}\mathbb{C}^{(3)}_{\delta^2,1}, \\ \nonumber
\mathbb{C}^{(3)}_{s^3,1} &=& \sfrac{21}{4}\mathbb{C}^{(3)}_{\delta,2} + \sfrac{45}{4}\mathbb{C}^{(3)}_{\delta,3}
                                                      - \sfrac{103}{8}\mathbb{C}^{(3)}_{\delta^2,1} -\sfrac{137}{16}\mathbb{C}^{(3)}_{\delta^2,2} 
                                                      + \sfrac{511}{72}\mathbb{C}^{(3)}_{\delta^3,1} -\sfrac{3}{4}\mathbb{C}^{(3)}_{s^2,2}, \\ \nonumber
\mathbb{C}^{(3)}_{st,1} &=&  \sfrac{7}{2}\mathbb{C}^{(3)}_{\delta,2} + \sfrac{9}{2}\mathbb{C}^{(3)}_{\delta,3}
                                                      - \sfrac{37}{12}\mathbb{C}^{(3)}_{\delta^2,1} -\sfrac{71}{24}\mathbb{C}^{(3)}_{\delta^2,2} 
                                                      + \sfrac{25}{12}\mathbb{C}^{(3)}_{\delta^3,1} -\sfrac{1}{2}\mathbb{C}^{(3)}_{s^2,2}, \\ \nonumber
\mathbb{C}^{(3)}_{\psi,1} &=& 2\mathbb{C}^{(3)}_{\delta,2} + 2\mathbb{C}^{(3)}_{\delta,3}
                                                      - \mathbb{C}^{(3)}_{\delta^2,1} -\sfrac{55}{42}\mathbb{C}^{(3)}_{\delta^2,2} 
                                                      + \mathbb{C}^{(3)}_{\delta^3,1} -\sfrac{2}{7}\mathbb{C}^{(3)}_{s^2,2}, \\ \nonumber
\mathbb{C}^{(3)}_{\delta s^2,1} &=&  \sfrac{7}{2}\mathbb{C}^{(3)}_{\delta^2,1} +\sfrac{7}{4}\mathbb{C}^{(3)}_{\delta^2,2} 
                                                      - \sfrac{17}{6}\mathbb{C}^{(3)}_{\delta^3,1}.
\eea 
Finally, we repeat here the expression for the bias tracer overdensity field given in terms of the $BoD$ operator basis~(eq.~(\ref{eq:delta_h_CoI_t})) more explicitly: 
\bea \label{eq:delta_h_CoI}
\delta_h(\vec k,t)&=& \tilde{c}_{\delta,1}(t) \; \Big( \mathbb{C}^{(1)}_{\delta,1}(\vec k,t)+\mathbb{C}^{(2)}_{\delta,1}(\vec k,t)+\mathbb{C}^{(3)}_{\delta,1}(\vec k,t) \Big)\\ \nonumber
&+& \tilde{c}_{\delta,2(2)}(t) \; \mathbb{C}^{(2)}_{\delta,2}(\vec k,t) + \tilde{c}_{\delta,2(3)}(t) \; \mathbb{C}^{(3)}_{\delta,2}(\vec k,t)\\ \nonumber
&+& \tilde{c}_{\delta,3}(t) \; \mathbb{C}^{(3)}_{\delta,3}(\vec k,t)
+ \tilde{c}_{\delta,3_{c_s}}(t) \; \mathbb{C}^{(3)}_{\delta,3_{c_s}}(\vec k,t)\\ \nonumber
&+& \tilde{c}_{\delta^2,1(2)}(t) \; \mathbb{C}^{(2)}_{\delta^2,1}(\vec k,t) + \tilde{c}_{\delta^2,1(3)}(t) \; \mathbb{C}^{(3)}_{\delta^2,1}(\vec k,t)\\ \nonumber
&+& \tilde{c}_{\delta^2,2}(t) \; \mathbb{C}^{(3)}_{\delta^2,2}(\vec k,t) + \tilde{c}_{s^2,2}(t) \; \mathbb{C}^{(3)}_{s^2,2}(\vec k,t)\\ \nonumber
&+& \tilde{c}_{\delta^3,1}(t) \; \mathbb{C}^{(3)}_{\delta^3,1}(\vec k,t)\\ \nonumber
&+& \tilde{c}_{\epsilon,1}(t) \; \mathbb{C}^{(3)}_{\epsilon,1}(\vec k,t) + \ldots
\eea
where the new bias coefficients $\tilde{c}_i$ are given in terms of the old ones from equation~(\ref{eq:CoI_all}): 
\bea \label{eq:c_tildas}
\tilde{c}_{\delta,1}&=&c_{\delta,1}\, , \\ \nonumber
\tilde{c}_{\delta,2(2)}&=&c_{\delta,2} + \sfrac{7}{2}c_{s^2,1}\,, \\ \nonumber
\tilde{c}_{\delta,2(3)}&=&c_{\delta,2} - \sfrac{7}{2}c_{s^2,1} +\sfrac{21}{4}c_{s^3,1} +\sfrac{7}{2}c_{st,1} + 2 c_{\psi,1}\,, \\ \nonumber
\tilde{c}_{\delta,3}&=&c_{\delta,3} +\sfrac{45}{4}c_{s^3,1} +\sfrac{9}{2}c_{st,1} + 2 c_{\psi,1}\,, \\ \nonumber
\tilde{c}_{\delta,3_{c_s}}&=&c_{\delta,3_{c_s}}\,, \\ \nonumber
\tilde{c}_{\delta^2,1(2)}&=&c_{\delta^2,1} - \sfrac{17}{6}c_{s^2,1}\,, \\ \nonumber
\tilde{c}_{\delta^2,1(3)}&=&c_{\delta^2,1} - \sfrac{17}{6}c_{s^2,1} -\sfrac{103}{8}c_{s^3,1} 
                                              -\sfrac{145}{25}c_{st,1} -\sfrac{97}{42} c_{\psi,1} + \sfrac{21}{4} c_{\delta s^2,1}\,, \\ \nonumber
\tilde{c}_{\delta^2,2}&=&c_{\delta^2,2} - \sfrac{137}{16}c_{s^3,1}, \\ \nonumber
\tilde{c}_{s^2,2}&=&c_{s^2,2}-\sfrac{3}{4}c_{s^3,1} -\sfrac{1}{2}c_{st,1}  -\sfrac{2}{7} c_{\psi,1}, \\ \nonumber
\tilde{c}_{\delta^3,1}&=&c_{\delta^3,1} + \sfrac{511}{72}c_{s^3,1} 
                                              -\sfrac{25}{12}c_{st,1} + c_{\psi,1} - \sfrac{17}{6} c_{\delta s^2,1}\ .
\eea

At last, we present an explicit expression for the third term of eq.~(\ref{eq:Phm}):
\begin{align}
\label{eq:P13_K}
3\; P_{11}(k;t,t) \int \frac{d^3q}{(2\pi)^3} \;& K^{(3)}_s(\vec k,-\q,\q) P_{11}(q;t,t) = \non \\
       \tilde{c}_{\delta,1}(t) \; 3\; &P_{11}(k;t,t) \int \frac{d^3q}{(2\pi)^3} \;\widehat{c}^{(3)}_{{\delta,1}}(\vec{k},-\q,\q) P_{11}(q;t,t) \non \\
     +\tilde{c}_{\delta,2(3)}(t) \; 3\; &P_{11}(k;t,t) \int \frac{d^3q}{(2\pi)^3} \;\widehat{c}^{(3)}_{{\delta,2}}(\vec{k},-\q,\q) P_{11}(q;t,t) \non \\
     +\tilde{c}_{\delta,3}(t) \; 3\; &P_{11}(k;t,t) \int \frac{d^3q}{(2\pi)^3} \;\widehat{c}^{(3)}_{{\delta,3}}(\vec{k},-\q,\q) P_{11}(q;t,t) \non \\
     +\tilde{c}_{\delta^2,1(3)}(t) \; 3\; &P_{11}(k;t,t) \int \frac{d^3q}{(2\pi)^3} \;\widehat{c}^{(3)}_{{\delta^2,1}}(\vec{k},-\q,\q) P_{11}(q;t,t) \non \\    
     +\tilde{c}_{\delta^2,2}(t) \; 3\; &P_{11}(k;t,t) \int \frac{d^3q}{(2\pi)^3} \;\widehat{c}^{(3)}_{{\delta^2,2}}(\vec{k},-\q,\q) P_{11}(q;t,t) \non \\
     +\tilde{c}_{\delta^3,1}(t) \; 3\; &P_{11}(k;t,t) \int \frac{d^3q}{(2\pi)^3} \;\widehat{c}^{(3)}_{{\delta^3,1}}(\vec{k},-\q,\q) P_{11}(q;t,t) \non \\
     +\tilde{c}_{s^2,2}(t) \; 3\; &P_{11}(k;t,t) \int \frac{d^3q}{(2\pi)^3} \;\widehat{c}^{(3)}_{{s^2,2}}(\vec{k},-\q,\q) P_{11}(q;t,t) \, ,
\end{align}
and for the second term in eq.~(\ref{eq:Phh}):
\begin{align}
2\int \frac{d^3q}{(2\pi)^3} \;\left[K^{(2)}_s(\vec{k}-\q,\q)\right]^2 & P_{11}(q;t,t)P_{11}( |\vec{k}-\vec{q}|;t,t) =  \non \\
      \tilde{c}_{\delta,1}^2(t)\; 2 &\int \frac{d^3q}{(2\pi)^3} \;
                     \left[\widehat{c}^{(2)}_{{\delta,1}}(\vec{k}-\q,\q)\right]^2P_{11}(q;t,t)P_{11}( |\vec{k}-\vec{q}|;t,t) \non \\
   +\tilde{c}_{\delta,2(2)}^2(t)\; 2 &\int \frac{d^3q}{(2\pi)^3} \;
                    \left[\widehat{c}^{(2)}_{{\delta,2}}(\vec{k}-\q,\q)\right]^2P_{11}(q;t,t)P_{11}( |\vec{k}-\vec{q}|;t,t) \non \\
   +\tilde{c}_{\delta^2,1(2)}^2(t)\; 2 &\int \frac{d^3q}{(2\pi)^3} \;
                     \left[\widehat{c}^{(2)}_{{\delta^2,1}}(\vec{k}-\q,\q)\right]^2P_{11}(q;t,t)P_{11}( |\vec{k}-\vec{q}|;t,t) \non \\
   +\tilde{c}_{\delta,1}(t)\tilde{c}_{\delta,2(2)}(t)\; 4 &\int \frac{d^3q}{(2\pi)^3} \;
                     \widehat{c}^{(2)}_{{\delta,1}}(\vec{k}-\q,\q) \widehat{c}^{(2)}_{{\delta,2}}(\vec{k}-\q,\q)P_{11}(q;t,t)P_{11}( |\vec{k}-\vec{q}|;t,t) \non \\
   +\tilde{c}_{\delta,1}(t)\tilde{c}_{\delta,2(2)}(t)\; 4 &\int \frac{d^3q}{(2\pi)^3} \;
                    \widehat{c}^{(2)}_{{\delta,1}}(\vec{k}-\q,\q)  \widehat{c}^{(2)}_{{\delta^2,1}}(\vec{k}-\q,\q)P_{11}(q;t,t)P_{11}( |\vec{k}-\vec{q}|;t,t) \non \\
   +\tilde{c}_{\delta,2(2)}(t) \tilde{c}_{\delta^2,1(2)}(t)\; 4 &\int \frac{d^3q}{(2\pi)^3} \;
                    \widehat{c}^{(2)}_{{\delta,2}}(\vec{k}-\q,\q)  \widehat{c}^{(2)}_{{\delta^2,1}}(\vec{k}-\q,\q)P_{11}(q;t,t)P_{11}( |\vec{k}-\vec{q}|;t,t)\ .
\label{eq:kkhaloterm}
\end{align}


\section{Results for one additional halo mass}
\label{app:bbins}

\begin{figure*}[t!]
   \begin{center}
   \hspace*{-0.5cm}
   \includegraphics[scale=0.5]{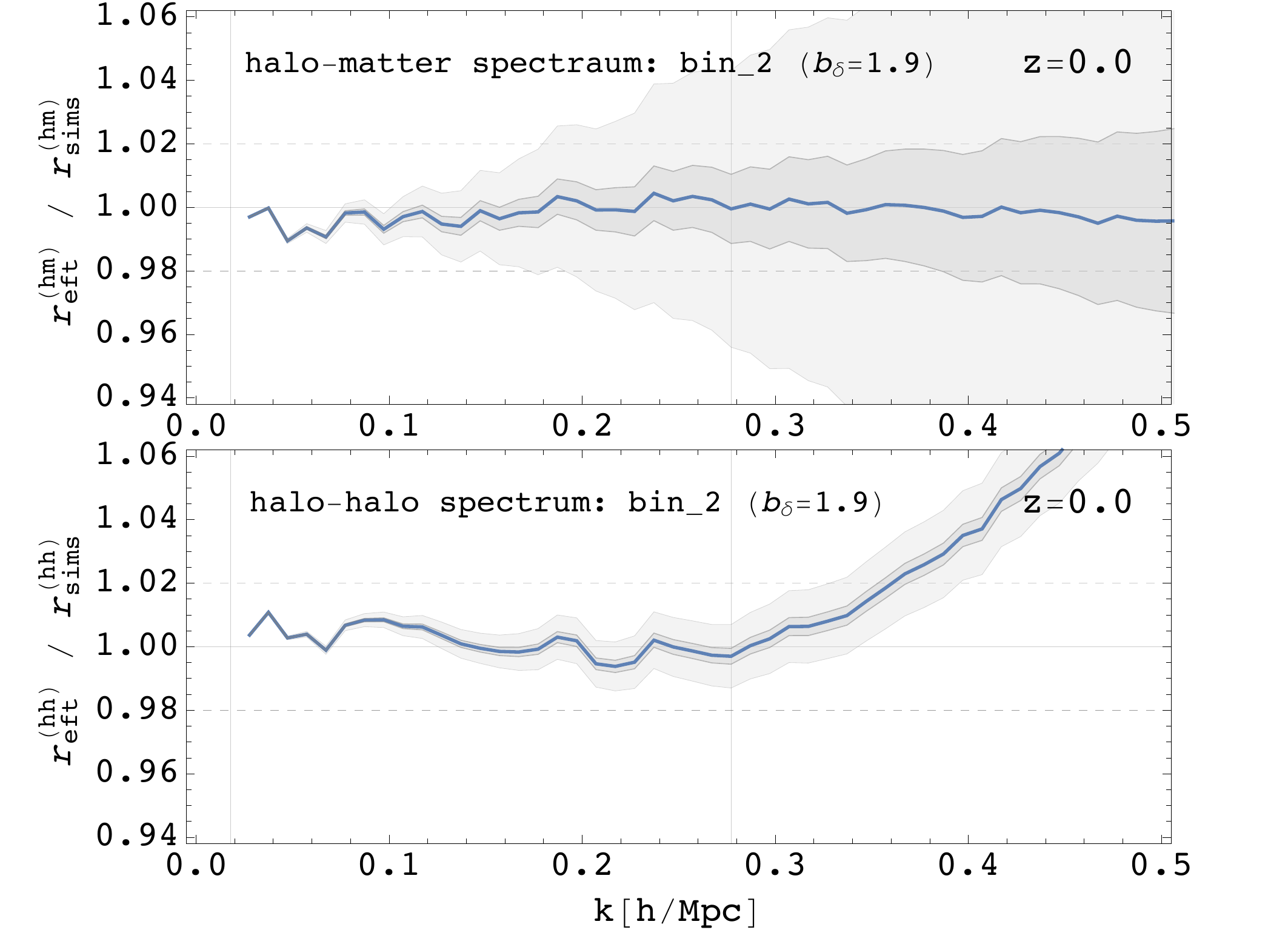}
   \end{center}
   \vspace*{-0.5cm}
   \caption{\small Same as Fig.~\ref{fig:PSbin0} for the higher halo mass bin2.
   }
   \label{fig:PSbin123}
\end{figure*}

In this appendix we present the rest of the power spectra (for higher mass bin2) and bispecta plots.
In figure \ref{fig:PSbin123} we show the halo-halo and halo-matter power spectrum results 
for this higher mass bin2, similar as in figure \ref{fig:PSbin0}.

\begin{figure*}[t!]
   \begin{center}
   \hspace*{-0.5cm}
   \includegraphics[scale=0.58]{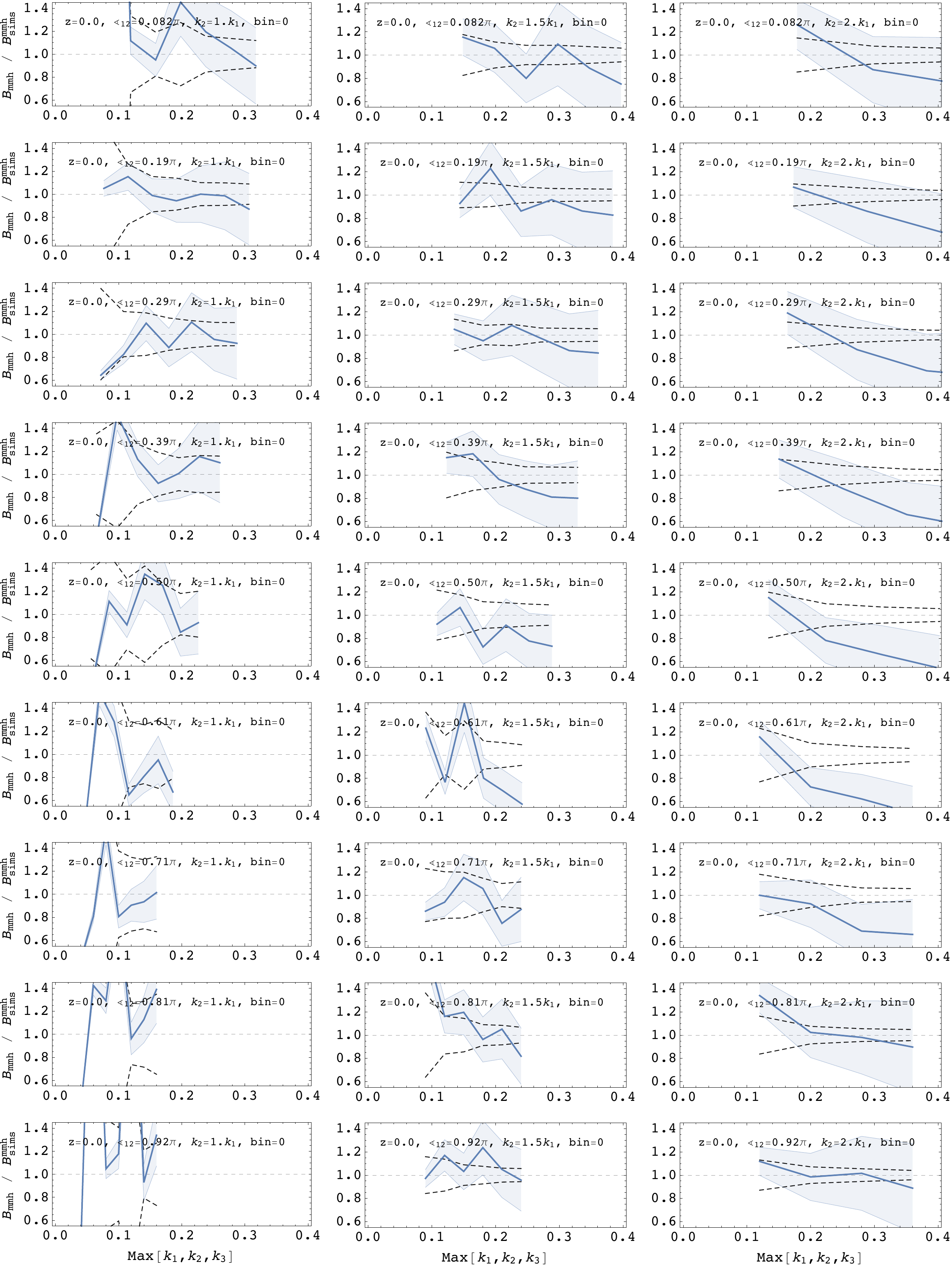}
   \end{center}
   \vspace*{-0.5cm}
   \caption{\small Halo-matter-matter bispectrum results for bin0 ($b_{\df 1}=1.0$).
   All the lines are divided by the simulations data. Blue line is the tree level theory 
   prediction. Black dashed lines represent the simulation error bars as given in equation~(\ref{eq:BErrror}),
   and gray lines represent the estimate of the theory errors given by the equations (\ref{eq:Bispec_Errors}).}
   \label{fig:BiMMHbin0}
\end{figure*}

\begin{figure*}[t!]
   \begin{center}
   \hspace*{-0.5cm}
   \includegraphics[scale=0.58]{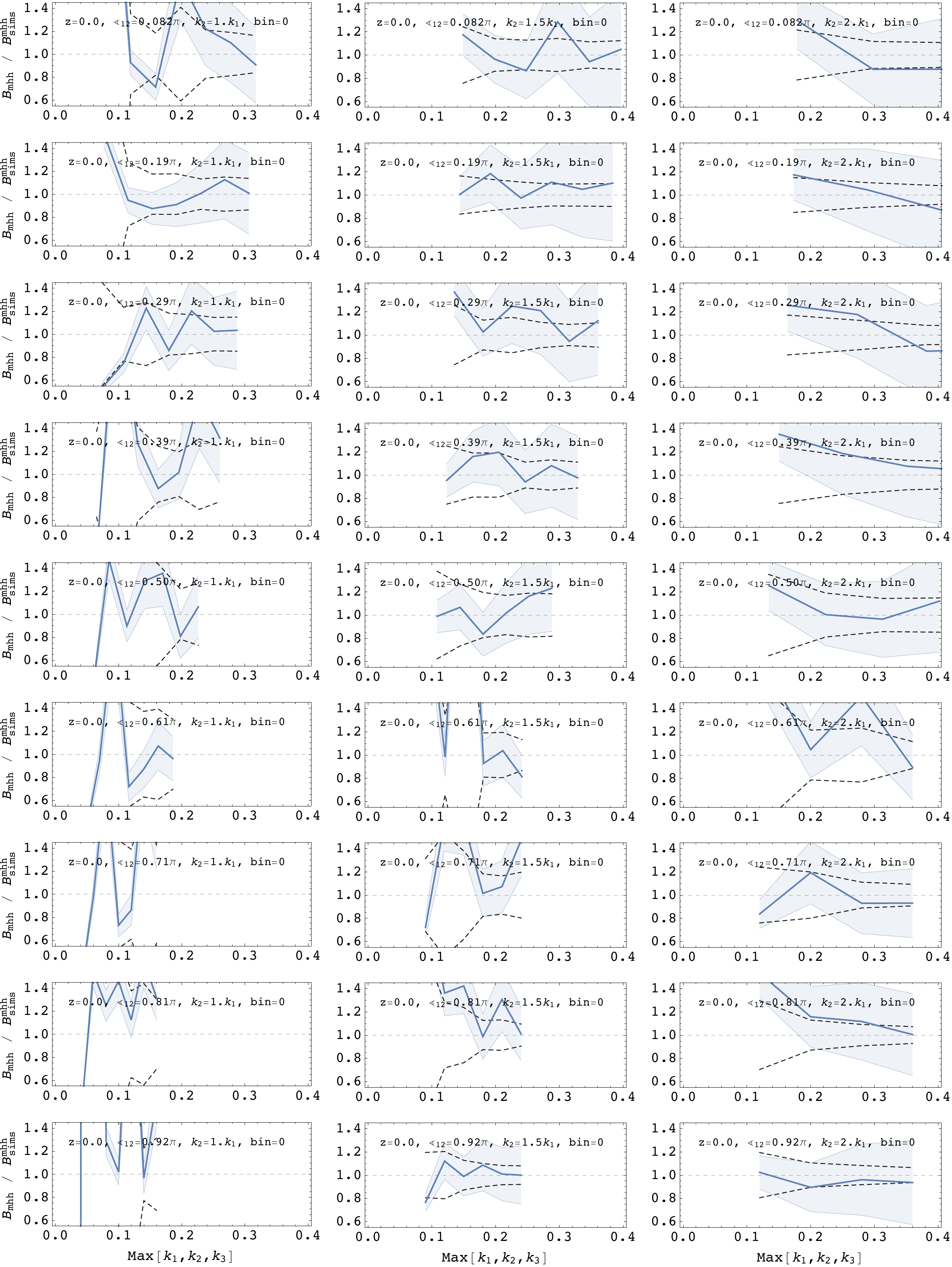}
   \end{center}
   \vspace*{-0.5cm}
   \caption{\small Halo-halo-matter bispectrum results bin0 ($b_{\df 1}=1.0$). 
   Lines are the same as in figure \ref{fig:BiMMHbin0}. 
   }
   \label{fig:BiMHHbin0}
\end{figure*}

\begin{figure*}[t!]
   \begin{center}
   \hspace*{-0.5cm}
   \includegraphics[scale=0.58]{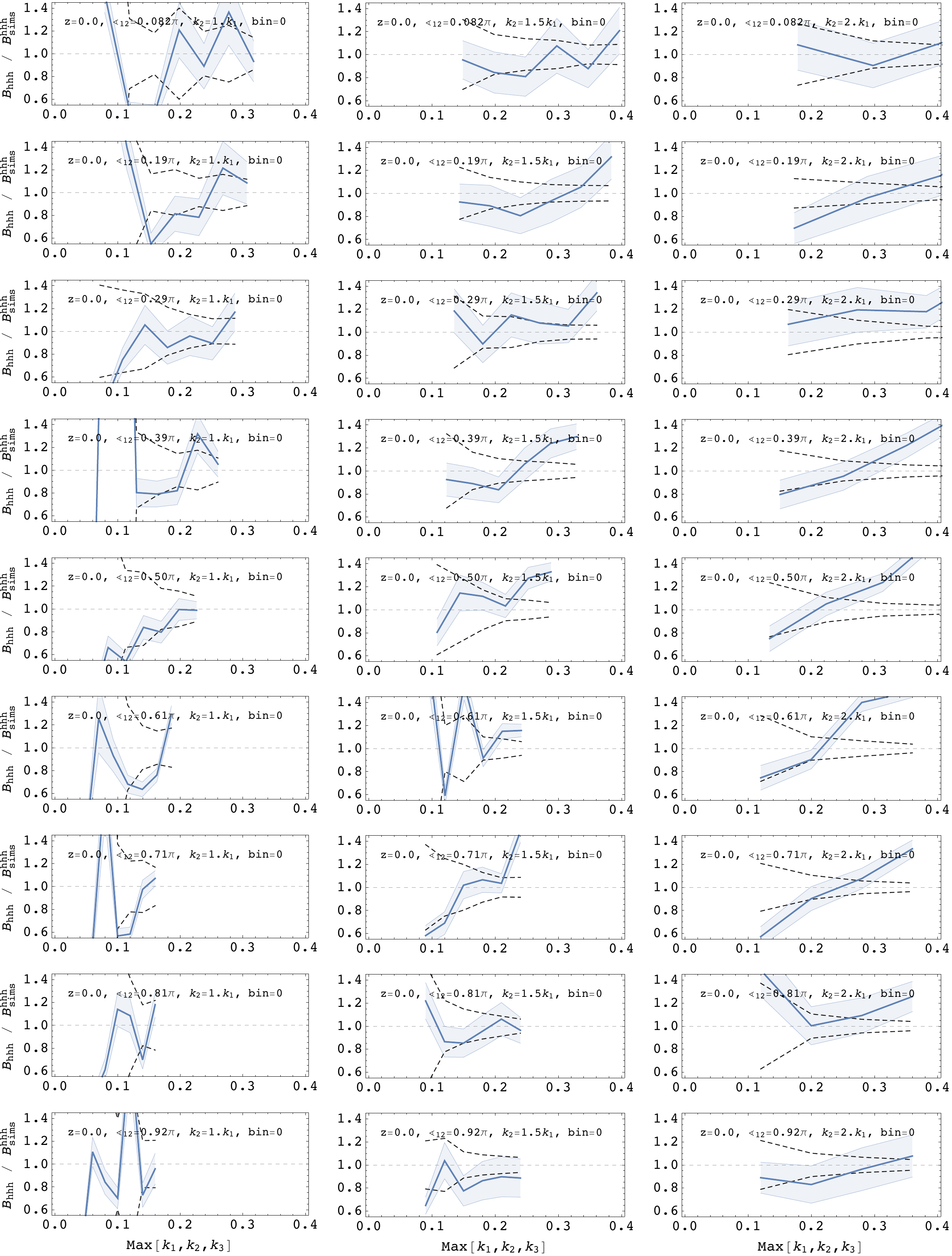}
   \end{center}
   \vspace*{-0.5cm}
   \caption{\small Halo-halo-halo bispectrum results bin0 ($b_{\df 1}=1.0$). 
   Lines are the same as in figure \ref{fig:BiMMHbin0}.
   }
   \label{fig:BiHHHbin0}
\end{figure*}

\begin{figure*}[t!]
   \begin{center}
   \hspace*{-0.5cm}
   \includegraphics[scale=0.58]{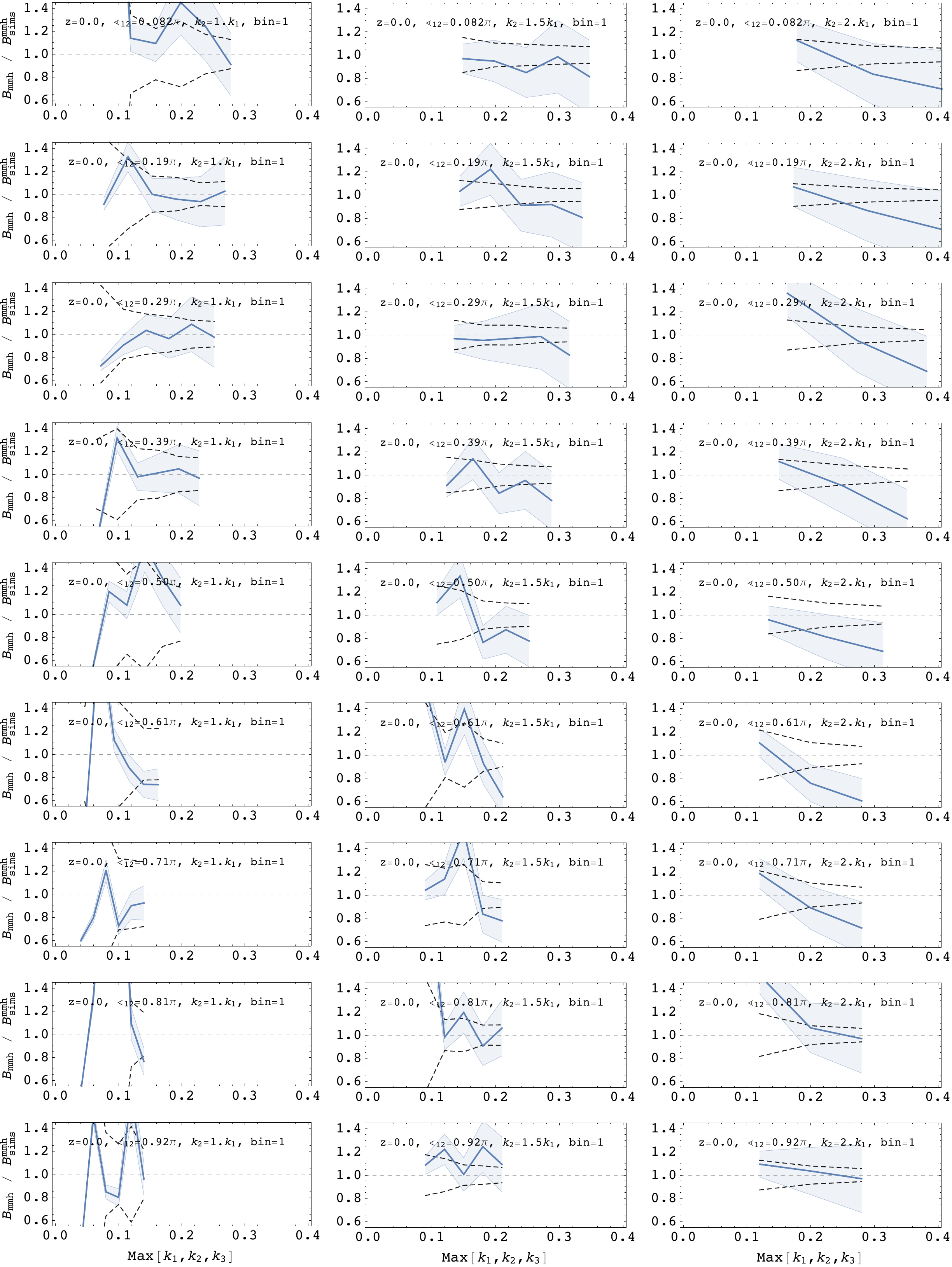}
   \end{center}
   \vspace*{-0.5cm}
   \caption{\small Halo-matter-matter bispectrum results bin1 ($b_{\df 1}=1.33$). 
   Lines are the same as in figure \ref{fig:BiMMHbin0}.
   }
   \label{fig:BiMMHbin1}
\end{figure*}

\begin{figure*}[t!]
   \begin{center}
   \hspace*{-0.5cm}
   \includegraphics[scale=0.58]{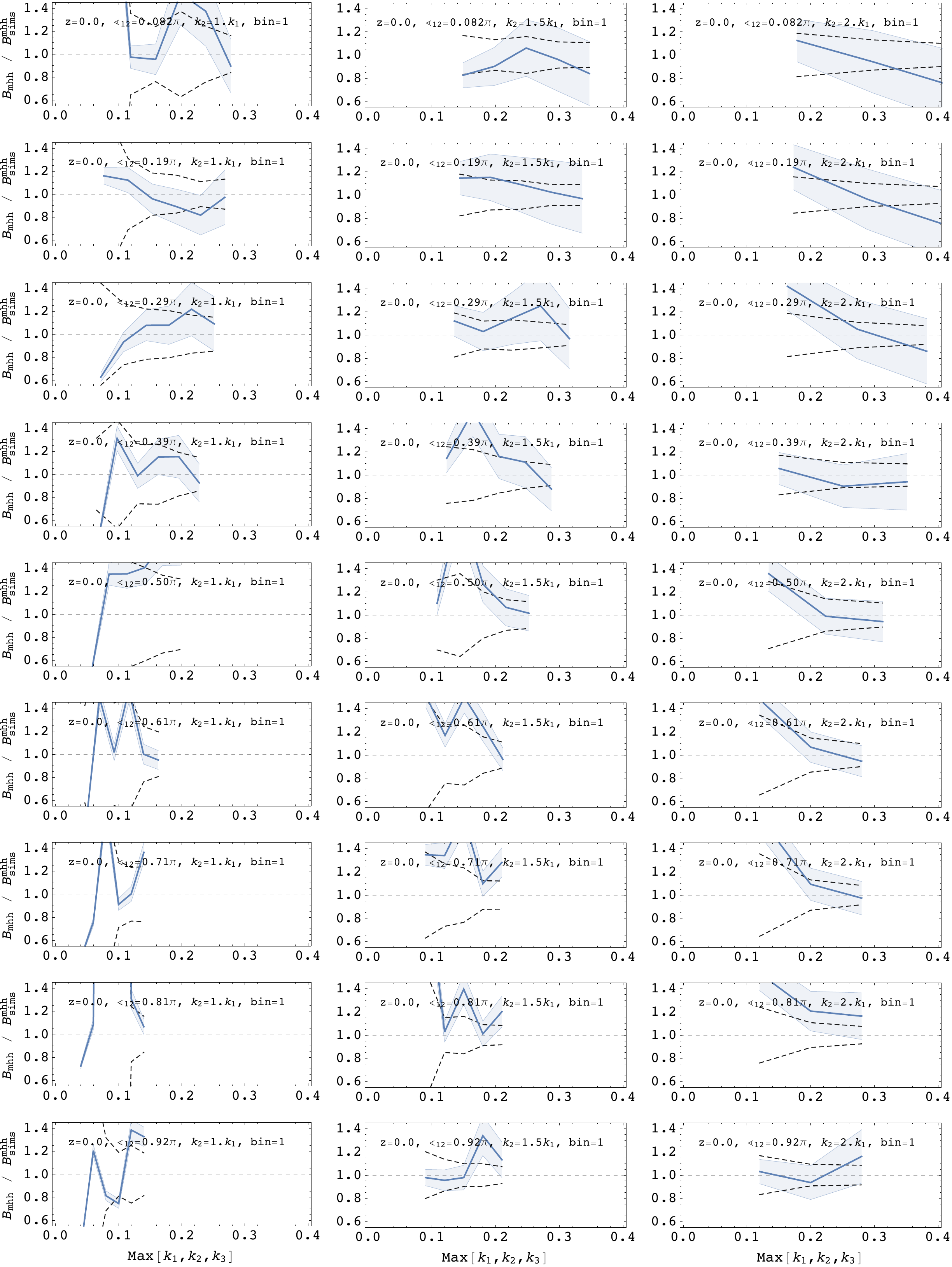}
   \end{center}
   \vspace*{-0.5cm}
   \caption{\small Halo-halo-matter bispectrum results bin1 ($b_{\df 1}=1.33$). 
   Lines are the same as in figure \ref{fig:BiMMHbin0}.
   }
   \label{fig:BiMHHbin1}
\end{figure*}

\begin{figure*}[t!]
   \begin{center}
   \hspace*{-0.5cm}
   \includegraphics[scale=0.58]{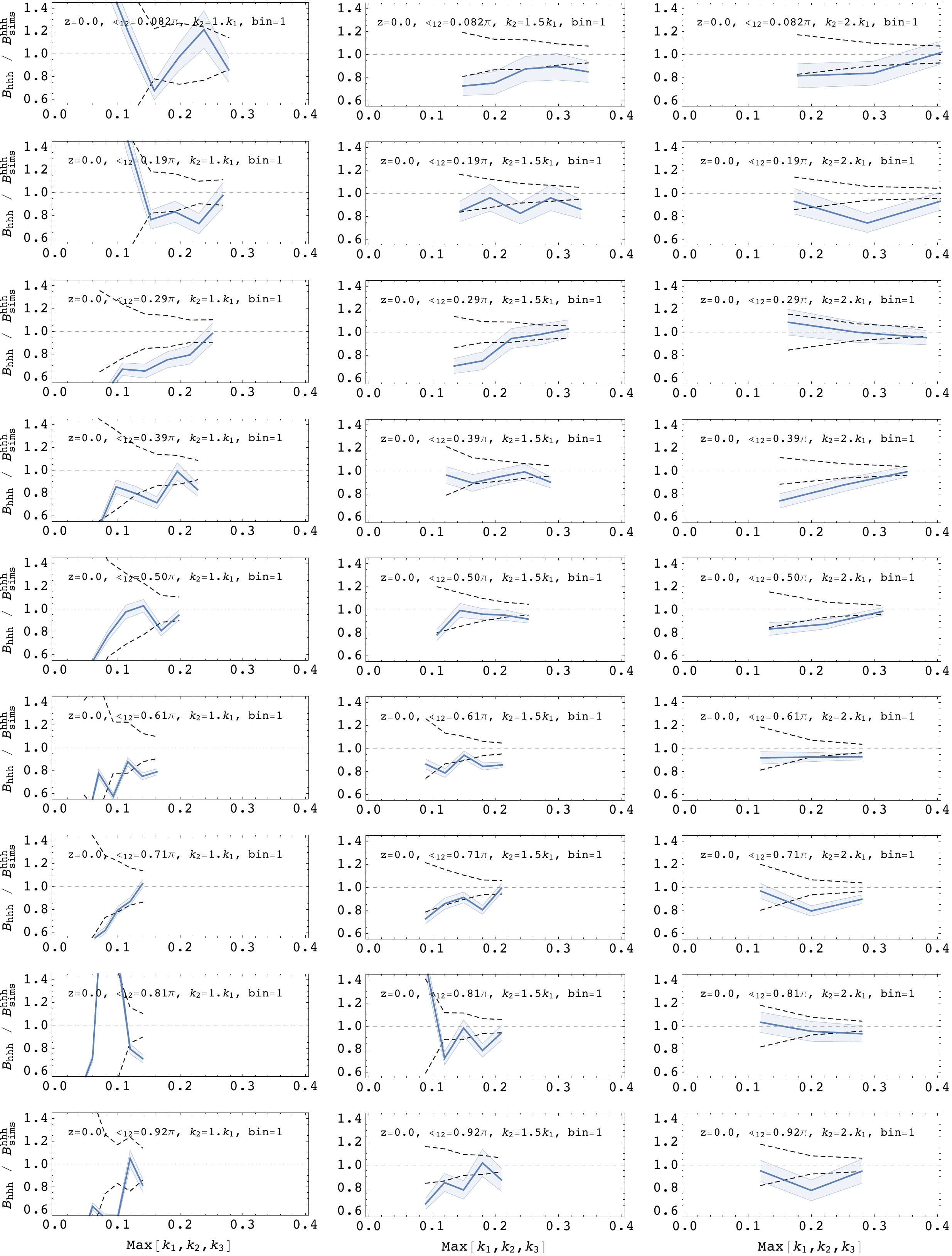}
   \end{center}
   \vspace*{-0.5cm}
   \caption{\small Halo-halo-halo bispectrum results bin1 ($b_{\df 1}=1.33$). 
   Lines are the same as in figure \ref{fig:BiMMHbin0}.
   }
   \label{fig:BiHHHbin1}
\end{figure*}

\begin{figure*}[t!]
   \begin{center}
   \hspace*{-0.5cm}
   \includegraphics[scale=0.58]{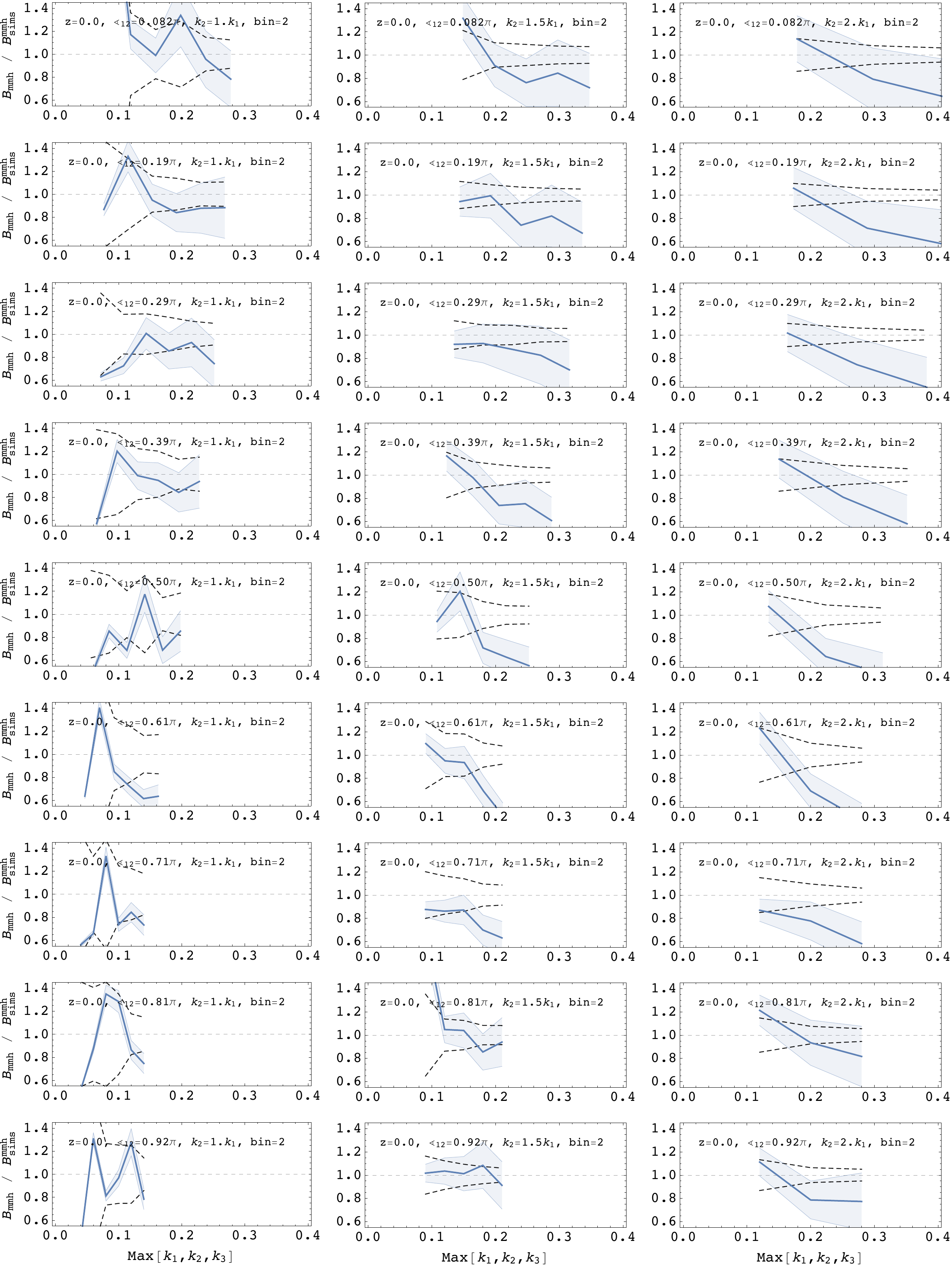}
   \end{center}
   \vspace*{-0.5cm}
   \caption{\small Halo-matter-matter bispectrum results bin2 ($b_{\df 1}=1.95$). 
   Lines are the same as in figure \ref{fig:BiMMHbin0}.
   }
   \label{fig:BiMMHbin2}
\end{figure*}

\begin{figure*}[t!]
   \begin{center}
   \hspace*{-0.5cm}
   \includegraphics[scale=0.58]{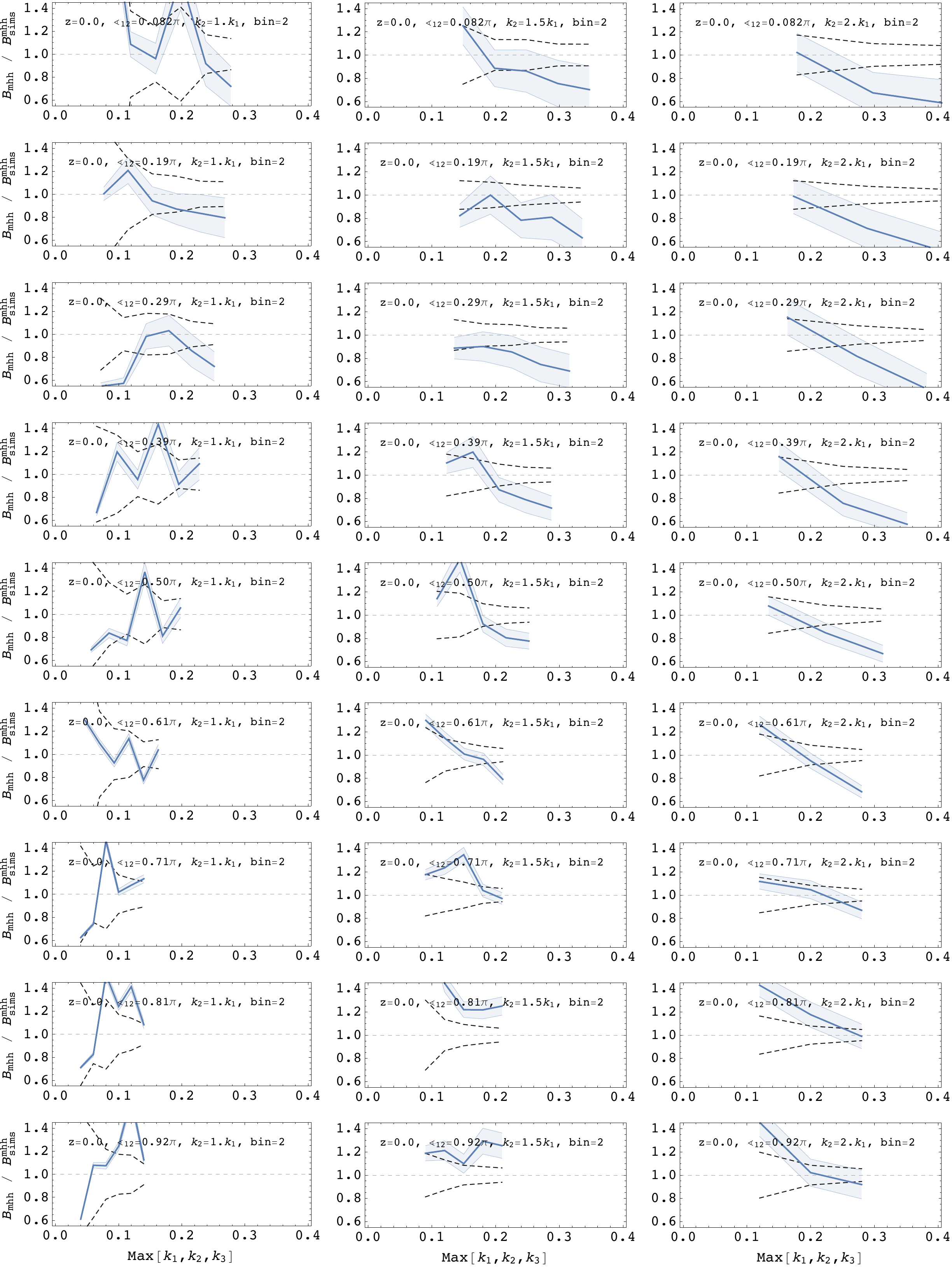}
   \end{center}
   \vspace*{-0.5cm}
   \caption{\small Halo-halo-matter bispectrum results bin2 ($b_{\df 1}=1.95$). 
   Lines are the same as in figure \ref{fig:BiMMHbin0}.
   }
   \label{fig:BiMHHbin2}
\end{figure*}

\begin{figure*}[t!]
   \begin{center}
   \hspace*{-0.5cm}
   \includegraphics[scale=0.58]{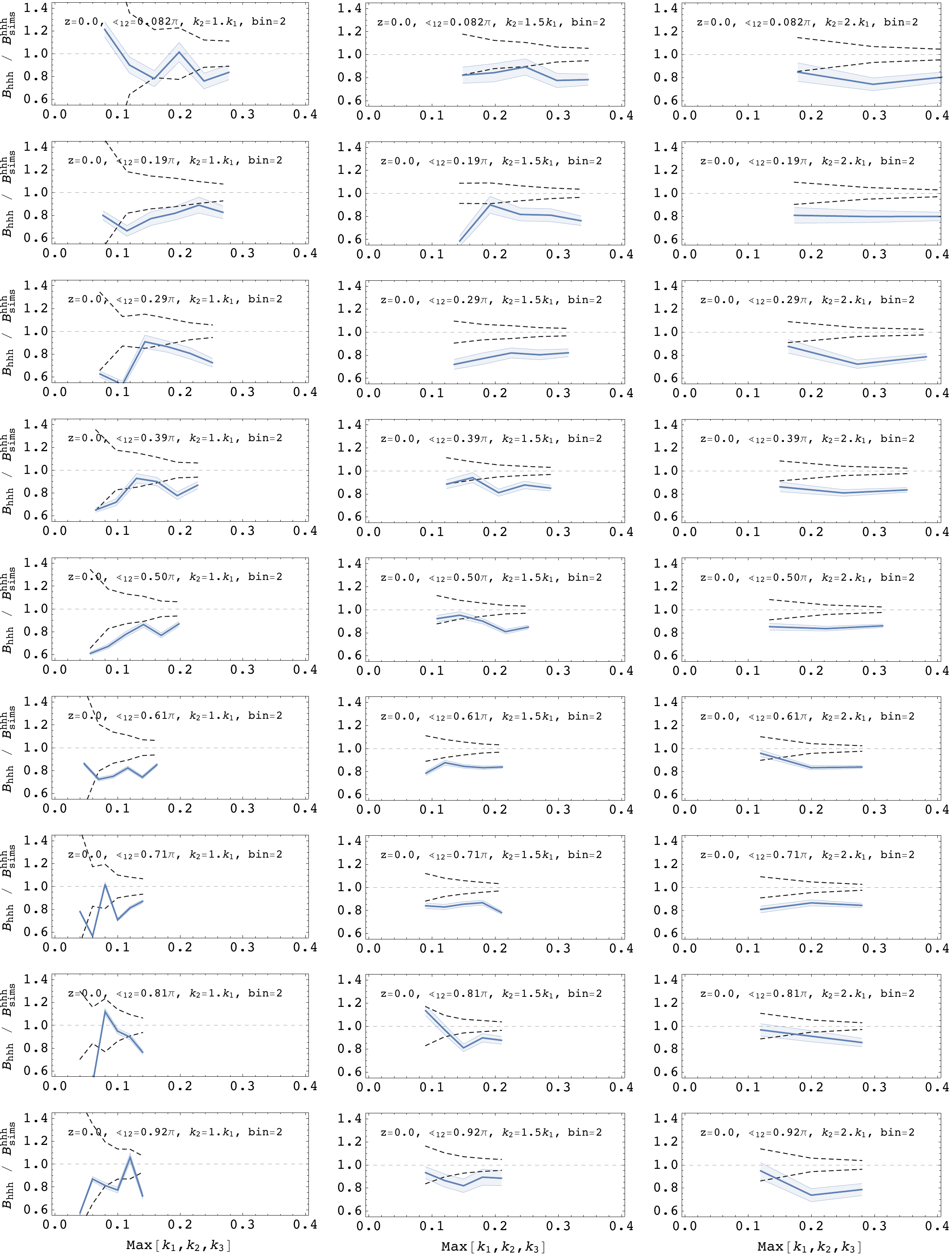}
   \end{center}
   \vspace*{-0.5cm}
   \caption{\small Halo-halo-halo bispectrum results bin2 ($b_{\df 1}=1.95$). 
   Lines are the same as in figure \ref{fig:BiMMHbin0}.
   }
   \label{fig:BiHHHbin2}
\end{figure*}

In the rest we present the figures for the bispectra halo-halo-halo, halo-halo-matter and halo-matter-matter
bispectra as given in the equations (\ref{eq:Bisspec}) with the best fitted parameters in table~\ref{tb:bias}.
bin0 results are given in figures \ref{fig:BiMMHbin0}, \ref{fig:BiMHHbin0} and \ref{fig:BiHHHbin0}.
bin1 results are given in figures \ref{fig:BiMMHbin1}, \ref{fig:BiMHHbin1} and \ref{fig:BiHHHbin1}.
bin2 results are given in figures \ref{fig:BiMMHbin2}, \ref{fig:BiMHHbin2} and \ref{fig:BiHHHbin2}.

The attentive reader will notice that some of the plots in e.g. Fig.~(8, 9, 10) noticeably deviate from unity. 
In interpreting these plots one should keep in mind the combined effect of the relatively large simulation 
error estimate due to cosmic variance (black dashed lines) and the fact that we model the bispectrum only 
up to tree level, and so the $k$-reach is limited. For the sake of completeness, in these plots we show a 
number of bispectrum configurations even though for these we are well outside the $k$-reach of our model, 
whilst for the corresponding longer scales predicted by the EFT there are no available simulation data. For 
these particular configurations then, a conspicuous deviation from unity is to be expected. In fact, as shown 
in the main text, the $p$-value of the fit of the theory to data is very large up to the $k$-reach of the theory.

\vfill

\bibliographystyle{JHEP}


\end{document}